\documentclass[preprintnumbers,superscriptaddress,amsmath,amssymb,nofootinbib,twocolumn,showpacs]{revtex4-1}

\usepackage{multirow}
\usepackage{dcolumn}
\usepackage{bm}
\usepackage{graphicx}
\usepackage{threeparttable}
\usepackage{float}

\usepackage[normalem]{ulem}

\usepackage{footmisc}

\usepackage{color}
\usepackage{natbib}
\usepackage{twoopt}

\newcommand{\revise}[1]{#1}

\newcommand{\repl}[2]{#2}

\newcommand{\rmv}[1]{}

\newcommand{\beq}{\begin{equation}}
\newcommand{\eeq}{\end{equation}}
\newcommand{\bsubeq}{\begin{subequations}}
\newcommand{\esubeq}{\end{subequations}}
\newcommand{\ben}{\begin{eqnarray}}
\newcommand{\een}{\end{eqnarray}}
\newcommand{\balign}{\begin{align}}
\newcommand{\ealign}{\end{align}}
\newcommand{\balt}{\begin{alignat}}
\newcommand{\ealt}{\end{alignat}}
\newcommand{\bi}{\begin{itemize}}
\newcommand{\ei}{\end{itemize}}
\newcommand{\nn}{\nonumber}

\newcommand{\ie}{\mbox{\it i.e.}}
\newcommand{\eg}{\mbox{\it e.g.}}

\newcommand{\citeeq}[1]{Eq.~(\ref{#1})}

\newcommand{\citeeqs}[1]{Eqs.~(\ref{#1})}

\newcommand{\citeeqp}[1]{Eq.~\ref{#1}}

\newcommand{\citesec}[1]{Sect.~\ref{#1}}

\newcommand{\citeapp}[1]{Appendix~\ref{#1}}

\newcommand{\citetab}[1]{Table~\ref{#1}}
\newcommand{\citefig}[1]{Fig.~\ref{#1}}

\newcommand{\dd}{\ifmmode {\rm d} \else {d}\fi}

\newcommand{\rhoc}{{\ifmmode {\rho_{\rm c}} \else $\rho_{\rm c}$\fi}}
\newcommand{\rhos}{{\ifmmode {\rho_{\rm s}} \else $\rho_{\rm s}$\fi}}
\newcommand{\rhotilde}{{\ifmmode \tilde{\rho} \else $\tilde{\rho}$\fi}}
\newcommand{\rs}{{\ifmmode {r_{\rm s}} \else $r_{\rm s}$\fi}}
\newcommand{\rt}{{\ifmmode {r_{\rm t}} \else $r_{\rm t}$\fi}}
\newcommand{\Msun}{\mbox{$M_{\odot}$}}
\newcommand{\msun}{\mbox{$M_{\odot}$}}

\newcommand{\mchi}{{\ifmmode m_\chi \else $m_\chi$\fi}}
\newcommand{\sigv}{{\ifmmode \langle \sigma v \rangle \else $\langle \sigma v \rangle$\fi}}

\newcommand{\angleave}[2]{\ifmmode \left \langle{#1} \right \rangle_{#2} \else $\left\langle{#1}\right\rangle_{#2}$ \fi}

\newcommand{\gevpcc}{{\ifmmode {\rm GeV/cm^{3}} \else ${\rm GeV/cm^{3}}$\fi}}

\newcommand{\lcdm}{{\ifmmode \Lambda{\rm CDM} \else $\Lambda{\rm CDM}$\fi}}

\def\aap{A\&A}%

\def\aj{AJ}%

\def\apj{ApJ}%
\def\apjl{ApJ}%
\def\apjs{ApJS}%
\def\araa{ARA\&A}%
\def\epjc{Europ.~Phys.~J.~C}%
\def\jcap{J. Cosmology Astropart. Phys.}%
\def\mnras{MNRAS}%
\def\na{New A}%
\def\nat{Nature}%
\def\pdu{Phys.~Dark~Univ.}%
\def\physrep{Phys.~Rep.}%
\def\prd{Phys.~Rev.~D}%
\def\prl{Phys.~Rev.~Lett.}%
\usepackage[pdftex,
  bookmarks=false,
  breaklinks=true,%
  colorlinks=true,%
  linkcolor=red,
  urlcolor=blue,
  citecolor=blue,
  pdfauthor={Facchinetti, Lavalle, and Stref},%
  pdftitle={Template for article in XXX}%
]{hyperref}

\DeclareGraphicsExtensions{.png,.pdf}
\graphicspath{{./}{FinalFigs/}}

\begin{document}

\title{Statistics for dark matter subhalo searches in gamma rays from a
  kinematically constrained population model: Fermi-LAT-like telescopes}

\author{Ga\'etan Facchinetti}
\email{gaetan.facchinetti@umontpellier.fr}
\altaffiliation[Current affiliation: ]
{Service de Physique Th\'eorique, Universit\'e Libre de Bruxelles, Boulevard du Triomphe, CP225, 1050 Brussels --- Belgium}
\affiliation{Laboratoire Univers \& Particules de Montpellier (LUPM),
  CNRS \& Universit\'e de Montpellier (UMR-5299),
  Place Eug\`ene Bataillon,
  F-34095 Montpellier Cedex 05 --- France}

\author{Julien Lavalle}
\email{lavalle@in2p3.fr}
\affiliation{Laboratoire Univers \& Particules de Montpellier (LUPM),
  CNRS \& Universit\'e de Montpellier (UMR-5299),
  Place Eug\`ene Bataillon,
  F-34095 Montpellier Cedex 05 --- France}

\author{Martin Stref}
\email{martin.stref@lapth.cnrs.fr}
\affiliation{LAPTh,
  Universit\'e Savoie Mont Blanc \& CNRS,
  Chemin de Bellevue,
  74941 Annecy Cedex --- France}

\begin{abstract}
Cold dark matter subhalos are expected to populate galaxies in numbers. If dark matter self-annihilates, these objects turn into prime targets for indirect searches, in particular with gamma-ray telescopes. Incidentally, the Fermi-LAT catalog already contains many unidentified sources that might be associated with subhalos. In this paper, we \revise{determine the probability for subhalos to} be identified as gamma-ray pointlike sources from their predicted distribution properties. We use a semi-analytical model for the Galactic subhalo population, which, in contrast to cosmological simulations, can be made fully consistent with current kinematic constraints in the Milky Way and has no resolution limit. The model incorporates tidal stripping \revise{effects from} a realistic distribution of baryons in the Milky Way. The same baryonic distribution contributes a diffuse gamma-ray foreground which adds up to that, often neglected in subhalo searches, \revise{generated} by the smooth dark matter and the unresolved subhalos. This configuration implies a correlation between pointlike subhalo signals and diffuse background. Based on this \revise{semi-analytical} modeling, we \revise{generate mock gamma-ray data assuming an idealized telescope resembling Fermi-LAT and perform a likelihood analysis} to estimate the current and future sensitivity to subhalos \revise{in the relevant parameter space}. We find a number of \revise{detectable} subhalos of order ${\cal O}(<1)$ for optimistic model parameters and a WIMP mass of 100~GeV, maximized for a cored host halo. This barely provides support to the current interpretation of several Fermi unidentified sources as subhalos. We also find \revise{it} more likely to detect the smooth Galactic halo itself before subhalos, should dark matter in the GeV-TeV mass range self-annihilate through $s$-wave processes.
\end{abstract}

\maketitle
\preprint{LUPM:20-025}
\section{Introduction}
\label{sec:intro}
While under experimental or observational pressure, the thermal dark matter (DM) scenario is still considered as appealing owing to its simple production mechanism and to the fact that it is within reach of current experiments. A typical realization amounts to assuming that DM is made of exotic particles with masses and couplings to standard model particles such that they can be produced from the hot plasma in the early universe, and to selecting model parameters for which DM is cold \cite{Peebles1982,BlumenthalEtAl1984,BertoneEtAl2018} and with a predicted cosmological abundance that matches with the one measured by cosmological probes \cite{LeeEtAl1977a,BondEtAl1982,BinetruyEtAl1984a,SrednickiEtAl1988}. If there is no matter-antimatter asymmetry in the dark sector, and if DM is driven to chemical equilibrium before freezing out, then weakly-interacting massive particles (WIMPs) arise as prototypical self-annihilating DM candidates, leading to a diversity of potentially observable signatures\footnote{One of the main theoretical supports for WIMPs was that it was independently motivated by solutions to the so-called electroweak hierarchy problem in particle physics. The fact that no new particles have been discovered at the LHC has strongly affected approaches to that issue, see \eg~\cite{Giudice2017}, and motivations for WIMPs are now mostly phenomenological \cite{ArcadiEtAl2017,LeaneEtAl2018}.} \cite{GunnEtAl1978,SilkEtAl1984,GoodmanEtAl1985,PrimackEtAl1988,JungmanEtAl1996,Feng2010}. In this article, we focus on indirect DM searches \cite{Bergstroem2000,Feng2010,LavalleEtAl2012} with gamma rays, and therefore assume that DM self-annihilates in DM halos nearly at rest and into standard model particles, producing gamma rays through direct emission, hadronization of the final states, or bremsstrahlung \cite{BringmannEtAl2012c}. This implicitly restricts the available WIMP parameter space to $s$-wave annihilation processes (typically mediated by pseudo-scalar interactions if DM is made of fermions), for which the annihilation rate does not depend on DM particle velocities. Other parts of the WIMP parameter space (\eg~scalar interactions) can still be probed by indirect detection techniques \cite{LiuEtAl2016,BoudaudEtAl2019a}, but are more efficiently so with direct detection \cite{LewinEtAl1996,JungmanEtAl1996,FreeseEtAl2013} and at colliders \cite{FairbairnEtAl2007,AbdallahEtAl2015,ArcadiEtAl2017}. We also restrict the target space by focusing on searches in the Milky Way (MW) only \cite{Strigari2013}.

The generically rather small scattering rate between WIMPs and the hot plasma in the early universe leads to a very small cutoff scale in the matter power spectrum, implying a typical mass ranging from $10^{-3}$-$10^{-12}$~\Msun\ for the first DM structures to collapse in the matter cosmological era \cite{SchmidEtAl1999,BoehmEtAl2001,ChenEtAl2001,HofmannEtAl2001,BerezinskyEtAl2003,GreenEtAl2004,Bertschinger2006a,BringmannEtAl2007a}. In the standard hierarchical picture of structure formation \cite{PressEtAl1974,BondEtAl1991a,LaceyEtAl1993}, these first minihalos, or subhalos, merge into larger DM halos, but a significant fraction of them survives tidal disruption and populates galactic halos in numbers today \cite{DiemandEtAl2005a,IshiyamaEtAl2010,BerezinskyEtAl2014,StrefEtAl2017,IshiyamaEtAl2020a}. These DM inhomogeneities have long be invoked as potential boosters of the DM annihilation rate in galaxies, enhancing the production of gamma rays and antimatter cosmic rays \cite{SilkEtAl1993,BergstroemEtAl1999a,LavalleEtAl2007,LavalleEtAl2008,PieriEtAl2011}. They could also enhance the gamma-ray power spectrum on specific angular scales \cite{AndoEtAl2006,FornasaEtAl2016}. They actually also represent interesting point-source targets for gamma-ray telescopes \cite{TasitsiomiEtAl2002,StoehrEtAl2003,AloisioEtAl2004a,PieriEtAl2005,KuhlenEtAl2008,AndersonEtAl2010,PieriEtAl2011,HuettenEtAl2016}, with essentially no X-ray nor radio counterparts (but see \cite{BaltzEtAl2004}). This possibility has generated a particular attention in the recent years as the Fermi-LAT satellite has enriched its catalog with many unidentified and unassociated sources \cite{FermiLATCollab2015,FermiLAT2019}, some being interpreted as potential DM subhalos \cite{BelikovEtAl2012,BertoniEtAl2015,SchoonenbergEtAl2016,MirabalEtAl2016,HooperEtAl2017,CaloreEtAl2017,CaloreEtAl2019,GlawionEtAl2019,CoronadoBlazquezEtAl2019,CoronadoBlazquezEtAl2019a}.

In this study, we take advantage of the recent analytical Galactic subhalo population model developed in Ref.~\cite{StrefEtAl2017} (SL17 henceforth)---see complementary analytical approaches in \eg~Refs~\cite{IshiyamaEtAl2020a,HiroshimaEtAl2018,BartelsEtAl2015,ZavalaEtAl2014a,Benson2012,vandenBoschEtAl2005a}. This model was built to be consistent with both structure formation theory \cite{BerezinskyEtAl2014,ZavalaEtAl2019a} and kinematic constraints on the MW similar to those discussed in Ref.~\cite{McMillan2017}. Some gamma-ray properties of this model were already derived in Ref.~\cite{HuettenEtAl2019} using the Clumpy code \cite{CharbonnierEtAl2012,HuettenEtAl2019a}, which aimed at comparing them with predictions from cosmological simulations \cite{KelleyEtAl2019} (so-called MW-like simulations, but obviously with DM and baryonic distributions that may significantly depart from the real MW), but without fully addressing the detectability of individual objects in a realistic diffuse foreground. This issue was partly covered in \cite{CaloreEtAl2019}. Here, we want to inspect the potential of Fermi-like gamma-ray telescopes to detect subhalos in such a model, but going farther than previous studies in the attention given to the contribution of DM annihilation itself to the diffuse background. The model includes a subhalo population, a smooth dark matter halo, and a baryonic distribution, all made consistent with kinematic constraints, and the gravitational tides that prune or disrupt subhalos are calculated from the very same components (see SL17 for details). This internal self-consistency leads to a spatial correlation between the subhalos, the smooth DM, and the baryonic content, which affects the observational properties of the former through the contribution to the diffuse emission of the latter. Baryons induce gravitational tides that deplete the subhalo population and select the most concentrated objects. Besides, they set the intensity of the Galactic gamma-ray foreground (mostly the pionic component), which also plays a role in the balance between diffuse emission and potential pointlike emissions from subhalos. Finally, assessing the detectability of subhalos should also account for the fact that the diffuse DM emission is also bounded by current constraints to be less than the level of Galactic foreground statistical fluctuations \cite{AbdoEtAl2010d,CirelliEtAl2010,BlanchetEtAl2012,FermiLATEtAl2012,BringmannEtAl2012c,EssigEtAl2013a,CirelliEtAl2015,FornasaEtAl2015,ChangEtAl2018}. This means that part of the naively available parameter space is actually already excluded, and this can be fully characterized in a complete model. We will show that self-consistently combining all these ingredients leads to interesting, though not necessarily optimistic, consequences in terms of subhalo detectability.

The paper develops according to a very pedestrian approach and is organized as follows. We begin by quickly reviewing our global Galactic model in \citesec{sec:model}. In \citesec{sec:stats}, we describe the parameter space of subhalos and the related statistical ensemble, from which derive the statistical properties of their gamma-ray emissivity presented in \citesec{sec:gamma}. We further discuss the detectability of DM subhalos in \citesec{sec:detect},  which is the main part of the paper, and where we pay a particular attention to the possible background configurations. In particular, we exploit a simplified statistical method and derive useful analytical results showing \eg~the consequence of imposing to detect subhalos {\em before} the smooth halo on the sensitivity, which we further confirm with a more sophisticated analysis based on a full likelihood method applied to mock data. In that way, we can place ourselves in the context of an idealized experiment resembling Fermi-LAT, and derive predictions for both current and future observations. We summarize our results and draw our conclusions in~\citesec{sec:summ}, to which we invite the expert reader to go directly, and provide additional technical details in the Appendices.
\section{Review of the subhalo population model}
\label{sec:model}
In this section, we motivate the need for a dynamically consistent model for the DM distribution in the MW which globally include both the subhalo population and the smooth Galactic halo--see a more detailed discussion in SL17.

It is well known that in the cold DM scenario, structure formation leads to a high level of self-similarity that translates into an almost universal shape for the dark halos over a large range of scales, close to a parametric Navarro-Frenk-White (NFW henceforth) profile, as found in cosmological simulations \cite{Zhao1996,NavarroEtAl1996a}. Such a (spherical) halo shape should characterize systems like the MW down to all pre-existing layers of inhomogeneities like subhalos, the latter also globally contributing to shaping the former. Increasing the spatial/mass resolution of cosmological simulations does not modify this picture, it only uncovers a larger population of smaller subhalos in their host halos, sharing similar morphologies \cite{BullockEtAl2001b,DiemandEtAl2008b,SpringelEtAl2008}---the overall profiles of the host halos remaining unaffected. Consistency therefore demands that the sum of the smooth DM component and its substructure be globally following an NFW profile (or any variant motivated by improved fitting formulae \cite{Einasto1965,NavarroEtAl2004,NavarroEtAl2010}, with possible alterations in the central regions due to baryonic feedback \cite{PontzenEtAl2012,GovernatoEtAl2012,DiCintioEtAl2014a}).

This actually implies a spatial correlation between the smooth halo and its substructure, the details of which are related to the accretion history and more importantly to the tidal stripping experienced by subhalos and induced both by the total gravitational potential of the host halo and by baryons (disk shocking, stellar encounters, etc.). This spatial correlation is expected to have some impact on the gamma-ray observability properties of subhalos as pointlike sources, because it translates into a correlation between the hunted sources and the background in which they lurk. Such a correlation was partly accounted for in \eg~Refs~\cite{PieriEtAl2011,CharbonnierEtAl2012}, but without realistic treatment of gravitational tides. At this stage, it is worth recalling that the global DM content of the MW is better and better constrained as the quality of stellar kinematic data improves. This implicitly translates into limits on the distribution of dark subhalos---except for those ``visible'' subhalos hosting stars and already identified as MW satellites.

Here, we take advantage of the SL17 analytical subhalo population model for the MW. This model is consistent with recent kinematic constraints on the MW, as it is constructed to recover the best-fit Galactic mass model found in \cite{McMillan2017} (McM17). Note that the McM17 best-fit model (which includes both DM and baryons) is itself consistent with more recent results (\eg~Refs~\cite{WeggEtAl2019,CautunEtAl2020}) based on analyses of big samples of RR Lyrae or red-giant stars with accurate proper motions inferred from the Gaia survey \cite{GaiaCollab2018,EilersEtAl2019,HoggEtAl2019}. In the SL17 subhalo population model, subhalo tidal stripping is determined from the detailed distributions of both DM and baryons derived in McM17. The total DM density profile $\rho_{\rm tot}$ is assumed to be spherical and a mixture of two components:
\ben
\label{eq:rhotot}
\rho_{\rm tot}(R) = \rho_{\rm sm}(R) + \rho_{\rm sub}(R)\,,
\een
where $R$ is here the distance to the Galactic center (GC), $\rho_{\rm sm}$ describes the smooth DM component, and $\rho_{\rm sub}$ describes the average mass density in the form of subhalos. More precisely, the latter can formally be expressed as
\ben
\label{eq:rhosub}
\rho_{\rm sub}(R) = \int \dd m_{\rm t}\,m_{\rm t} \frac{\dd n(R)}{\dd m_{\rm t}}\,,
\een
where $n(R)$ is the number density of subhalos and the integral runs over the tidal mass $m_{\rm t}$---all this will be properly defined later. Kinematic data set constraints on $\rho_{\rm tot}$, and therefore, though more indirectly, on $n$. The SL17 model assumes that if subhalos were hard spheres, they would simply track the smooth component, and then $\rho_{\rm sub}$ would be proportional to $\rho_{\rm sm}$. Further calculating the effect of tidal stripping allows us to determine how DM initially in $\rho_{\rm sub}$ migrates to $\rho_{\rm sm}$, a leakage that increases in strength toward the inner parts of the MW where the gravitational potential gets deeper and where the baryonic disk is located. The SL17 model also predicts the spatial dependence of the subhalo concentration distribution function and of the mass function as a consequence of gravitational tides. All this is in perfect qualitative agreement with what is found in cosmological simulations with \cite{DOnghiaEtAl2010,ZhuEtAl2016} and without baryons \cite{DiemandEtAl2008b,SpringelEtAl2008,MolineEtAl2017}.

The main modeling aspects to bear in mind before discussing the gamma-ray properties of subhalos
are the following:
\bi
\item The total DM halo of \citeeq{eq:rhosub} is described either as a spherical NFW halo or as a cored halo, whose parameters are given in \citeapp{app:subs}, and which are both consistent with current kinematic constraints.
\item We assume inner NFW profiles for subhalos, and consider initial mass and concentration functions inferred from standard cosmology (before tidal stripping).
\item The final spatial distribution of subhalos follows the overall DM profile in the outskirts of the MW, but gets suppressed in the central regions of the MW as an effect of gravitational stripping---there is no simple parametric form to describe the smooth and subhalo components together, since the latter depend on the details of tidal stripping: they are predicted from the model.
\item Tidal effects make the final mass and concentration functions fully intricate and spatially dependent; they cannot be factorized out and the SL17 model accounts for this physical intrication.
\item Gravitational tides prune more efficiently the less concentrated subhalos, hence the more massive objects.
\item The tidal subhalo mass $m_{\rm t}$ (tidal radius $r_{\rm t}$) is generically much smaller than the mass $m_{200}$ (the virial radius $r_{200}$) it would have in a flat cosmological background---the actual minimal mass can therefore be much smaller than the minimal mass considered for subhalos in terms of $m_{200}$ (this will depend on the tidal disruption criterion discussed around \citeeqp{eq:disrupt}).
\item the baryonic content of the model comprises a multicomponent axisymmetric disk (thick and thin disks of stars and gas) and a spherical bulge; all these components are taken into account for the gravitational tides, but only the gaseous component is considered to model the regular Galactic diffuse gamma-ray emission.
\ei

In the next section, we discuss the statistical properties of subhalos, which are inherited from their cosmological origin.

\section{The subhalo population statistical ensemble}
\label{sec:stats}
In this section, we review the internal properties of subhalos and fully characterize their statistical ensemble. This will later translate into \repl{statistical}{observable} gamma-ray properties.
\subsection{Structural properties of subhalos and distribution functions}
\label{ssec:pdfs}
Here, we introduce the basic definitions inherent to subhalos, which are rather standard \cite{LavalleEtAl2008,CharbonnierEtAl2012}. We assume a spherical NFW inner profile for subhalos, defined as
\ben
\label{eq:nfw}
\rho(x) = \rhos \times \left\{ g(x)\equiv x^{-1}(1+x)^{-2} \right\}\,,
\een
where \rhos\ is the scale density, and the scale variable $x\equiv r/\rs$ expresses the distance $r$ to the subhalo center in units of the scale radius \rs, and where the dimensionless parametric function $g$ is explicitly defined as an NFW profile, though it needs not be the case. Note that $g$ encodes all the details of the profile, such that switching to another profile simply amounts to changing $g$. In the following, we use $\rho(x)$ and $\rho(r)$ interchangeably, letting the reader adapt the definition accordingly. The integrated mass reads
\ben
\label{eq:mass}
m(x) = 4\,\pi\, \rs^3\, \rhos\,\left\{ \mu(x)\equiv \int_0^{x}\dd x'\,x'^2\,g(x') \right\}\,,
\een
where we define the dimensionless mass $\mu(x)$ that encodes the morphological details of the inner profile. Again, we use $m(x)$ and $m(r)$ interchangeably in the following.

A subhalo is conventionally defined from its mass on top of a flat background density and its concentration. It is common practice to adopt $m_{200}\equiv m(r_{200})$ for the initial subhalo mass definition. This corresponds to the mass contained inside a radius $r_{200}$, often called virial radius, over which the subhalo has an average density of 200 times the critical density $\rhoc\equiv 3\,H_0^2\,M_{\rm P}^2/ 8\pi$, where $H_0$ is the Hubble parameter value today, and $M_{\rm P}$ is the Planck mass\footnote{The use of ``virial'' quantities $m_{200}$ and $r_{200}$ can be misleading in the context of subhalo phenomenology. Indeed, the actual mass and radius of a subhalo embedded in the gravitational potential of the MW (assuming spherical symmetry still holds) are the tidal ones, which depend on the tidal stripping it has experienced along its orbit---roughly speaking, the local gravitational potential and the number of disk crossings and stellar encounters along the orbit. Therefore, these virial quantities are only useful to determine the subhalo inner properties, once the mass-concentration relation is fixed.}. In the following, we use $H_0=68\,{\rm km/s/Mpc}$. The scale parameters of subhalos are then entirely defined once the concentration parameter $c_{200}\equiv r_{200}/\rs$ is fixed. The latter is not really a physical parameter since it formally depends on the cosmological background density, but tells us how dense the subhalo is inside \rs. Since smaller subhalos have formed first in a denser universe, the concentration is a decreasing function of the mass. In this paper, we use the SL17 model as derived from the concentration-mass relation given in Ref.~\cite{Sanchez-CondeEtAl2014}, to which we associate a log-normal distribution function ($p_{c}(c_{200})$, used below in \eg~\citeeq{eq:pdf}), with a variance set (in dex) to $\sigma_c^{\rm dex}=0.14$. To simplify the notations, we further use $m$ for $m_{200}$ and $c$ for $c_{200}$, unless specified otherwise.

Although the mass $m$ and the concentration $c$ fix the internal properties of a subhalo, the
only relevant physical parameters are actually \rhos\ and \rs, and more importantly the tidal
radius \rt. We also introduce
\ben
x_{\rm t}\equiv \rt/\rs\,,
\een
its dimensionless version. Subhalos are indeed not moving in a flat background. Tidal radii are actually difficult to determine since they depend on the details of all gravitational effects felt by subhalos along their orbits in the host halo. The SL17 model precisely provides us with a prediction of subhalos' tidal radii which depend on their structural properties, their position in the halo, and on the details of the DM and baryonic components featuring the MW. Therefore, the real mass and extension of a subhalo are not $m_{200}$ nor $r_{200}$, but instead
\bsubeq
\label{eq:tides}
\ben
\text{the tidal radius}\;\; & \rt (m,c,R) \leq r_{200}\,,\\
\text{and the tidal mass}\;\; & m_{\rm t}=m(\rt)\leq m\,,
\een
\esubeq
where the dependence of the tidal radius on the subhalo structural properties and on its average position $R$ in the MW has been made explicit. It is important to keep in mind that the tidal extension of a subhalo is usually much smaller than $r_{200}$, which may strongly decrease the subhalo gamma-ray luminosity with respect to a naive estimate using $r_{200}$. The SL17 model further proposes a criterion for tidal disruption, which is expressed as a lower limit in $x_{\rm t}$. This can be understood as the fact that tidal stripping can be so efficient that the remaining subhalo core has not enough binding energy left to survive, and gets disrupted. In the following, we will mainly use two different disruption thresholds according to the following rule:
\ben
\label{eq:disrupt}
\text{tidal disruption}\;\forall x_{\rm t}< \epsilon_{\rm t} =
\begin{cases}
  1\;\text{(fragile subhalos)}\\
  0.01\;\text{(resilient subhalos)}
\end{cases} \,.
\een
The fragile case refers to a criterion found in early simulation studies of tidal stripping \cite{HayashiEtAl2003}, while the latter case accounts for the fact that the disruption efficiency found in simulations is very likely overestimated due to the lack of resolution and to spurious numerical effects \cite{BoschEtAl2018}. It can actually be reasonably conceived that the very inner parts of subhalos, which are also very dense, could actually survive tidal stripping for a very long time, simply as a consequence of adiabatic invariance \cite{Weinberg1994,GnedinEtAl1999b}. One of the advantages of the SL17 model is that we can really check the impact of the disruption efficiency on gamma-ray predictions by tuning the disruption parameter $\epsilon_{\rm t}$. Including further evolution of the structural properties themselves is possible in principle \cite{PenarrubiaEtAl2010,DrakosEtAl2017}, but it is actually not straightforward to scale that up to a population study. We will therefore just assume a hard cut of the subhalo density profile at the tidal radius, which can be considered as an optimistic assumption in terms of gamma-ray emissivity. Self-consistently accounting for tidal stripping is anyway already a significant improvement with respect to many past studies.

Beside the individual properties of subhalos, the SL17 model also provides the population's global properties, which amounts to define a probability distribution function (pdf) for subhalos. Assuming subhalos are independent from each other, the subhalo number density per unit of (virial) mass can be expressed as
\ben
\frac{\dd n(R,m)}{\dd m} = N_{\rm tot} \int_{1}^{\infty} \dd c\,\hat{p}_{\rm t}(R,m,c)\,,
\label{eq:dndm}
\een
where the integral runs over concentration, $N_{\rm tot}$ is the total number of subhalos in the MW, which will be discussed later below \citeeq{eq:dndmt}, and the global pdf $\hat{p}_{\rm t}$ is given
by
\ben
\label{eq:pdf}
\hat{p}_{\rm t}(R,m,c) &=&
\frac{ \theta( x_{\rm t}(R,m,c)- \epsilon_{\rm t}) }{K_{\rm t}}\\
&\times& p_V(R) \times p_{m}(m)\times p_{c}(c)\,.\nn
\een
In these equations, $m=m_{200}$ stands for the virial (fictitious) mass in a flat background, $c=c_{200}$ is the concentration parameter, and $K_{\rm t}$ allows for the normalization to unity over the whole parameter space defined by the product of the volume element $4\,\pi\,R^2\dd R$ with the concentration element $dc$, the reference mass element $\dd m$, and the associated pdfs. All pdfs $p$'s above are normalized to unity over their own individual range. Tidal disruption, despite its quite simple implementation in the form of a step function $\theta()$, induces an intrication of the individual pdfs. Moreover, since the dimensionless tidal radius $x_{\rm t}$ depends on all parameters, the same holds true for the tidal mass: a subhalo with a given $m$ can obviously have a different $m_{\rm t}$ depending on its concentration and position in the MW.

For the ``fictitious-mass'' function $p_m(m)$, we adopt a power law for simplicity,
\ben
p_m(m) = K_m (m/m_0)^{-\alpha}\,,
\een
where $K_m$ and $m_0$ are dimensionful constants that allow us to normalize the mass function to unity over the full subhalo mass range. More involved functions can actually be used, but it turns out that the extended Press-Schechter formalism, reflecting the state-of-the-art analytical formalism in this framework \cite{PressEtAl1974,BondEtAl1991a,LaceyEtAl1993,ShethEtAl2001,Zentner2007}, gives a mass function close to a power law of index $\alpha\sim 1.95$---see \citefig{fig:mass_func} for illustration. We will therefore use values of 1.9 and 2 as reference cases. The real(tidal)-mass function, in contrast, also depends on position, and can be written
\ben
\label{eq:global_pt}
\hat{p}_{m_{\rm t}}(m_{\rm t},R) &=& 
\int \dd m \,p_m(m)\int \dd c\,p_{c}(c)\\
&&\times \theta( x_{\rm t}(R,m,c)- \epsilon_{\rm t})\nn\\
&&\times\delta(m-m_{\rm t}(R,m,c))\,.\nn
\een
This expression makes it clear that the tidal mass function is spatially dependent not only because of tidal disruption (in the step function), but also because of tidal stripping (in the delta function). In the SL17 model, surviving subhalos are more stripped and more concentrated as they are found closer to the central regions of the MW. More precisely, tidal stripping acts as a high-pass filter by moving upward a threshold in the concentration distribution function (for a given mass), leading to a strong depletion of the subhalo population as one approaches the central Galactic regions. This effect is genuinely observed in cosmological simulations, and usually parametrically modeled as an additional radial dependence in the median mass-concentration relation (see \eg~Refs~\cite{PieriEtAl2011,MolineEtAl2017}). In the SL17 model it is not parametrized but predicted from the constrained distributions of the Galactic components.

\begin{figure}[!t]
\centering
\includegraphics[width = 0.495\textwidth]{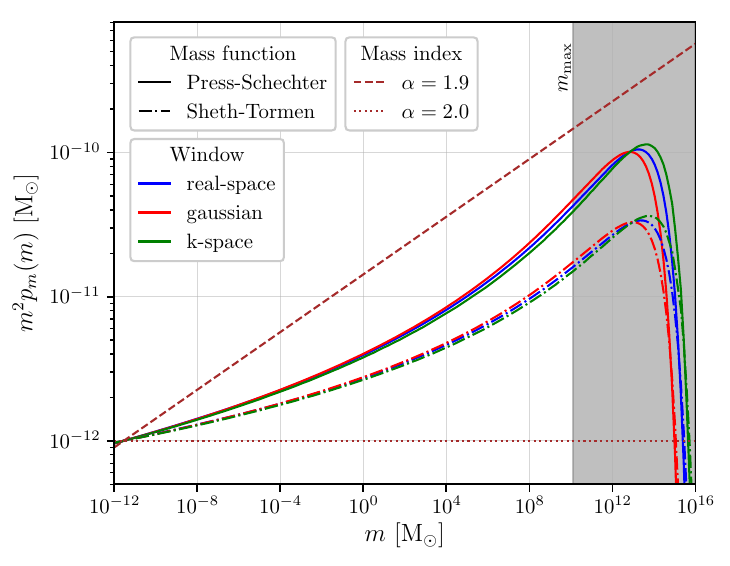}
\caption{\small ``Fictitious'' virial mass function rescaled by $m_{200}^2$   as a function of $m_{200}$ at redshift $z = 0$, assuming a cutoff mass $m_{\rm min} = 10^{-12}$ $\rm M_\odot$. The Press-Schechter and Sheth-Tormen mass functions are calculated using Planck best-fit cosmological parameters \cite{PlanckCollab2020} for different window filters and fall in all cases between the power-law functions of indices $\alpha = 1.9$ and $\alpha = 2.0$.  The gray band corresponds to halos too massive to be accounted for MW subhalos. }
\label{fig:mass_func}
\end{figure}

The SL17 subhalo spatial distribution is built upon assuming that if subhalos were hard spheres, they would simply follow the global DM profile, as is the case for ``particles'' in cosmological simulations. Therefore, the hard-sphere spatial distribution of the total population of subhalos (including the disrupted ones) is simply
\ben
p_V(R) = \frac{\rho_{\rm tot}(R)}{M_{\rm halo}}\,,
\een
where $M_{\rm halo}$ is the total DM mass in the assumed extent of the MW halo. However, tidal stripping and disruption strongly distorts that hard-sphere distribution, and the actual one only describing surviving subhalos has to integrate the disrupted ones out. It can be written as
\ben
\hat{p}_V(R) &=& \int \dd m\int \dd c\, \hat{p}_{\rm t}(R,m,c)\\
&\neq& p_V(R)\nn\,,
\een
where $\hat{p}_{\rm t}$ is the global pdf that includes tidal stripping, given in \citeeq{eq:pdf}. The whole population of subhalos is then described from its number density per unit (tidal) mass as follows,
\ben
\frac{\dd n(R,m_{\rm t})}{\dd m_{\rm t}} &=& N_{\rm tot} \int \dd m \int \dd c\,
\hat{p}_{\rm t}(R,m,c)\nn\\
&\times&\delta(m-m_{\rm t}(R,m,c))\,.
\label{eq:dndmt}
\een
Note that $N_{\rm tot}$, the total number of subhalos, can be normalized according to different choices. A possibility is to normalize it from the number of dwarf galaxy satellites in the relevant mass range \cite{LavalleEtAl2008} (correcting for sky and efficiency completion), from merger-tree arguments \cite{LaceyEtAl1993,ParkinsonEtAl2008}, or similarly from a global subhalo mass fraction also in a given mass range $\Delta_m$ \cite{PieriEtAl2011}. We adopt the normalization of SL17 that matches the Via Lactea II DM-only simulation results \cite{DiemandEtAl2008b}, and conventionally sets the {\em fictitious} mass fraction\footnote{It is called {\em fictitious} mass fraction because it was calibrated in such a way that each surviving subhalo should carry its full fictitious mass $m$ in the mass integral, even though its real mass $m_{\rm t}$ is generically smaller.} to $\tilde{f}_{\rm sub}\sim 10$\% (for $\Delta_m$ taken in the heavy tail of the subhalo mass range, which is very well resolved in simulations). Then
\ben
N_{\rm tot} = \frac{K_{\rm t}}{\widetilde{K}_{\rm t}}
\frac{\tilde{f}_{\rm sub}\,M_{\rm halo}}{\widetilde{\langle m\rangle}_{\Delta_m}}\,,
\een
where $K_{\rm t}$ is the global pdf normalization introduced in \citeeq{eq:pdf}, and the tilde indicates quantities for which baryonic tides are unplugged---see Ref.~\cite{StrefEtAl2017} for details.

It is instructive to calculate the expected number of subhalos that might fall in the mass range of satellite dwarf galaxies in this model, which we give in \citetab{tab:dwarfs} for different assumptions in the mass function index and in the tidal disruption efficiency. We see that the number of objects more massive than a typical threshold of $10^8\msun$ \cite{PenarrubiaEtAl2010} is of order $\sim 100$, consistent with current observations.

Finally, we show the radial distribution of the number density of subhalos for a mass function index $\alpha=1.9$ in the left panel of \citefig{fig:fpl}, where we have considered both the fragile and resilient subhalos, several values of minimal cutoff mass, and started from two different assumptions for the global Galactic halo---an NFW or a cored halo, both consistent with kinematic constraints \cite{McMillan2017}.

\begin{table}[h]
  \caption{\small Number of subhalos with a virial mass $m$, or with physical mass
    $m_{\rm t}$ greater than $10^8M_\odot$ inside a Galactic radius $R_{\rm max} = 250$ $\rm kpc$,
    for fragile ($\epsilon_{\rm t} = 1$) and resilient ($\epsilon_{\rm t} = 10^{-2}$) subhalos.}
  \label{tab:dwarfs}
\begin{ruledtabular}
\begin{tabular}{c|cc}
 &\multicolumn{2}{c}{$N_{\rm sub}(m_{200}>10^8M_\odot)|N_{\rm sub}(m_{\rm t}>10^8M_\odot)$} \\
Initial mass index  & $\epsilon_{\rm t} = 10^{-2}$ & $\epsilon_{\rm t} = 1$ \\
\hline
  $\alpha = 1.9$   & $322|133$  &   $268|130$ \\
  $\alpha = 2.0$ &  $278|108$ &   $232|106$ \\
\end{tabular}
\end{ruledtabular}
\end{table}
\section{Gamma rays from subhalos: a statistical description}
\label{sec:gamma}
In this section,  we relate the gamma-ray properties of subhalos to their internal properties. This will fully characterize the statistical properties of their gamma-ray emission, an important step before rigorously determining their detectability properties.
\subsection{Subhalo luminosity}
\label{ssec:emiss}
Since we consider DM annihilation in subhalos, it is convenient to define an intrinsic luminosity or emissivity function (in units of squared mass per volume),
\ben
\xi(r,m,c) &=& 3\left\{ \xi_{\infty}\equiv \frac{4\,\pi}{3}\,\rs^3\,\rho_s^2\right\}\\
&\times& \int^x_0\dd x'\,x'^2\,g^2(x')\,,\nn
\een
where $x'=r'/\rs$, and where we have introduced a reference luminosity $\xi_{\infty}$ which is such that for an NFW profile $\xi_{\infty}=\lim_{r\to\infty} \xi(r)$ and
\ben
\label{eq:sub_lum}
\xi(2\,r_s) = \frac{26}{27}\,\xi_{\infty} = 0.963\,\xi_{\infty} \approx \xi_{\infty}\,.
\een
The tidal luminosity of a given object depends only on its position, viral mass, and concentration, which we can express as
\ben
\xi_{\rm t}(R,m,c)=\xi(\rt(R,m,c),m,c)\,.
\een
For simplicity, we fix the ``luminosity'' size of a subhalo to
\ben
\label{eq:l_radius}
\begin{cases}
  r_{\rm t} & \text{if}\;r_{\rm t}<2\,r_s\\
  2\,r_s\;\; & \text{if}\;r_{\rm t}\geq 2\,r_s\,.
  \end{cases}
\een
This defines the spatial/angular extension of a subhalo in the gamma-ray sky. It will be used when discussing pointlike subhalos in \citesec{ssec:point}.
\subsection{Gamma-ray fluxes and $J$~factors}
Here we introduce our conventions to deal with gamma-ray fluxes. For a target seen by an observer on Earth, we use the common distance-longitude-latitude triplet (Galactic coordinates), $\vec{s}=(s,l,b)$, such that in the direct Cartesian frame attached to the observer and defined by the unit vectors $(\vec{e}_x,\vec{e}_y,\vec{e}_z)$, where $\vec{e}_y$ points to the GC and $\vec{e}_x$ is also attached to the Galactic plane,
\ben
\vec{s} = s \,(\cos b\, \sin l\,\vec{e}_x + \cos b\, \cos l\,\vec{e}_y + \sin b\,\vec{e}_z)\,. 
\een
The GC is therefore located at $\vec{R}_\odot=(0,R_\odot,0)$, where $R_\odot$ is the distance of the Sun to the GC, such that the target distance $R$ to the GC is simply
\ben
\label{eq:psi}
R^2(s,l,b) &=& (\vec{s}-\vec{R}_\odot)^2\\
&=&  s^2+ R_\odot^2 -
2\,s\,R_\odot\,\left\{ \cos\psi\equiv \cos b\, \cos l \right\}\nn\,,
\een
where we have introduced the angle $\psi=(\vec{s},\vec{R}_\odot)$ between the line of sight sustaining the target and the observer-GC axis. Since the SL17 model is spherically symmetric, the averaged amplitude of the gamma-ray flux induced by DM annihilation is fully specified by $\psi$.

Gamma rays accumulate inside a cone characterized by the angular resolution of the telescope, so the spherical MW volume element associated with the spatial distribution of subhalos $4\,\pi\,R^2\dd R$ (see \citesec{ssec:pdfs}) has to be replaced by the conical volume element about the line of sight
\ben
\label{eq:dVlos}
s^2 \dd\Omega\,ds = s^2\,\sin\theta\,\dd\theta\,\dd \phi \,ds\,,
\een
where $\theta$ is the polar angle defining the aperture about the line of sight, and $\phi$ the azimuthal angle. The distance $R$ of the target to the GC then acquires an extra dependence in $\theta$ and $\phi$ which amounts to replace
\ben
\label{eq:psi_to_thetaphi}
\cos\psi \longrightarrow (\cos\psi\,\cos\theta-\sin\psi\,\cos\theta\,\sin\phi)
\een
in \citeeq{eq:psi}. In practice, conical volume integrals are performed over the resolution angle under consideration.

We can now write the gamma-ray flux induced by DM annihilation along the line of sight of angle $\psi$ [equivalently all corresponding pairs $(l,b)$ in Galactic coordinates]:
\ben
\frac{\dd \phi_{\gamma,\chi}(E,\psi)}{\dd E \dd\Omega} =
\frac{{\cal S}_\chi(\mchi,E)}{4\,\pi}\int_0^{s_{\rm max}(\psi)} \dd s\, \rho^2_\chi(s,\psi)\,,\nn\\
\label{eq:dphidEdO}
\een
where $\rho_\chi$ denotes any DM mass density profile under consideration, and $s_{\rm max}(\psi)\approx R_{200} + R_\odot\,\cos\psi$ is the distance to the virial border of the halo in the $\psi$ direction. We have introduced a spectral function,
\ben
\label{eq:spectral_func}
    {\cal S}_\chi(\mchi,E)\equiv
    \frac{\delta_\chi\,\sigv}{2\,m_\chi^2}\frac{\dd N_\gamma(E)}{\dd E}\,,
\een
that carries all the WIMP-model-dependent information, namely the particle mass \mchi, its total $s$-wave annihilation cross section into photons \sigv, and the differential photon spectrum $\dd N_\gamma/\dd E$, which sums up the contributions of all relevant annihilation channels to the photon budget. Parameter $\delta_\chi=1$ (1/2) for scalar DM or Majorana (Dirac) fermionic DM.

Integrating this flux over a solid angle $\delta\Omega_{\rm r}=\delta\Omega(\theta_{\rm r})$, where $\theta_{\rm r}$ is a fixed resolution angle, we can define a first version of the usual $J$~factor \cite{BergstroemEtAl1998a} as follows:
\ben
\label{eq:dphidE_from_jpsi}
\frac{\dd \phi_{\gamma,\chi}(E,\psi,\theta_{\rm r})}{\dd E} =  {\cal S}_\chi(\mchi,E) \,
J_\psi(\theta_{\rm r})\,,
\een
that is
\ben
\label{eq:def_jpsi}
J_\psi(\theta_{\rm r})&\equiv& \frac{1}{4\,\pi}\int_{\delta\Omega_{\rm r}}
\dd\Omega \,j_\psi(\psi,\theta,\phi)\\
\text{with}\,j_\psi(\psi,\theta,\phi) &\equiv&\int_0^{s_{\rm max}} \dd s\, \rho^2_\chi(s,\psi,\theta,\phi)
\nn
\een
This $J$~factor carries the dimensions of a squared mass per (length)$^5$ and may slightly differ from other conventions found in the literature. Note that in the general case, an {\em experimental} resolution angle $\theta_{\rm r}$ depends on energy, hence the $J$~factor as defined above. We will account for this energy dependence whenever relevant.

Following up with practical declensions, the flux averaged over the resolution angle $\theta_{\rm r}$ in the $\psi$ direction is simply
\ben
\label{eq:dphidEdO_from_jpsi}
\angleave{\frac{d \phi_{\gamma,\chi}(E,\psi)}{\dd E\,\dd\Omega}}{\delta\Omega_{\rm r}}
&=& {\cal S}_\chi(\mchi,E) \, {\cal J}_\psi(\theta_{\rm r})\\
\text{with}\;{\cal J}_\psi(\theta_{\rm r})&\equiv&
\frac{J_\psi(\theta_{\rm r})}{\delta \Omega_{\rm r}}\,,
\een
where we implicitly assume a flat and maximal collection efficiency over $\theta_{\rm r}$. This angular average of the $J$~factor, ${\cal J}_\psi$, is directly related to the gamma-ray flux per solid angle provided by experimental collaborations in diffuse gamma-ray studies.

Finally, we introduce a last variant of the $J$~factor, more directly related to the real measurements performed by experiments:
\bsubeq
\label{eqs:jpsi_obs}
\ben
\overline{J_\psi} (\Delta E)&\equiv &
\frac{\int_{\Delta E} \dd E \,{\cal A}(E)\,{\cal S}_\chi(E)\,J_\psi(\theta_{\rm r}(E)) }
     {\Delta E \, \overline{\cal A S_{\chi}}}\,,\\
\overline{\cal J_\psi} (\Delta E) &\equiv & \frac{\int_{\Delta E} \dd E \,{\cal A}(E)\,{\cal S}_\chi(E)\,
  {\cal J}_\psi(\theta_{\rm r}(E)) }{\Delta E \, \overline{\cal A S_{\chi}}}\,,\\
\overline{\cal A S_{\chi}}(\Delta E) &\equiv & \frac{1}{\Delta E}
\int \dd E \,{\cal A}(E)\,{\cal S}_\chi(E)\\
&=& \frac{\sigv}{2\,m_\chi^2}
\frac{ \left\langle {\cal N}_\gamma {\cal A}\right\rangle_{\Delta E}}{\Delta E}\,,\nn
\een
\esubeq
where $\Delta E$ is an energy range to be specified and ${\cal A}$ is an effective experimental collection area. The latter should depend both on the energy and the angle with respect to the pointing direction, but for simplicity we assume a flat and maximal angular acceptance within the resolution angle $\theta_{\rm r}$, which can itself depend on energy. We have also introduced the number of photons per annihilation ${\cal N}_\gamma$ in the energy range $\Delta E$. These experiment-averaged definitions will allow us to formulate the observational sensitivity more accurately. Note that when the resolution angle does not depend much on energy within $\Delta E$, then $\overline{J}\simeq J$ and $\overline{\cal J}\simeq {\cal J}$. Independently, if the line-of-sight integral does not vary much within the resolution angle, whatever large may the latter be, then $\overline{\cal J}\simeq {\cal J}$---this is typically the case at reasonable angular distance from the Galactic center. Finally, one can easily convince oneself that for a pointlike object, $\overline{\cal J^{\rm pt}}={\cal J}^{\rm pt}$ (see \citesec{sssec:pt_subhalos}).

\subsection{Diffuse emission from the smooth and subhalo components}
\label{ssec:diffuse}
The total averaged DM contribution to the gamma-ray flux is the sum of the smooth contribution, the global subhalo contribution, and the cross-product (\eg~\cite{CharbonnierEtAl2012,StrefEtAl2017}). It can be expressed as
\ben
\frac{\dd \phi_{\gamma,\chi}(E,\psi,\theta_{\rm r})}{\dd E} &=&
{\cal S}_\chi(E)\left\{ J_\psi^{\rm diff}\equiv  J_\psi^{\rm sm} +J_\psi^{\rm sub} +J_\psi^{\rm cross} \right\}
\,,
\label{eq:jdifftot}
\een
where we have introduced the total diffuse contribution $J_\psi^{\rm diff}$, which is the sum of
\bsubeq
\label{eqs:jpsi_ave}
\ben
J_\psi^{\rm sm} &=& \frac{1}{4\,\pi}\int_{\delta\Omega_{\rm r}}
\dd\Omega\int_0^{s_{\rm max}(\psi)} \dd s\, \rho^2_{\rm }(s,\psi)\\
J_\psi^{\rm sub} &=& \frac{1}{4\,\pi}\int_{\delta\Omega_{\rm r}}
\dd\Omega\int_0^{s_{\rm max}(\psi)} \dd s\, \int \dd m \frac{\dd n(s,\psi)}{\dd m}\nn\\ &&\times \int \dd c\, \xi_{\rm t} \,
\theta(x_{\rm t}-\epsilon_{\rm t})\\
J_\psi^{\rm cross} &=& \frac{1}{2\,\pi}\int_{\delta\Omega_{\rm r}}
\dd\Omega\int_0^{s_{\rm max}(\psi)} \dd s\, \int \dd m \frac{\dd n(s,\psi)}{\dd m} \nn\\&& \times\int \dd c \,m_{\rm t}\,
\rho_{\rm sm}(s,\psi) \, \theta(x_{\rm t}-\epsilon_{\rm t})\,.
\een
\esubeq
All these terms characterize the DM contribution to diffuse gamma rays. Note that in the averaged subhalo contribution $J_\psi^{\rm sub}$, featuring the differential subhalo number density $\dd n$ given in \citeeq{eq:dndm}, we have actually integrated the contribution of all subhalos assuming that they are pointlike (\ie~their tidal radii are contained in the solid angle characterized by the resolution $\theta_{\rm r}$)---hence the presence of the full $\xi_{\rm t}$ luminosity function. This is formally an approximation, but a very accurate one in fact because the number of pointlike objects is much larger than the extended ones in the resolution angles we will consider (see \citesec{sssec:nsub}). The Heaviside function allows us to integrate only over those subhalos which have not been destroyed by gravitational tides in our model.

\subsection{Pointlike subhalos}
\label{ssec:point}
Here we give a practical definition to the concept of pointlike subhalo. To avoid any confusion, we emphasize here that this notion applies to both resolved and unresolved sources, in the observational sense (\ie~above and below background).

\subsubsection{Definition}
We start with a geometric definition (see \eg~Refs~\cite{BuckleyEtAl2010,CharbonnierEtAl2012}). A subhalo located at a distance $s$ from the observer is considered as pointlike if most of its luminosity is contained in the resolution angle $\theta_{\rm r}$ assumed for the telescope, \ie\
\ben
\frac{\min(r_{\rm t},2\,r_s)}{s}\leq \sin(\theta_{\rm r})\,,
\een
where we have used the luminosity radius introduced in \citeeq{eq:l_radius}, and based on that $96\%$ of the luminosity is contained within $2\,r_{s}$ for NFW (sub)halos [see \citeeq{eq:sub_lum}]. Trading the scale radius for a combination of the virtual (virial) mass $m$ and the concentration $c$, this inequality relation for the tidal radius becomes an inequality relation for the (virial, not tidal) mass, reading
\ben
\label{eq:mpl}
m \leq m_\text{pt}^\text{max}(s,c,x_{\rm t})\equiv \frac{4\,\pi}{3}(200\,\rho_c)\left\{ \frac{c\,s\,\sin(\theta_{\rm r})}{\min(x_{\rm t},2)}\right\}^3\,.
\een
This relation only tells us that the probability for a subhalo to be pointlike increases with its concentration, its distance to the observer, or a combination of both. It allows us to define a maximal mass $m_\text{pt}^\text{max}$ that depends on that distance and on the subhalo properties. Remember that the dimensionless tidal radius $x_{\rm t}$ is a function of position and concentration in our model, $x_{\rm t}(R(s,\psi),c)$. That can further be rephrased in terms of virial (virtual) radius as
\ben
r_{200}(m) &\leq& \frac{c\,s\,\sin(\theta_{\rm r})}{\min(x_{\rm t},2)}
\approx \frac{c\,s\,\theta_{\rm r}}{2}\nn\\
\Leftrightarrow r_s(m,c) &\lesssim & \frac{s\,\theta_{\rm r}}{2}\,,
\een
Since we only consider resolution angles such that $\sin(\theta_{\rm r})\sim \theta_{\rm r} \ll 1$, we can see that the size of a pointlike subhalo is always much smaller than its distance to the observer.
\subsubsection{Number of pointlike subhalos}
\label{sssec:nsub}
It is instructive to compute the fraction $f_\psi^\text{pt}$ of pointlike subhalos lying in the solid resolution angle $\delta\Omega_{\rm r}$ in any direction $\psi$ in the sky. Given the subhalo parameter space introduced in \citesec{ssec:pdfs} and the definition introduced in the previous paragraph, then
\ben
\label{eq:fpl1}
f_\psi^\text{pt} &=&
\frac{\int_{ \left\{ m \leq m_\text{pt}^\text{max} (s,c) \right\} } d\hat \sigma\, \hat p_{\rm t}(R(s,\psi),m,c)}
     {\int_{ \left\{ m \leq m_\text{max}\right\} } d\hat \sigma\,\hat p_{\rm t}(R(s,\psi),m,c)}\\
\label{eq:fpl2}
\text{with}\; d\hat \sigma & \equiv & s^2\,ds\,\sin\theta\,\dd\theta\,\dd \phi\,\dd m\,\dd c\,.
\een
We have used \citeeq{eq:dVlos} to define the full phase-space volume element $d\hat \sigma$ about the line of sight. It is easy to understand that $f_\psi^\text{pt}\simeq 1$ for all angles $\psi$ and for the resolution angles we consider, just because the volume where most subhalos would appear as extended is strongly confined around the observer. This is shown in the right panel of \citefig{fig:fpl}, where we have evaluated this fraction (more precisely $1-f_\psi^\text{pt}$) numerically as a function of the line-of-sight angle $\psi$ for different assumptions on the minimal subhalo mass, the initial mass index $\alpha$, and the tidal disruption efficiency $\epsilon_{\rm t}$.

\begin{figure*}[!t]
\centering
\includegraphics[width = 0.495\textwidth]{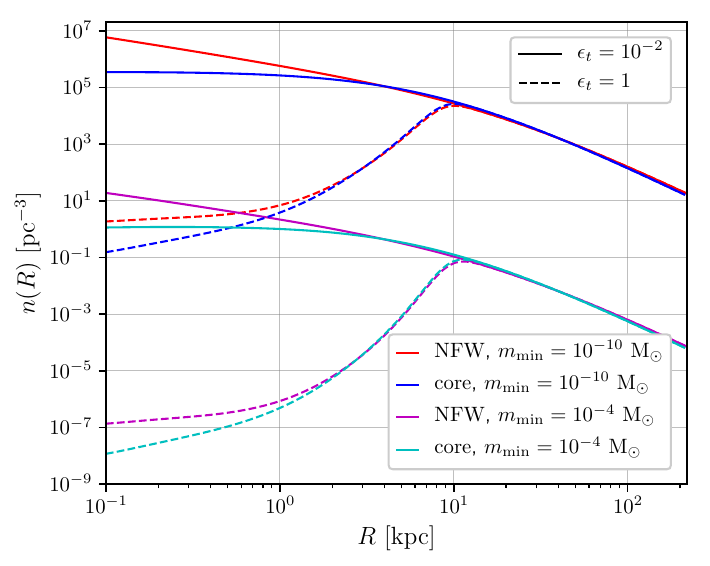}
\includegraphics[width = 0.495\textwidth]{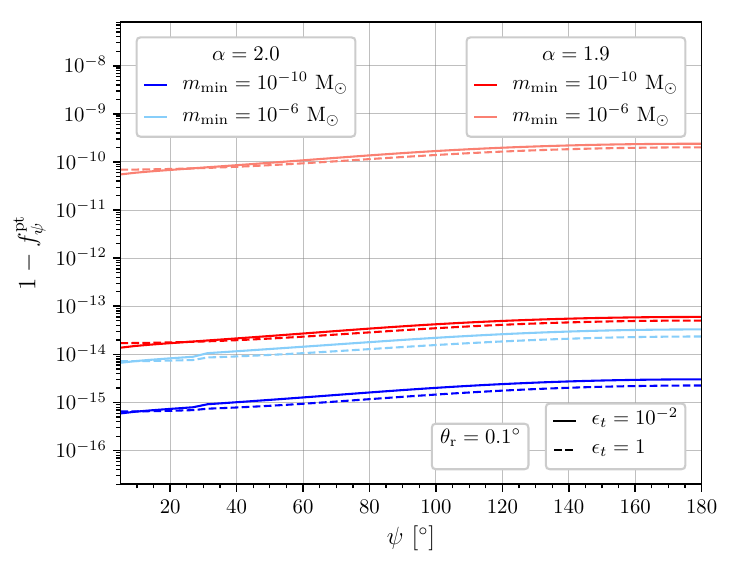}
\caption{\small {\bf Left panel:} Radial distribution of the number density of subhalos assuming a function mass slope of $\alpha=1.9$, different lower cutoff masses, for subhalos either resilient ($\epsilon_{\rm t}=0.01$) or fragile ($\epsilon_{\rm t}=1$)  against tidal disruption. {\bf Right panel:} Total fraction of extended subhalos per solid angle as a function of the line-of-sight angle $\psi$, for different mass functions and cutoff masses. As expected, the fraction of pointlike subhalos $f_\psi^\text{pt}\simeq 1$, such that the fraction of extended objects $(1-f_\psi^\text{pt})\ll 1$.}
\label{fig:fpl}
\end{figure*}

\subsubsection{$J$~factor for a single object}
\label{sssec:pt_subhalos}
If a subhalo of mass $m_{{\rm t},i}$ is pointlike, located at a distance $s_i\gg r_{\rm t}$, its $J$~factor $J^{\rm pt}_{\psi,i}$ should actually account for the fact that its occupancy volume $\delta V_i$, assumed centered about the line of sight and characterized by an angular radius equal to or smaller than the considered experimental resolution angle, contains both the subhalo density and the smooth halo density. This should lead to
\ben
J^{\text{pt}/\delta V_i}_{\psi,i}= J^\text{pt}_{\psi,i}
+ J^{\rm sm}_{\psi,i} + J^{\rm cross}_{\psi,i}\,,
\een
where
\bsubeq
\label{eq:jpsi_pl_th}
\ben
J^\text{pt}_{\psi,i} &\equiv& J^{\rm pt}_{\psi,i}(m,c,s_i)
= \frac{\xi_{\rm t}}{4\,\pi\,s_i^2}\\
J^{\rm sm}_{\psi,i} &\equiv&
\frac{1}{4\,\pi} \int \dd\Omega\int_{\vec{s}\in\delta V_i} \dd s\, \rho_{\rm sm}^2(R(s,\psi))\nn\\
&\simeq& \frac{\rho^2_{\rm sm}(R(s_i,\psi))\delta V_i}{4\,\pi\,s_i^2}\\
J^{\rm cross}_{\psi,i} &\equiv&
J^{\rm cross}_{\psi,i} (m,c,s_i) \nn\\
&=& \frac{1}{2\,\pi} \int \dd\Omega\int_{\vec{s}\in\delta V_i} \dd s\, \rho_{\rm sm}(R(s,\psi))\,\rho(s)\nn\\
&\simeq& \frac{\rho_{\rm sm}(R(s_i,\psi))\,m_{{\rm t},i}}{2\,\pi\,s_i^2}\,.
\een
\esubeq
The smooth contribution $J^{\rm sm}_{\psi,i}$ is actually already included in the foreground contribution of the smooth halo, so we can formally remove it. Besides, since the DM mass density at the border of the subhalo is always such that $\rho(r_{{\rm t},i})>\rho_{\rm sm}(R(s_i,\psi))$ as a consequence of tidal stripping \cite{StrefEtAl2017}, we always have $J^\text{pt}_{\psi,i}\gg J^{\rm cross}_{\psi,i}\gg J^{\rm sm}_{\psi,i}$. Therefore, in the following, we only consider
\ben 
J^{\text{pt}/\delta V_i}=J^\text{pt}_{\psi,i}= J^\text{pt}_i 
\label{eq:jpsi_pl}
\een
for the $J$~factor associated with a pointlike subhalo, which is precise at the subpercent level. Note that for a point source, we also have $J^\text{pt}_i=\overline{J^\text{pt}_i}$, where $\overline{J}$, defined in \citeeq{eqs:jpsi_obs}, involves an average over the experimental acceptance. The associated gamma-ray flux is simply given by
\ben
\label{eq:pt_flux}
\frac{\dd \phi_{\gamma,i}(E)}{\dd E} =  {\cal S}_\chi(\mchi,E) \, J^\text{pt}_i\,,
\een
consistently with \citeeq{eq:dphidE_from_jpsi}.
\subsubsection{Statistical properties of pointlike subhalo $J$~factors}
In order to assess the possibility of detecting subhalos as pointlike sources, we have to derive the full statistical properties of $J^\text{pt}_\psi$. They are obviously related to the properties of subhalos themselves, which are encoded in the global pdf $\hat p_{\rm t}$ introduced in \citeeq{eq:global_pt}. However, now, the parameter space becomes limited by the maximal mass $m_\text{pt}^\text{max}$ attainable by a pointlike object, and defined in \citeeq{eq:mpl}. Actually, given a resolution angle $\theta_{\rm r}$ and a line-of-sight angle $\psi$, the differential probability $d{\cal P}_J^\text{pt}$ for a subhalo to have a $J$~factor equal to $J^0$ can be formally expressed as
\ben
\frac{d{\cal P}_J^\text{pt}}{dJ_\psi}(J_\psi^0) &=& \int_{ \left\{ m \leq m_\text{pt}^\text{max} (s,c) \right\} }
    d\hat \sigma\,   \hat p_{\rm t}(R(s,\psi),m,c)\nn\\
    &&\times \,\delta(J_{\psi}(s(R,\psi),m,c) - J_\psi^0)\,.
\label{eq:prob_jpsi0}
\een
The volume element $d\hat \sigma$ about the line of sight was introduced in \citeeq{eq:fpl2}. One can then define the integrated probability to have a $J$~factor larger than some value as
\ben
{\cal P}_J^\text{pt} (J_{\psi}^\text{pt}\geq J_\psi^0) &=&
\int_{ \left\{ m \leq m_\text{pt}^\text{max} (s,c) \right\} } d\hat \sigma\,  
\hat p_{\rm t}(R(s,\psi),m,c)\nn\\
&&\times\, \theta(J_{\psi}^\text{pt}(s(R,\psi),m,c) - J_\psi^0)\nn\\
&=& \int_{J_{\psi}^0}^{\infty}dJ'\,\frac{d{\cal P}_J^\text{pt}}{dJ_\psi} (J') \,.
\label{eq:prob_jpsi}
\een
Note that ${\cal P}_J^\text{pt} (J_{\psi}^\text{pt}\geq0)<1$ because it defines the probability in the $\psi$ direction only. It normalizes to unity only after integration over the full sky. In the left panel of \citefig{fig:prob_jpsi_pl}, we show the shapes of these pdfs assuming line-of-sight angles of $\psi=20^{\circ}$ and $90^{\circ}$, the former being optimal for subhalo searches and the later possibly minimizing the foreground. We also considering two minimal virial subhalo masses, $m_{\rm min}=10^{-10}M_\odot$ and $10^{-4}M_\odot$, for a conservative initial mass function index $\alpha=1.9$. Here, the subhalo population is embedded in a global NFW halo. We also anticipate as a green vertical band a range of threshold $J$~factors that expresses the sensitivity of a Fermi-like experiment for 100~GeV DM particles annihilating in $\tau^+\tau^-$ in an observation time of 10~yr. This will be discussed extensively in \citesec{sec:detect}, notably in \citesec{ssec:sensitivity}.

This plot illustrates the nontrivial dependence of the ${\cal P}_J^\text{pt}$ on the pointing angle, characterized by a sharp decrease beyond a given $J$ at small angles, which can be associated with the ring structure arising within $\sim 50^\circ$ from the GC (we shall discuss this in more details later when reaching \citefig{fig:nsub_dmbg}). This transition just reflects the position of the peak in the number density arising the inner regions of the MW, close to the solar circle, as shown in \citefig{fig:fpl}. This peak corresponds to the region where tidal effects start depleting the subhalo population beyond the peak of the concentration pdf associated with the smallest objects, hence the dramatic decline of subhalos inward. On the other hand, around this peak is where subhalos are still both numerous enough and highly concentrated. One can integrate subhalos over this peak within $\sim 50^\circ$ from the GC (corresponding to a height of $\sim 10$~kpc from the GC), which explains this particular feature in ${\cal P}_J^\text{pt}$. Much less important than it seems is the difference of probability amplitude between $m_{\rm min}=10^{-10}M_\odot$ and $10^{-4}M_\odot$, which only comes from the fact that the total number of subhalos scales like $\propto 1/m_{\rm min}$ (hence the 6 orders of magnitude between the amplitudes); once rescaled by the total number of subhalos, the pdfs actually match with one another very well (except, obviously, for the very low $J^0_\psi$ tail, not appearing in the plot).

The right panel of \citefig{fig:prob_jpsi_pl} shows the same results in terms of the number of pointlike subhalos with $J$~factors larger than a threshold $J_0$ as a function of $J_0$, still for a subhalo population embedded in a global NFW halo. We report the number distributions obtained with different line-of-sight angles $\psi$, and in the bottom frame, we also indicate the relative difference when assuming subhalos embedded either in an NFW or in a cored global DM halo. We see that the global cored DM halo configuration generically leads to more visible subhalos, Except in the range of $J\in\sim[10^{18},10^{19}]$~GeV$^2$/cm$^5$, which just reflects the fact that the sharp decrease in ${\cal P}_J^\text{pt}$ for a cored host halo occurs at lower values of $J$.

\begin{figure*}[!t]
\centering
\includegraphics[width = 0.495\textwidth]{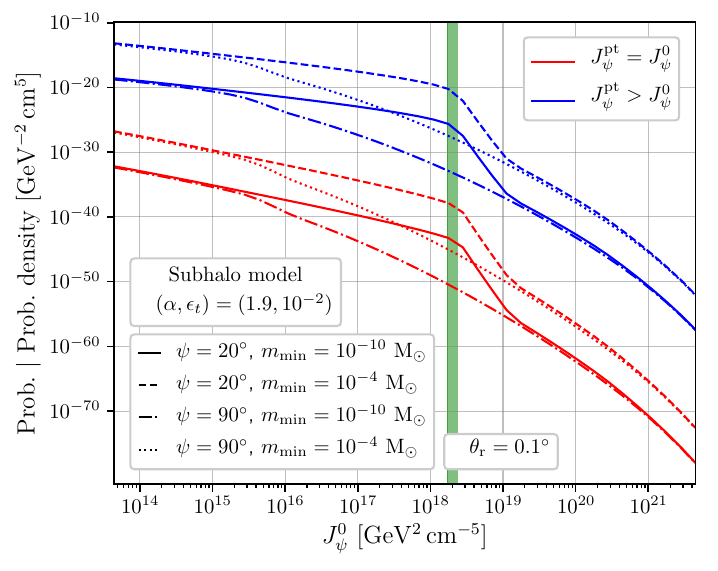}
\includegraphics[width = 0.495\textwidth]{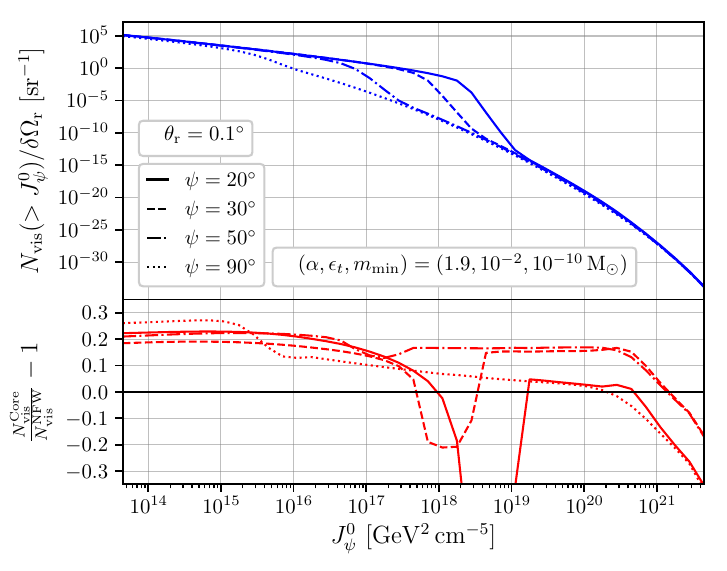}
\caption{\small {\bf Left panel:} Probability distribution functions $d{\cal P}_J^\text{pt} (J)/dJ_\psi$ (differential) and ${\cal P}_J^\text{pt} (J_{\psi}^\text{pt}\geq J)$ (integrated), for a resolution angle $\theta_{\rm r}=0.1^{\circ}$, line-of-sight angles $\psi=20^{\circ}$ (intermediate latitude) and $\psi=90^{\circ}$ (Galactic pole), and a subhalo population configuration of resilient subhalos with $(\alpha,m_{\rm min},\epsilon_{\rm t})= (1.9,10^{-10}-10^{-4},0.01)$ embedded in an NFW Galactic halo. The green vertical thick line gives the typical sensitivity for a Fermi-LAT-like experiment calculated for an observation time of 10~yr. {\bf Right panel:} Corresponding distribution of the number of subhalos with $J$~factors larger than $J_0$ as a function of $J_0$, for different line-of-sight angles. The bottom part of the plot shows the relative difference when using a subhalo population model embedded in a cored Galactic halo.}
\label{fig:prob_jpsi_pl}
\end{figure*}

From these pdfs, one can also calculate the $n$th moments of the $J$~factors (including the mean value with $n=1$) using
\ben
\label{eq:jpsi_moments}
\langle (J_{\psi}^\text{pt})^n \rangle = \int_0^\infty dJ\, (J)^n\,
\frac{d{\cal P}_J^\text{pt}}{dJ_\psi} (J)\,.
\een
\section{Detectability of subhalos as pointlike sources}
\label{sec:detect}
This section enters the prevailing discussion of the paper: assessing the detectability of pointlike subhalos. To proceed, we need to carefully define what are the main foregrounds or backgrounds (generically {\em background} henceforth) to any potential detection. In most past studies, the main background considered was the ``baryonic'' contribution to the $\gamma$-ray flux. This comprises the diffuse Galactic emission \revise{(DGE)} induced by interactions of cosmic rays with the interstellar gas or radiation (pion production, bremsstrahlung, and inverse Compton processes) and by unresolved conventional Galactic sources, and the isotropic diffuse extragalactic background. A lot of effort has been invested in describing the sensitivity of current gamma-ray experiments to exotic pointlike sources based on as accurate as possible models of such Galactic and extragalactic backgrounds, inferred from both phenomenological cosmic-ray modeling, or from more agnostic template fitting methods \cite{FermiLATCollab2015,SchoonenbergEtAl2016,CaloreEtAl2017,CaloreEtAl2019,FermiLAT2019}. Here, by contrast, we consider a very simplified model of baryonic background, and instead focus our attention onto another background component often neglected, \ie~the one induced by DM annihilation itself, which is made up of contributions from the smooth halo and from unresolved subhalos. That DM background has already been defined in \citesec{ssec:diffuse}.

We further want to place ourselves within the framework of an idealized Fermi-LAT-like experiment, in which we assume that a diffuse emission has been measured in predefined regions of interest (ROIs), which is consistent, while not perfectly, with the baryonic foreground (hence limiting the diffuse DM contribution to statistical or systematic fluctuations at maximum). This will allow us to set \revise{idealized} limits on the contribution of DM annihilation to the diffuse emission, hence on the annihilation cross section, which also impacts on the detectability of DM subhalos.

In \citesec{ssec:background}, we provide the details of our background model. In \citesec{ssec:pdet}, we describe the statistics of the number of pointlike subhalos contributing a flux above a given threshold. In \citesec{ssec:sensitivity}, we review the full statistical analysis we perform to infer the sensitivity to pointlike subhalos in our idealized framework. We start with a simplified statistical reasoning (see \citesec{sssec:simple_stats}), which allows us to derive useful analytical results for the threshold flux of subhalo detection as a function of time and annihilation cross section. Most notably, we derive useful time-independent asymptotic limits arising in the case of infinite observational time, which correspond to the most optimistic case for the detection of DM subhalos. Finally, we generate mock data and apply a complete likelihood analysis (i) to mimic the current Fermi data analysis, (ii) to qualitatively validate the aforementioned simplified statistical reasoning, and (iii) to get more definite results for the detectability of subhalos. We discuss these results in \citesec{sssec:likelihood}.

\subsection{Baryonic background model}
\label{ssec:background}
We consider two types of contributions to the diffuse background that may shield DM subhalos as individual sources: one coming from DM annihilation itself, already discussed in \citesec{ssec:diffuse}, and another one coming from conventional astrophysical processes, dubbed baryonic background \revise{(including both the DGE and the isotropic background)}. To maximize the self-consistency of our study, we base our DGE baryonic background model on the same ingredients used to determine the tidal stripping induced by the baryonic disk, \ie~those included in the Galactic mass model derived from kinematic data in Ref.~\cite{McMillan2017}. They consist of the spatial distributions for the atomic and molecular interstellar gas. We remind the reader that our goal is to have a realistic modeling of the background, though not necessarily a precise one. Indeed, we shall not discuss the Fermi data themselves, but instead provide a realistic insight as to what to expect to find in them in terms of any putative subhalo contribution.

For space-borne observatories like Fermi-LAT \cite{FermiLATEtAl2009}, the genuine background includes many different astrophysical contributions, as shortly stated above. However, for simplicity, we restrict ourselves to the pion decay contribution induced by the interactions of cosmic rays with the interstellar gas, which is the dominant DGE one in the $1-100$~GeV energy range that we consider \cite{AckermannEtAl2012}. There are of course other contributions (e.g. leptonic), but the spatial distribution of their amplitudes should not change much with respect to the pion decay one---\revise{we will play with the overall normalization of the ``pionic'' background for a better match, but this will anyway not be critical in our analysis}. We add by hand the isotropic diffuse emission assumed to be of extragalactic origin, for which we simply consider the spectrum derived in Ref.~\cite{AckermannEtAl2012}. In the following, we only consider gamma-ray energies above 1 GeV, to avoid modeling issues with the pion bump at $\sim 100$~MeV.

Consistently with our Galactic mass model, we can predict the relative intensity of the pionic emission by convoluting of a cosmic-ray flux, assumed homogeneous in the MW for simplicity, and the spatial-dependent hydrogen number density, $n_{\rm ism}$. The latter can be expressed as
\ben
n_{\rm ism}(\vec{x}) = n_{\rm H}(\vec{x}) + 2\,n_{\rm H_2}(\vec{x})
= \frac{\rho_{\rm H}(\vec{x})}{m_{\rm H}} + 2\,\frac{\rho_{\rm H_2}(\vec{x})}{m_{\rm H_2}}\,,\nn\\
\een
where indices H and H$_2$ refer to atomic and molecular hydrogen, respectively, $m_{\rm H/H_2}$ being their masses, and where, consistently with the SL17 subhalo model, we take the associated gas mass densities $\rho$'s from McM17. Further integrating this density along the line of sight, within a resolution solid angle $\delta\Omega_{\rm r}$, we get
\ben
\angleave{\frac{\dd \phi_\pi(E,l,b)}{\dd E\,\dd\Omega}}{\delta\Omega_{\rm r}} &=&
\frac{f_\pi(E)}{4\,\pi\,\delta\Omega_{\rm r}}
\int_{\delta\Omega_{\rm r}} \dd\Omega \int \dd s\,n_{\rm ism}(s,l,b)\nn\\
&\simeq &\frac{f_\pi(E)}{4\,\pi} \int \dd s\,n_{\rm ism}(s,l,b)\,,
\label{eq:pion_bckg}
\een
where $l$ and $b$ are the longitude and latitude, respectively. The spectral function $f_\pi(E)$ is taken as a power law \revise{over three energy ranges},
\begin{align}
f_\pi(E) = \sum_{i=1}^3 & \theta \left(E-E_{{\rm max}}^{(i-1)}\right) \theta\left( E_{{\rm max}}^{(i)}-E\right) \nn\\
& \times f_{0}^{(i)}\,  \left[ \frac{E}{1\,{\rm GeV}} \right]^{-\gamma_{{\rm b}}^{(i)}}\,,
\label{eq:dge_bckg_spectrum}
\end{align}
\revise{where the normalization coefficients $f_{0,i}$ and spectral indices $\gamma_{{\rm b},i}$ are tuned to give a decent fit to the pionic contribution estimated in Ref.~\cite{AckermannEtAl2012}. Starting from a threshold energy $E_{\rm min}=E_{{\rm max},0}=1$~GeV, these parameters read:}
\ben
\revise{
\begin{bmatrix}
  \displaystyle \frac{E_{{\rm max}}^{(i)}}{\displaystyle \rm GeV}\\
  \displaystyle \gamma_{{\rm b}}^{(i)}\\
  \displaystyle\frac{f_{0}^{(i)}}{ 10^{-27}{\rm GeV^{-1}s^{-1}}}
\end{bmatrix}
  = 
\begin{cases}
(i=1)\rightarrow\begin{bmatrix}1.4\\2.27\\6.69\end{bmatrix}\\
(i=2)\rightarrow\begin{bmatrix}2.3\\2.59\\7.45\end{bmatrix}\\
(i=3)\rightarrow\begin{bmatrix}100\\2.72\\8.31\end{bmatrix}
\end{cases}
}
\label{eq:bckg_parameters}
\een

\revise{The latitudinal profiles of this pionic gamma-ray flux background model integrated over two energy ranges, [1.6-13]~GeV and [13-100]~GeV are shown in \citefig{fig:bckg_profile} as solid red curves (top and bottom panels, respectively), for both the central and anticentral Galactic regions (left and right panels, respectively), and are compared with the ones inferred from the Fermi-LAT data and taken from Ref.~\cite{AckermannEtAl2012} (dashed red curves for the pionic contribution, and dashed blue for the total DGE). We also show our pionic background model rescaled by a constant factor in the range 1.5-2.5 (redish shaded bands), and the corresponding residuals with respect to the total DGE inferred from the Fermi-LAT data. We see that our DGE model of both the Fermi-LAT reconstructed pionic emission and of the total DGE are reasonably recovered both in the central regions and in the outskirts of the MW, with errors in amplitude fluctuating by a factor of $\sim 2$. This angular gradient is realistic enough for our study. Since we want to remain on the optimistic side regarding the detection of subhalos, we adopt a rescaling factor $\alpha_{\pi\to{\rm DGE}}$ such that our background DGE model does not exceed current data, and therefore fix it to 1.5 from now on. We have checked that our results are qualitatively not sensitive to slight changes around this value.}

\begin{figure*}[!t]
\centering
\includegraphics[width = 0.49\textwidth]{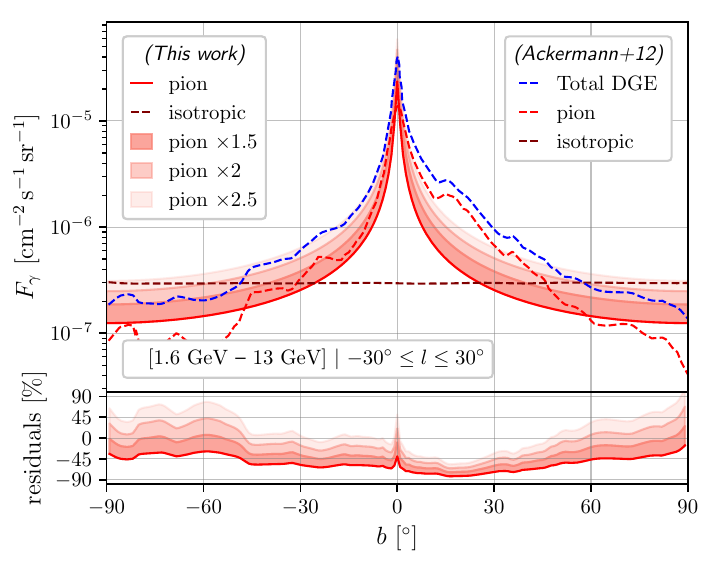}
\includegraphics[width = 0.49\textwidth]{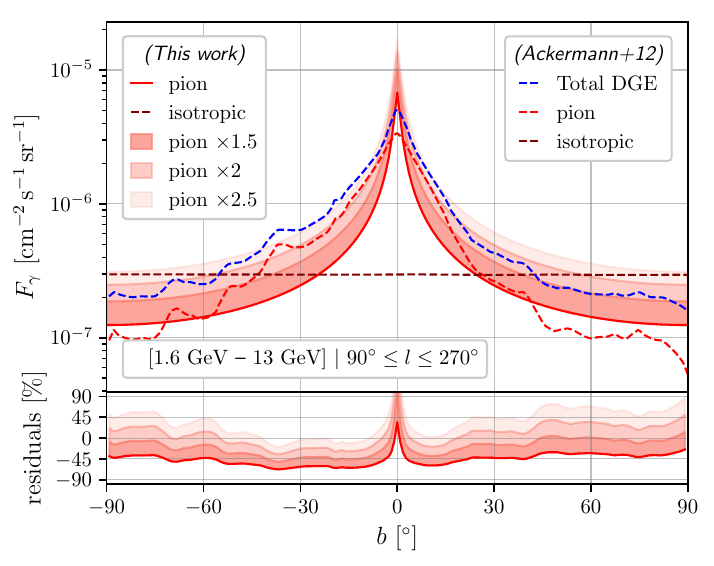}
\includegraphics[width = 0.49\textwidth]{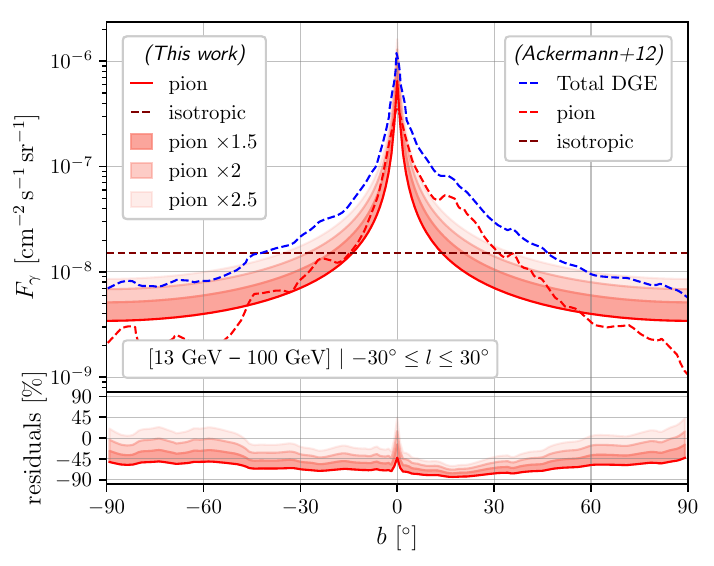}
\includegraphics[width = 0.49\textwidth]{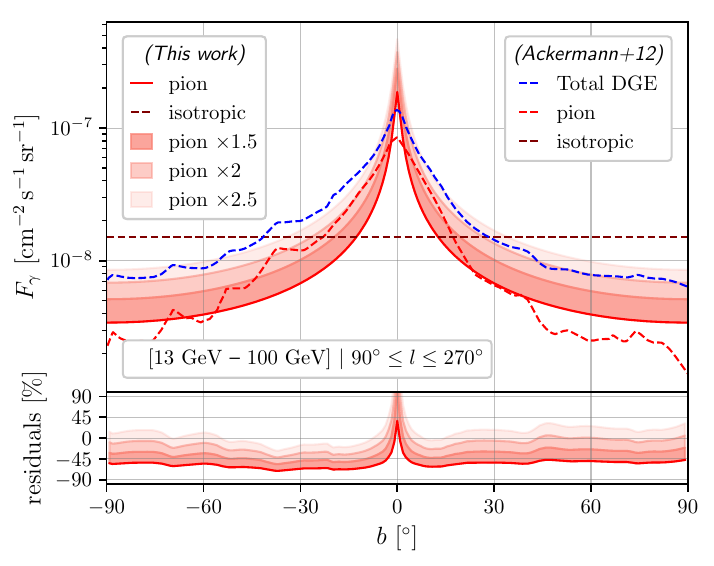}
\caption{\small {\bf Left panels:} Latitudinal profiles calculated from the flux given \citeeq{eq:pion_bckg} and integrated over two energy ranges, [1.6-13] GeV (top left panel) and [13-100] GeV (bottom left panel), and averaged in the inner galactic longitudinal range $-30^{\circ}\leq l \leq 30^{\circ}$. The model (solid curves) is compared with the Fermi data (dashed red curves for the pionic emission, dashed blue curves for the total DGE). At the bottom of  each plot, we the residuals of our rescaled pionic model (redish shaded bands) with respect to the total DGE inferred from the data. We also report the isotropic gamma-ray flux (brown dashed curves), for which our model is tuned to the one inferred from the data. {\bf Right panels:} Same as in the left panel, but averaged in the outer longitudinal range $90^{\circ}\leq l \leq 270^{\circ}$.}
\label{fig:bckg_profile}
\end{figure*}

Our full baryonic background flux is then given by
\ben
\label{eq:full_bckg}
\angleave{\frac{\dd \phi_{\rm b}(E,l,b)}{\dd E\,\dd\Omega}}{\delta\Omega_{\rm r}} &=& \alpha_{\pi\to{\rm DGE}}\,
\angleave{\frac{\dd \phi_\pi(E,l,b)}{\dd E\,\dd\Omega}}{\delta\Omega_{\rm r}} \\
&& + \angleave{\frac{\dd \phi_{\rm iso}(E,l,b)}{\dd E\,\dd\Omega}}{\delta\Omega_{\rm r}}\,,\nn
\een
\revise{where $\phi_{\rm iso}$ is the isotropic component that we directly extract from Ref.~\cite{AckermannEtAl2012}. We have explicitly introduced the tuning parameter $\alpha_{\pi\to{\rm DGE}}$, \revise{which will be further altered later} to mimic additional systematic uncertainties or missing sources of background.}

\revise{To conservatively match with the constrained DGE, we assume $\alpha_{\pi\to{\rm DGE}}=1.5$ unless specified otherwise. This allows our DGE background model never to exceed the genuine one, especially at large latitudes where mismodeling errors in the real data analysis are expected to be less important. This is at the cost of underestimating the DGE background by up to a factor of $\sim 2$ in some regions of the sky, which we will comment on in the final discussion but which anyway positions our forthcoming mock data analysis in the rather optimistic configuration as far as subhalo detection is concerned.}

\subsection{Number of subhalos above threshold and associated probability}
\label{ssec:pdet}
Before entering the details of the determination of the sensitivity to pointlike subhalos in our idealized model, hence of the detection threshold in terms of gamma-ray flux, it is useful to describe how we can translate a sensitivity estimate into a number of observable subhalos and associated probability. For given DM particle mass, annihilation cross section and channel, the gamma-ray flux is fully determined by the $J$~factor (see \citeeqp{eq:pt_flux}). Therefore the sensitivity to pointlike subhalos can be expressed in terms of a minimal $J$~factor that we call $J_{\rm min}$. Since the background is not isotropic, $J_{\rm min}=J_{\rm min}(l,b) = J_{\rm min}^{(l,b)}$.

The integrated probability for a pointlike subhalo to have a $J$~factor larger than $J_{\rm min}^{(l,b)}$ in the direction characterized by the angle $\psi(l,b)$ such that $\cos\psi=\cos b\,\cos l$ is given in \citeeq{eq:prob_jpsi}, for a resolution solid angle $\delta \Omega_{\rm r}$---see also \citefig{fig:prob_jpsi_pl}. We can further integrate this probability over the full sky, accounting for the fact that $J_{\rm min}$ depends on the pointing angle. We get
\ben
P_{\rm vis}^\text{pt} = \int db\,\cos b\int dl\, P_J^\text{pt} (J_{\psi}^\text{pt}\geq J_{\rm min}^{(l,b)})
\een
Here, $P_{\rm vis}^\text{pt}$ is normalized by construction in such a way that it is 1 for $J_{\rm min}^{(l,b)}=0$. From now on, we denote this probability $p$.

Given a total number of pointlike subhalos $N_{\rm pt}\simeq N_{\rm tot}$, the probability to detect $k$ subhalos is given by the binomial probability
\ben
P(k|N_{\rm pt}) = \binom{N_{\rm pt}}{k} \,p^k\,(1-p)^{N_{\rm pt}-k}\,.
\een
\revise{Since} in realistic situations we expect $k\ll N_{\rm pt}$ and $N_{\rm pt}\gg 1$, we can use the Poissonian limit of the previous equation,
\ben
\label{eq:prob_k}
P(k|N_{\rm pt}) \simeq \frac{\nu^k}{k!} e^{-\nu}\; \text{with}\; \nu \equiv N_{\rm pt}\,p\,.
\een
Therefore the probability to observe at least $n$ objects is given by
\ben
\label{eq:prob_geqn}
P(\geq n|N_{\rm pt}) \simeq 1 - \sum^{n-1}_{i=0} P(i|N_{\rm pt})\,.
\een
We can further consider the cumulative of the probability given in \citeeq{eq:prob_k} by promoting $k$ to a real number $x$, such that
\ben
P_x(x|N_{\rm pt}) = e^{-\nu} \sum_{k=0}^{\lceil x\rceil-1}\frac{\nu^k}{k!} =
\frac{\Gamma(\lceil x\rceil,\nu)}{\Gamma(\lceil x\rceil)}\,,
\een
where the $\Gamma$ functions in the denominator and in the numerator are the standard and incomplete gamma functions, respectively. We can then define a confidence interval at $100(1-c)\%$ that $x$ be measured in the range $[N_c^-,N_c^+]$ by solving
\ben
\frac{c}{2} = \frac{\Gamma(N_c^-+1,\nu)}{\Gamma(N_c^-+1)} =
1 - \frac{\Gamma(N_c^+,\nu)}{\Gamma(N_c^+)}\,.
\label{eq:n_unc}
\een
In the following, we use this formalism to determine the number of subhalos that could be observed with a Fermi-LAT-like observatory. The fundamental quantity that should now be characterized is the minimal $J$~factor, $J_{\rm min}^{(l,b)}$, that we address below.

\subsection{Sensitivity to pointlike subhalos}
\label{ssec:sensitivity}
\subsubsection{Specifications of our virtual Fermi-LAT-like instrument and of our DM benchmarks}
\label{sssec:specs}
Since we wish to address the potential of Fermi-LAT or any other similar experiment to detect subhalos, we first have to fix the main specifications that will be used to make predictions. These specifications need not match exactly those of Fermi-LAT, but need to be close enough to be quantitatively realistic\footnote{Details can be found on the dedicated \href{https://www.slac.stanford.edu/exp/glast/groups/canda/lat_Performance.htm}{Fermi-LAT} webpage.}. We do not seek for percent precision, but rather order 1 precision in terms of subhalo searches. We can therefore simplify the experimental characteristics such that they can be manipulated with ease at the level of calculations. Consequently, in the following, unless specified otherwise, we assume:
\bi
\item a search energy window\footnote{We restrict ourselves to a limited energy range where the effective area is constant. A maximum of 100~GeV allows a reach in WIM mass of $\sim 300$~GeV ($\sim 2$ TeV) for an annihilation in $\tau^+\tau^-$ ($b\bar b$) \cite{BergstroemEtAl1998a,CirelliEtAl2011,CaloreEtAl2017}.} of [1-100]~GeV with a flat effective area ${\cal A}$ of 0.9~m$^2$, and a field of view of 1/5 of the sky (consistent with the acceptance of $\sim 2.3$~m$^2$sr quoted in \cite{AtwoodEtAl2013,BruelEtAl2018}, and with the exposure of 2.7-$4.5\times 10^{11}$~cm$^2$s $=0.86$-1.43~m$^2$yr quoted in the fourth Fermi catalog and corresponding to 8~yr of data taking \cite{FermiLAT2019});
\item two benchmark resolution angles of $\theta_{\rm r}=0.1^\circ$ and $1^\circ$, with the latter to very roughly address the search for extended subhalos;
\item a uniform coverage of the sky.
\ei

For WIMP DM, we assume a default canonical $s$-wave thermal annihilation cross section
fixed to $\sigv = 3\times 10^{-26}\,{\rm cm^3/s}$ (neglecting changes with the WIMP mass, see \eg~\cite{CerdenoEtAl2012,SteigmanEtAl2012}), and consider the $b\bar b$ or $\tau^+\tau^-$ annihilation channels using the spectral tables provided in Ref.~\cite{CirelliEtAl2011}.

\subsubsection{A simplified but helpful warm-up statistical analysis}
\label{sssec:simple_stats}
We start with a very simple statistical method based on On-Off event number counting \cite{LiEtAl1983}. Given the gamma-ray fluxes for a pointlike source and associated background, we can very roughly define the sensitivity in terms of rudimentary Poisson statistics \cite{LiEtAl1983,ZhangEtAl1990,MattoxEtAl1996,CousinsEtAl2008}. For a subhalo of index $i$ located at position $\vec{s}_i$ in the observer's frame, and characterized by an angle $\psi_i$ and Galactic coordinates $(s_i,l_i,b_i)$, with $\cos\psi_i=\cos b_i\,\cos l_i$, we can estimate the number of gamma-ray events $N_\gamma^i$ collected in an arbitrary energy range $\Delta E$ by a telescope of time-area efficiency set by the effective collection area  ${\cal A}$ and and observation time ${\cal T}_{\rm obs}$. Neglecting for simplicity dependencies other than on energy for the effective collection area, this number of events reads
\ben
\label{eq:nsignal}
N_\gamma^i(l_i,b_i,\Delta E) &=&
\Delta E \,\left\langle \frac{\dd R^i}{\dd E}\right\rangle {\cal T}_{\rm obs}^i\,,
\een
with
\ben
\label{eq:def_EvtRate_sig}
\left\langle \frac{\dd R^i}{\dd E}\right\rangle&\equiv& 
\overline {\cal A S}_\chi(\mchi,\sigv,\Delta E)\,J_i\\
&=& \revise{\frac{\sigv}{2\,m_\chi^2}}  \frac{ \left\langle {\cal N}_\gamma {\cal A}\right\rangle_{\Delta E}}{\Delta E}\,J_i\,.\nn
\een
We have introduced the differential event rate $\dd R/\dd E$. The flux factor $J_i$ is given by \citeeq{eq:jpsi_pl}, and the spectral function ${\cal S}_\chi$ by \citeeq{eq:spectral_func}, with the effective collection area ${\cal A}$. Since this expression is for a point source, $J_i$ needs not be modified by the average over the experimental acceptance [see discussion below \citeeq{eq:jpsi_pl_th}].

Similarly, the number of background events is given by
\ben
\label{eq:nbg}
N_\gamma^{\rm bg}(l_i,b_i,\Delta E) &=&
\Delta E \,\left\langle \frac{\dd R^{\rm bg}}{\dd E}\right\rangle {\cal T}_{\rm obs}^{\rm bg}\,,
\een
with the background rate averaged over $\Delta E$
\ben
\left\langle \frac{\dd R^{\rm bg}}{\dd E}\right\rangle &\equiv& \frac{1}{\Delta E}
\int_{\Delta E}\dd E \int_{\delta\Omega_{\rm r}(E)} \dd\Omega \,
\frac{\dd \phi_\gamma^{\rm bg}(E,l_i,b_i)}{\dd E\,\dd\Omega} \, {\cal A}(E) \nn\\
&\simeq & \frac{\pi\,\theta_{\rm r}^2}{\Delta E}
\int_{\Delta E}\dd E \, \frac{\dd \phi_\gamma^{\rm bg}(E,l_i,b_i)}{\dd E\,\dd\Omega} \, {\cal A}(E)\,.
\label{eq:bg_evt_rate}
\een
Again, we have assumed that the angular efficiency is flat and maximal within the energy-dependent angular resolution $\theta_{\rm r}(E)$ of the instrument, such that $\Theta(\theta_{\rm r}(E)-\theta)$ can be traded for the solid angle domain $\delta\Omega_{\rm r}(E)$. The latest approximated equation assumes a vanishingly small energy-independent resolution angle and that the background flux varies by less than a statistical fluctuation within this angle. In that case the angular integral factorizes out, giving $2\,\pi(1-\cos\theta_{\rm r})\simeq \pi\,\theta_{\rm r}^2$. In the following, we actually neglect the energy dependence of $\theta_{\rm r}$ for the sake of simplicity, and because it has negligible impact on our results (it would have impact in studies of the Galactic center emission).

Without loss of generality, a pointlike source can be detected (or resolved, equivalently) when the number of signal events becomes larger than some threshold number $n_\sigma$ times the Poissonian fluctuation of background events, assuming the same exposure for both the signal and background. This can be expressed as
\ben
\label{eq:detection}
\frac{N_\gamma^i(l_i,b_i,\Delta E)}{\sqrt{N_\gamma^{\rm bg}(l_i,b_i,\Delta E)}}> n_{\sigma}\,.
\een
We can actually artificially absorb any exposure difference between the target and reference background in the number of fluctuations $n_\sigma$, which should then be thought of as an effective threshold number of order $\sim 1-10$ \cite{LiEtAl1983}. In the classical case of exact Poisson statistics with equal on- and off-source exposure, a detection threshold corresponds to $n_\sigma\geq 5$. From the above equation, we can define a minimal $J$~factor for a pointlike subhalo to be detected as follows:
\ben
\label{eq:def_jmin}
J_{\rm min}^{(l,b)}(\Delta E,\mchi,\sigv) &=& \frac{n_{\sigma} }{{\cal T}_{\rm obs}}
\frac{\sqrt{N_\gamma^{\rm bg}(l,b,\Delta E)}}
     {\overline {\cal A S}_\chi(\mchi,\sigv,\Delta E)}\\
     &=& \frac{n_\sigma}{\sqrt{{\cal T}_{\rm obs}}}
     \frac{2\,m_\chi^2}{\sigv}
     \frac{\sqrt{\Delta E \left\langle \frac{\dd R^{\rm bg}}{\dd E}\right\rangle}}
     {\langle{\cal N}_\gamma{\cal A}\rangle_{\Delta E}}\,.\nn
\een
This equation explicitly shows that the pointing-direction dependence of $J_{\rm min}^{(l,b)}$ is only set by that of the background. Even though obvious, this is an important point because in essence, this means that the most visible point-source subhalos (relative to background) may have different internal properties depending on the pointing direction, and are not necessarily the most intrinsically luminous in a background-free setting (detection probability does not necessarily correspond to luminosity probability). The dependence in $\sigv$ is rather trivial at first sight since $J_{\rm min}^{(l,b)}$ simply linearly increases as the annihilation cross section decreases. A quick inspection of the right panel of \citefig{fig:prob_jpsi_pl}, which shows the exponentially decreasing number of subhalos as a function of some threhold in $J$, already tells us that increasing a bit $J_{\rm min}^{(l,b)}$ can actually have a dramatic impact on the number of visible subhalos: if constraints on \sigv\ get stronger and stronger, the probability to detect subhalos is going to shrink accordingly, but exponentially. However, we will see below that this is less trivial if the constraint is set from the analysis of the diffuse Galactic emission, and if one insists on detecting subhalos before the smooth halo.

Eventually, one can translate $J_{\rm min}^{(l,b)}$ in terms of a threshold flux
\ben
\label{eq:def_phimin}
\phi_{\rm min}^{(l,b)}(\Delta E) &=& \int_{\Delta E} \dd E\,{\cal S}_\chi(\mchi,E)\, J_{\rm min}^{(l,b)}\\
&\propto & \sigv\,J_{\rm min}^{(l,b)}\nn\,,
\een
where the integral is performed over an arbitrary energy range $\Delta E$.
\subsubsection{Impact of different background configurations}
\label{sssec:diff_bgs}
The composite nature of the background affects the behavior of the sensitivity to pointlike subhalos. Here we inspect several background configurations still in the framework of the simplified statistical method introduced above. We first consider subhalo searches neglecting the baryonic foreground and accounting only for the smooth DM and unresolved subhalos background emission. Then we do the contrary, \ie~neglecting the diffuse DM contribution and considering only baryons. Finally, we study a more realistic background case including both the baryonic and diffuse DM contributions, and further derive the conditions for a subhalo to be detected {\em before} the diffuse DM component. As we will see, the latter configuration gives rise to asymptotic conditions that do depend neither on the annihilation cross section nor on the observation time. That result will actually be recovered by means of a more sophisticated statistical analysis resembling that used by the Fermi Collaboration.

\paragraph{\bf DM-only background model:}

Neglecting the baryonic background is obviously not realistic, but this allows us to figure out quickly where the most visible subhalos should concentrate in the sky, notably if the smooth halo were to be discovered first. These are not necessarily the most intrinsically luminous, since they still have to contrast with the background. However, in this case, the background is the lowest possible, \ie\ induced by DM itself (both the smooth halo and unresolved subhalos). That background configuration also leads to a dependence of the sensitivity to pointlike subhalos on the annihilation cross section different from the baryonic background case, which \revise{would rather characterize} subhalo searches after the detection of the smooth halo. In the DM-only case, the number of background events is given by
\ben
N_\gamma^{\rm bg}(l,b,\Delta E) &=& N_\gamma^{\rm bg/dm}(l,b,\Delta E)  =
N_\gamma^{\rm diff}(l,b,\Delta E) \nn\\
\label{eq:bg_dm}
&=&
\frac{\sigv}{2\,m_\chi^2}\left\langle {\cal N}_\gamma {\cal A}\right\rangle
\, \overline{J_\psi^{\rm diff}} {\cal T}_{\rm obs}^{\rm diff}\,,
\een
which implies
\ben
\label{eq:jmin_dmbg}
J_{\rm min}^{(l,b)}=J_{\rm min}^\psi & \propto & \frac{\theta_{\rm r}}{\sqrt{\sigv T_{\rm obs}}}\,
\Leftrightarrow \phi_{\rm min}^\psi \propto  \theta_{\rm r}\,\sqrt{\frac{\sigv}{T_{\rm obs}}}.
\een
The number of background events is therefore similar to that of signal events defined in \citeeq{eq:nsignal}, except for the $J$~factor of the diffuse DM component $\overline{J_\psi^{\rm diff}}$, defined in \citeeqs{eq:jdifftot} and \eqref{eqs:jpsi_obs}. Note that for an energy-independent resolution angle and a flat angular acceptance $\overline{J}_\psi^{\rm diff}= J_\psi^{\rm diff}$. Since the diffuse DM background is itself proportional to \sigv, the threshold $J$~factor $J_{\rm min}$ given in \citeeq{eq:def_jmin} scales like $1/\sqrt{\sigv T}$, and \revise{no longer} like $1/(\sigv\sqrt{T})$, which only holds when the background is independent of the DM annihilation rate. Consequently, paradoxically enough, even though the sensitivity to pointlike subhalos increases as \sigv\ increases ($J_{\rm min}$ decreases---see the right panel of \citefig{fig:prob_jpsi_pl}), the pointlike flux sensitivity $\phi_{\rm min}$ actually degrades because of the brighter background. The additional factor of $\theta_{\rm r}$ arises from the assumption that the diffuse background varies by less than a statistical fluctuation within the resolution angle of the instrument, see \citeeq{eq:bg_evt_rate}. That assumption essentially holds while not pointing toward the Galactic center, and \revise{implies} that both the subhalo and flux sensitivities degrade ($J_{\rm min}$ and $\phi_{\rm min}$ increases) when the resolution angle increases simply as a consequence of collecting more background photons.

\begin{figure*}[!th]
\centering
\hrule
{NFW Galactic halo - pointlike subhalos in diffuse DM-only background}\\
\includegraphics[width = 0.41\textwidth]{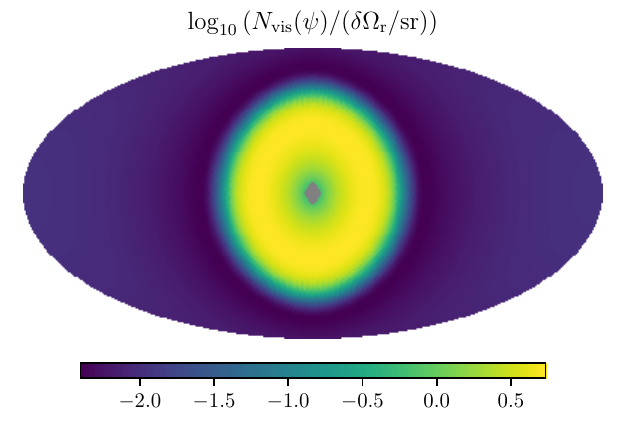}
\includegraphics[width = 0.41\textwidth]{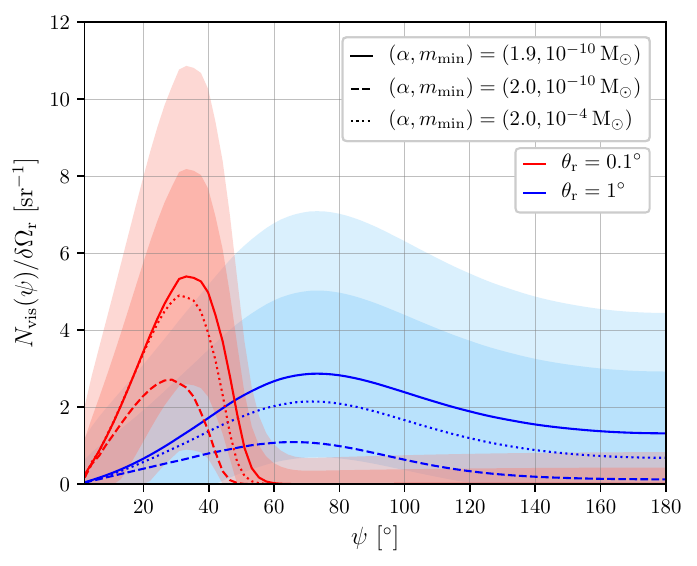}
\hrule
{Cored Galactic halo - pointlike subhalos in diffuse DM-only background}\\
\includegraphics[width = 0.41\textwidth]{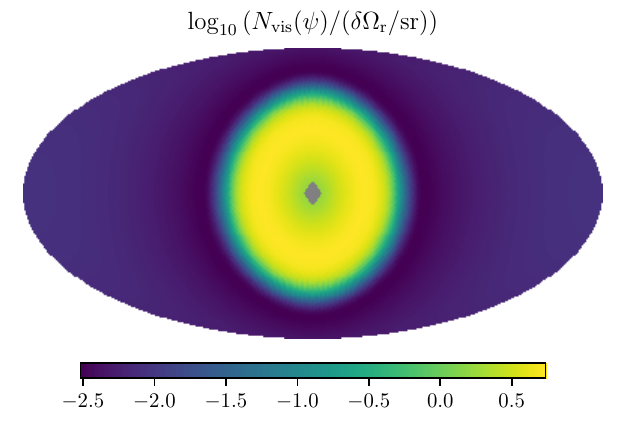}
\includegraphics[width = 0.41\textwidth]{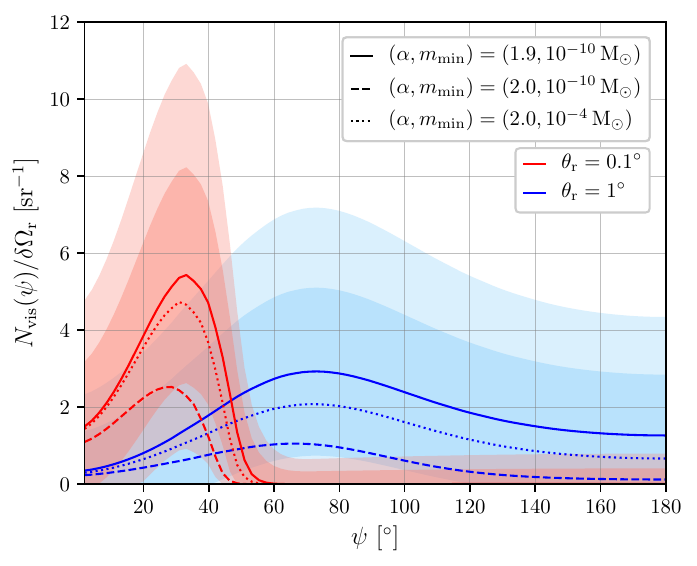}
\hrule
{Summary}\\
\includegraphics[width = 0.5\textwidth]{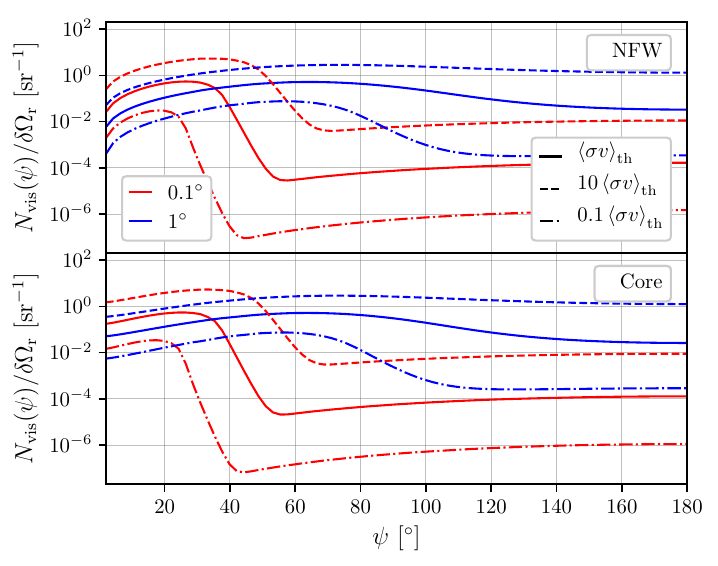}
\caption{\small {\bf Top left panel}: Sky map of the effective number of subhalos per solid angle unit in a DM-only background---assuming a WIMP mass of 100~GeV annihilating to $\tau^+\tau^-$ with $\sigv=3\times 10^{-26}$~cm$^3/$s, a gamma-ray energy range $1-100$~GeV, and a subhalo population configuration $(\alpha,m_{\rm min}/M_\odot,\epsilon_{\rm t})=(1.9,10^{-10},0.01)$ embedded in an NFW Galactic halo. {\bf Top right panel}: Associated angular distribution (with 95\% confidence band), with two angular resolutions $\theta_{\rm r}=0.1^\circ$ and 1$^\circ$, and several configurations for the subhalo population ranging in $(\alpha,m_{\rm min}/M_\odot)\in(1.9-2,10^{-10}-10^{-4})$. {\bf Middle left and right panels:} Same as above for subhalos embedded in a cored Galactic halo. {\bf Bottom panel}: Same as upper right panels, summarizing the angular distribution behavior for both the NFW (upper half) and cored Galactic halo (lower half), and for several annihilation cross sections around the canonical baseline $\sigv_{\rm th}=3\times 10^{-26}\,{\rm cm^3/s}$.}
\label{fig:nsub_dmbg}
\end{figure*}

In the left panels of \citefig{fig:nsub_dmbg}, we show sky maps of the effective number of visible subhalos per solid angle unit. They are computed using the nominal subhalo population model self-consistently embedded either within an NFW Galactic halo (top panels) or in a cored one (middle panels), and pointlike subhalos were defined by taking a resolution angle of $\theta_{\rm r}=0.1^\circ$. Although we consider the DM-only induced diffuse gamma-ray background for the moment, the subhalo population model still includes baryonic tidal stripping. The model parameters are set to $(\alpha,m_{\rm min}/M_\odot,\epsilon_{\rm t})=(1.9,10^{-10},0.01)$. We further assume WIMPs of 100~GeV annihilating into $\tau^+\tau^-$ with the canonical annihilation cross section, and restrict the spectral gamma-ray window to the [1-100]~GeV energy range---we define ``visible'' by demanding $n_\sigma\geq 3$ in \citeeq{eq:def_jmin}, taking an observation time of 10~yr. With this setup, we get $<1$ photon in the virtual detector, \revise{so the discussion here is only qualitative, and numbers should only be compared relatively between one another}. We see that visible subhalos concentrate in a ring around the Galactic center, whose width and peak actually depend on the subhalo sensitivity $J_{\rm min}^\psi$.

The right panels of \citefig{fig:nsub_dmbg} show the corresponding angular distributions as \revise{functions} of the line-of-sight angle $\psi$. \revise{They also show} the results obtained with a broader resolution angle of $\theta_{\rm r}=1^\circ$, as well as the impact of changing the mass slope $\alpha$ (1.9 or 2) and the minimal virial mass ($10^{-10}$ or $10^{-4}$~\msun)---the shaded areas correspond to the 68\% and 95\% statistical uncertainties, and are derived according to \citeeq{eq:n_unc}. It appears from these angular projections that in both NFW and cored Galactic halos, potentially visible subhalos for $\theta_{\rm r}=0.1^\circ$ are concentrated in a ring about the GC extending up to $\psi\sim 50^\circ$ with a peak around $\psi\sim 30^\circ$ (reddish curves). It also appears that a larger resolution angle of $\theta_{\rm r}=1^\circ$ drastically changes this angular distribution (blueish curves) due to two different effects: (i) as seen from \citeeq{eq:jmin_dmbg}, the sensitivity degrades simply as the detector integrates more background photons; (ii) changing the resolution angle allows bigger (hence intrinsically more luminous) subhalos to become point sources, and bigger subhalos are more \revise{efficiently destroyed} by gravitational tides in the central Galactic regions. As an outcome, increasing the angular window for individual subhalo searches has the effect of shifting the angular distribution to much larger values of $\psi$ (larger latitude, longitude, or both)---with a very flattened peak now around $\psi\sim 70^\circ$. The precise angular distribution of visible subhalos strongly depends on that of the diffuse background. The latter is affected by unresolved subhalos at large angles, which makes it important to include them \revise{as an additional background contribution}.

In contrast, changing the global DM halo from an NFW (top panels) to a cored profile (middle panels) does not significantly affect these features, except for enlarging the peaks toward \revise{lower} angles and slightly flattening them as well (\revise{there is} less diffuse background in the central regions, but also slightly less subhalos within the halo scale radius). Notice that in the DM-only background configuration, there are more visible subhalos in an NFW Galactic halo than in a cored one. This will actually be reversed when the baryonic foreground is added, which will degrade the sensitivity toward the central Galactic regions. A summary plot of the DM-only background case is presented in the bottom panel of \citefig{fig:nsub_dmbg}, where the level of background and subhalo sensitivity are varied by tuning \sigv\ instead---see \citeeq{eq:jmin_dmbg}.

Such trends are consistent with the Monte Carlo results obtained in \cite{HuettenEtAl2019}, which instead describe the distribution of the brightest point-source subhalos as a function of distance to the observer. We stress that these are not necessarily the most visible when contrasted with the diffuse background. Our analytical calculations have the advantage of very easily covering the full dynamical range and as many model configurations as necessary, in a very short CPU time.

In the right panels of \citefig{fig:nsub_dmbg}, we also explore the impact of changing the main subhalo population model parameters by taking different combinations within $(\alpha,m_{\rm min}/M_\odot,\epsilon_{\rm t})= (1.9-2,10^{-10}-10^{-4},0.01)$. It is well known that varying the minimal virial subhalo mass $m_{\rm min}$ has only significant (nonlogarithmic) impact for $\alpha>1.9$ (see \eg~\cite{LavalleEtAl2008,PieriEtAl2011,CharbonnierEtAl2012}). Therefore, we vary $m_{\rm min}$ only for $\alpha=2$. This self-consistently keeps the global Galactic halo profile (sum of all components) unchanged once it has been fixed (NFW or cored halo) in the SL17 model, and therefore remains consistent with kinematic constraints by construction. We see that $\alpha=1.9$ results in significantly more visible pointlike subhalos than $\alpha=2$. This might look surprising because the number of subhalos is much larger in the latter case, for a given $m_{\rm min}$. However, there are two compensating effects: (i) there are relatively bigger subhalos (hence more luminous) in the $\alpha = 1.9$ case because the mass function is less steep, and (ii) the diffuse background induced by unresolved subhalos (equivalently the boost factor) is larger in the $\alpha=2$ case. The impact of the unresolved subhalo contribution to the diffuse background can actually be evaluated by changing $m_{\rm min}$ from $10^{-10}$ to $10^{-4}\,M_\odot$, in the $\alpha=2$ case. This shrinks the total number of subhalos (hence that of unresolved) by orders of magnitude ($N_{\rm tot}\propto m_{\rm min}^{1-\alpha}$), but that depletion concerns only subhalos in the range $10^{-10}$-$10^{-4}\,M_\odot$, which are not massive enough to detach from the background. Therefore, increasing $m_{\rm min}$ in this mass range only reduces the DM-induced diffuse background emission, leading to more visible subhalos. \revise{One should still bear in mind that on general grounds}, increasing $m_{\rm min}$ corresponds to decreasing \mchi\ \cite{GreenEtAl2004,BringmannEtAl2007a}.

Finally, it would be tempting to discuss the absolute numbers of detectable subhalos read off from the angular distribution plots. Caution is of order though, since these numbers are for the moment based on the very rudimentary statistical analysis defined in \citeeq{eq:detection}, and the observation configuration used is such that there is $<1$ photon detected. A more refined statistical method will be presented later, but will actually not qualitatively change these results. Anyway, we already see from the right panels of \citefig{fig:nsub_dmbg} that even when turning the baryonic background off, the expected number of visible subhalos is or order ${\cal O}(1)$, which only slowly varies with \sigv\ and \revise{observation} time, as shown in \citeeq{eq:jmin_dmbg}.

\paragraph{\bf Baryon-only background model:}
Considering only the baryonic foreground is a common practice to estimate the sensitivity to pointlike subhalos (\eg~\cite{HooperEtAl2017,CaloreEtAl2019}), and amounts here to plug the foreground fluxes defined in \citesec{ssec:background} into \citeeq{eq:nbg}, such that
\ben
\label{eq:bg_cr}
N_\gamma^{\rm bg}(l,b,\Delta E) &=& N_\gamma^{\rm bg/cr}(l,b,\Delta E)\,,
\een
where the subscript cr stands for ``cosmic rays'' (we neglect unresolved conventional astrophysical sources here).

In the absence of DM-induced background, the sensitivity to pointlike subhalos simply scales like
\ben
J_{\rm min}^{(l,b)} \propto \frac{\theta_{\rm r}}{\sigv\,\sqrt{T_{\rm obs}}} \Leftrightarrow
\phi_{\rm min}^{(l,b)} \propto \frac{\theta_{\rm r}}{\sqrt{T_{\rm obs}}}\,,
\een
where we see that the flux sensitivity ($\phi_{\rm min}^{(l,b)}$) has the standard scaling in time, and does not depend on \sigv\ anymore as expected (it is fixed by the baryonic background within $\Delta E$); as for the sensitivity to subhalos ($J_{\rm min}^{(l,b)}$), it does obviously depend on \sigv. Therefore, the reach in terms of $J_{\rm min}^{(l,b)}$ improves faster with \sigv\ than in the DM-only background case---see \citeeq{eq:jmin_dmbg}. This has consequences in the determination of the number of visible subhalos, since the pdf of the $J$~factor is a sharp function of $J$---see \citefig{fig:prob_jpsi_pl}. However, one should bear in mind the \revise{previous approximate result that if detected after the diffuse DM component, in which case the latter adds up to the background, then} the dependency in \sigv\ becomes much shallower.

The corresponding sensitivity map of visible subhalos is shown in \citefig{fig:nsub_baryons_and_dmbaryons} (top left panel). To increase the contrast, we have masked a region defined by \revise{$\psi<20^\circ$} in the middle top panel. In the right top panel, we show the sky map obtained for $J_{\rm min}^{(l,b)}$, which defines the sensitivity map to pointlike subhalos, after masking the region $|b|<5^\circ$ where most of the conventional \revise{DGE} and of the Galactic sources concentrate, and which is less suited for subhalo searches. These maps have been derived from a full likelihood analysis performed on mock data, which will be extensively discussed later, but would be qualitatively the same if derived from the simplified statistical analysis introduced above. Further comparing \revise{them} with the maps of \citefig{fig:nsub_dmbg} still on the qualitative level (they have been inferred from a different map of $J_{\rm min}^{(l,b)}$ set by the DM-only background), we see a similar concentration of visible subhalos in the central regions of the MW, except for the degraded sensitivity in the disk. The sensitivity to subhalos is less attenuated toward the very center because the increasing smooth halo contribution to the background \revise{has been unplugged here}. The angular distribution of visible subhalos is not shown, but has similar trends as in \citefig{fig:nsub_dmbg}, except for the different angular dependence of the background, and the fact that it is independent from \sigv\ (the angular peak would be at lower angle).

\begin{figure*}[!th]
\centering
\hrule
{NFW Galactic halo - Visible pointlike subhalos assuming CR-induced background only}\\
\includegraphics[width = 0.325\textwidth]{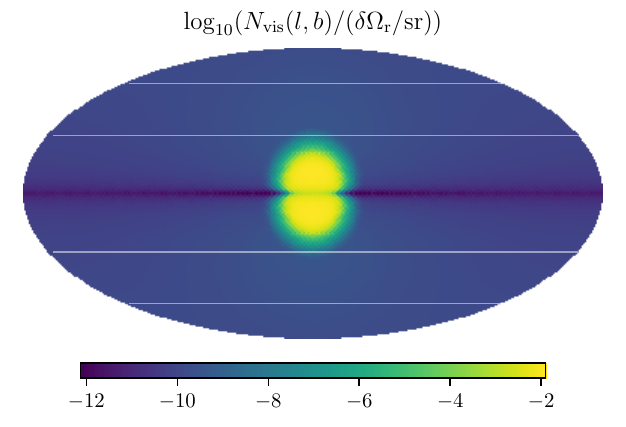}
\includegraphics[width = 0.325\textwidth]{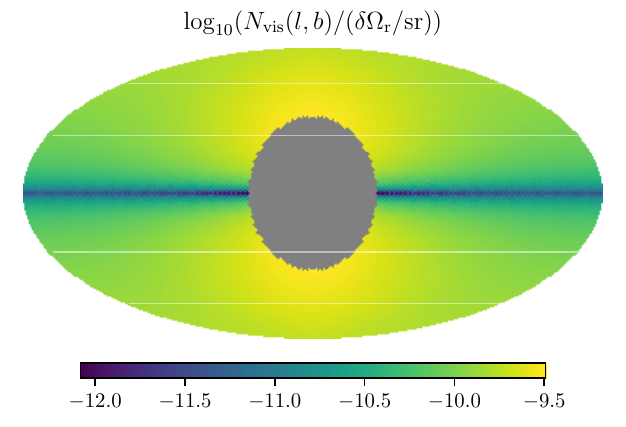}
\includegraphics[width = 0.325\textwidth]{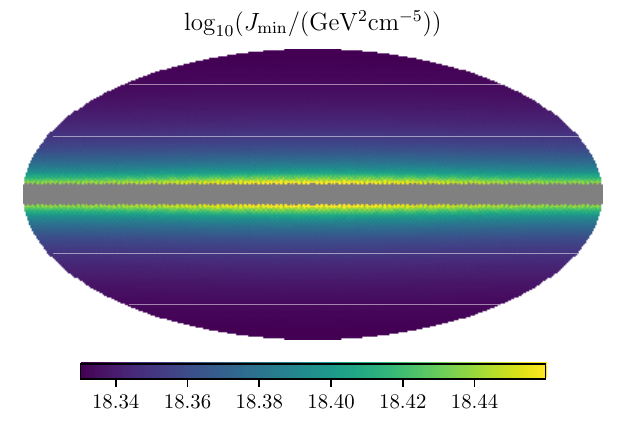}
\hrule
{NFW Galactic halo - Visible pointlike subhalos assuming DM+CR-induced background}\\
\includegraphics[width = 0.325\textwidth]{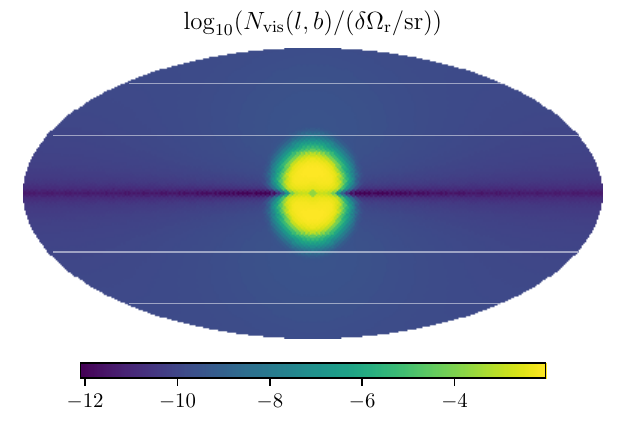}
\includegraphics[width = 0.325\textwidth]{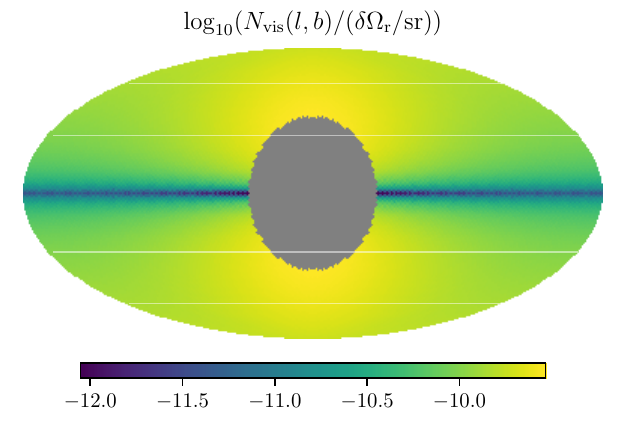}
\includegraphics[width = 0.325\textwidth]{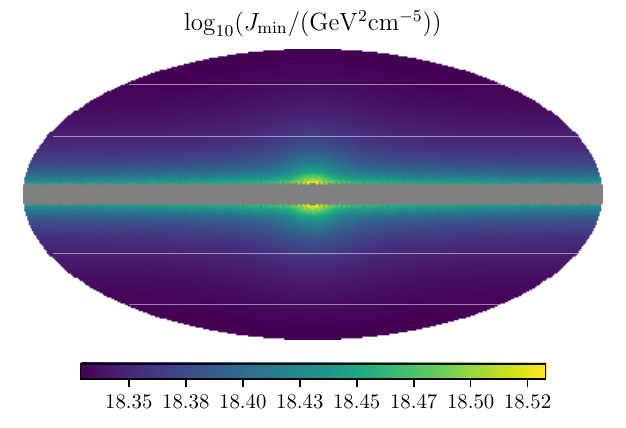}
\caption{\small Sky maps of the visible subhalos assuming a WIMP mass of 100~GeV annihilating into $\tau^+\tau^-$, and a subhalo population embedded in an NFW halo with parameters $(\alpha,m_{\rm min}/M_\odot,\epsilon_{\rm t})=(1.9,10^{-10},0.01)$. The annihilation cross section is fixed to the 3-$\sigma$ limit associated with the diffuse DM contribution. The detector configuration assumes a resolution angle of $0.1^\circ$, an observation time of 10~yr. The point-source sensitivity derives from a full likelihood analysis performed on mock data with parameters $(\alpha_{\rm b},\sigma_{\rm b})=(1.3,0.1)$ in ROIs of $0.2^\circ \times 0.2^\circ$, covering a region of $2.2^\circ \times 2.2^\circ$, and run over 5 logarithmic bins in the [1-100]~GeV energy range. Lines on maps indicate latitudes of $|b|=30^\circ,60^\circ$. {\bf Top panels}: Baryonic background only. {\bf Bottom panels}: \revise{Both baryonic and DM (smooth halo+unresolved subhalos) backgrounds.} {\bf Left panels}: Full sky. {\bf Middle panels}: Same sky map with central region $\psi<20^\circ$ masked to increase contrast. {\bf Right panels}: sky map of $J_{\rm min}^{(l,b)}$---sensitivity to pointlike subhalos---with $|b|<5^\circ$ masked.}
\label{fig:nsub_baryons_and_dmbaryons}
\end{figure*}

\paragraph{\bf Complete DM+baryon background model:}
\label{par:dM_bar_bg}
%
%
%

Finally, we consider a more realistic background model in which both the diffuse DM contribution and the baryonic foreground are included. The number of background events is now given by
\ben
N_\gamma^{\rm bg}(l,b,\Delta E) = N_\gamma^{\rm bg/cr}(l,b,\Delta E)+
N_\gamma^{\rm bg/dm}(l,b,\Delta E)\,,\nn\\
\een
where the number of DM-induced background events has been defined in \citeeq{eq:bg_dm}, and that of standard astrophysical processes in \citeeq{eq:bg_cr}.

To make this configuration even more realistic, we need to account for the fact that in the absence of departure from the background hypothesis, which is the current situation \cite{FermiLATEtAl2012,FornasaEtAl2015,ChangEtAl2018}, there are actually independent constraints on \sigv. Therefore, especially in the context of a consistent subhalo model in which all components of the MW are dynamically linked together, the sensitivity to subhalos inherently correlates with the sensitivity to the diffuse DM contribution. This needs to be properly considered.

The constraint on the diffuse DM contribution can be expressed as a limit on the annihilation cross section that derives, in this preliminary simplified statistical analysis, from the condition
\ben
\label{eq:lim_diff}
\frac{N_\gamma^{\rm diff}(l,b,\Delta E)}{\sqrt{N_\gamma^{\rm bg}(l,b,\Delta E)}}< \tilde n_{\sigma}\,,
\een
where $ \tilde n_{\sigma} = {\cal O}(1)$ can be considered as an effective number of background fluctuations below which the number of diffuse signal events must be confined to remain consistent with the background-only hypothesis. In the classical case of Poisson statistics, a $\sim$95\% ($\sim$99\%) confidence-level (C.L.) limit is usually set with $\tilde n_{\sigma} =2$ (3). Since current statistical tools in gamma-ray data analyses are well more advanced, as we shall see later, this number is only to be taken as indicative here. Assuming that $N_\gamma^{\rm bg/cr}\gg \tilde n_\sigma^2 > 1$, and that the diffuse DM signal remains unseen after an observational time $\tilde{\cal T}$, the above inequality becomes
\ben
\overline{\cal A S_\chi}\,\overline{J_\psi^{\rm diff}}\, \tilde{\cal T}< \tilde n_\sigma\,
\sqrt{\tilde{\cal T}\,\Delta E\,\left\langle \frac{\dd R^{\rm bg/cr}}{\dd E}\right\rangle}\,,
\een
where we have used Eqs.~[\eqref{eqs:jpsi_obs}, \eqref{eq:jdifftot}, and \eqref{eq:bg_evt_rate}]. This translates into an upper bound on the cross section:
\ben
\label{eq:sigvmax}
\sigv_{\rm max} &=&
\frac{2\,m_\chi^2\,\tilde n_\sigma}
     {\sqrt{\tilde{\cal T}}\,\left\langle {\cal N}_\gamma {\cal A} \right\rangle}\,
     \underset{(l_c,b_c)}{\rm min}\left\{
     \frac{\sqrt{\Delta E \,\left\langle \frac{\dd R^{\rm bg/cr}}{\dd E} \right\rangle}}
          {\overline{J_\psi^{\rm diff}}} \right\}\,.\nn\\
\een
We emphasize that the minimum appearing above within braces is uniquely determined for a given configuration of DM and baryonic foreground. It is found at Galactic coordinates $(l_c,b_c)$ (and may have replicates by symmetry). The scaling with \mchi\ is not fully explicit here, since the number of photons ${\cal N}_\gamma$ also depends on \mchi, almost $\propto\sqrt{\mchi}$ for a large variety of annihilation final states \cite{BergstroemEtAl1998a}; hence $\sigv_{\rm max}\overset{\propto}{\sim} m_\chi^{3/2}$.

\begin{figure*}[!th]
  \centering
  \hrule
  Limits on \sigv\ for a reference NFW Galactic halo\\
\includegraphics[width = 0.485\textwidth]{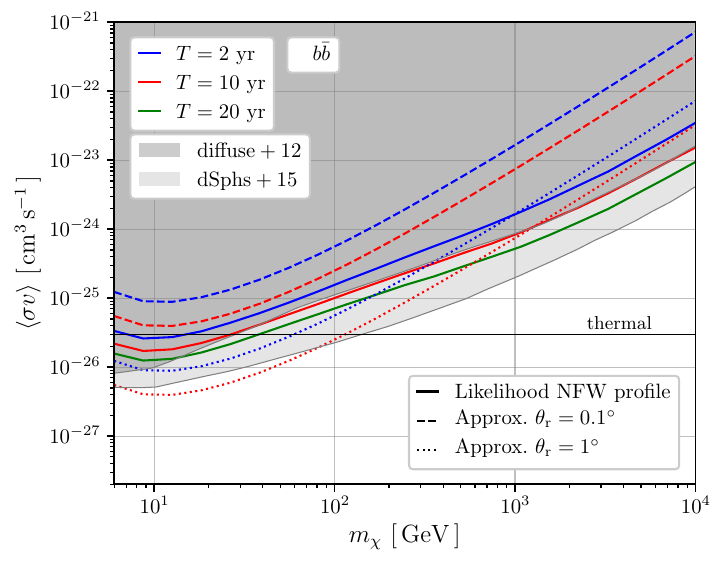}
\includegraphics[width = 0.485\textwidth]{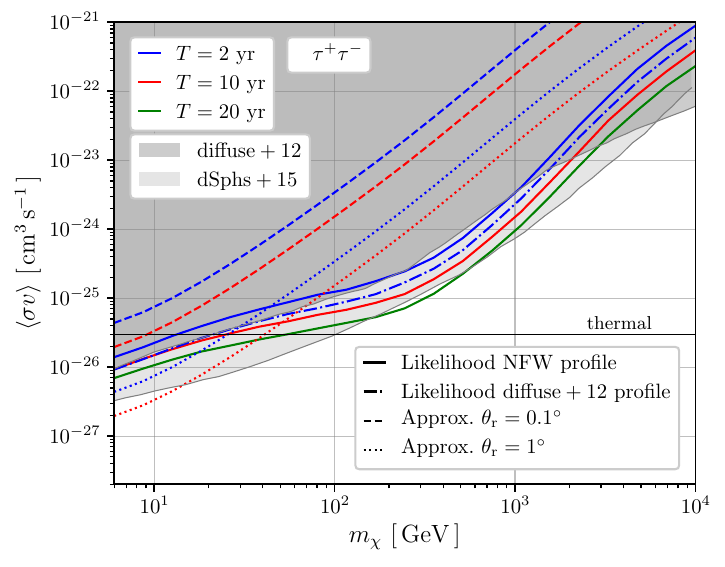}
  \hrule
  Limits on \sigv\ for a reference cored Galactic halo\\
\includegraphics[width = 0.485\textwidth]{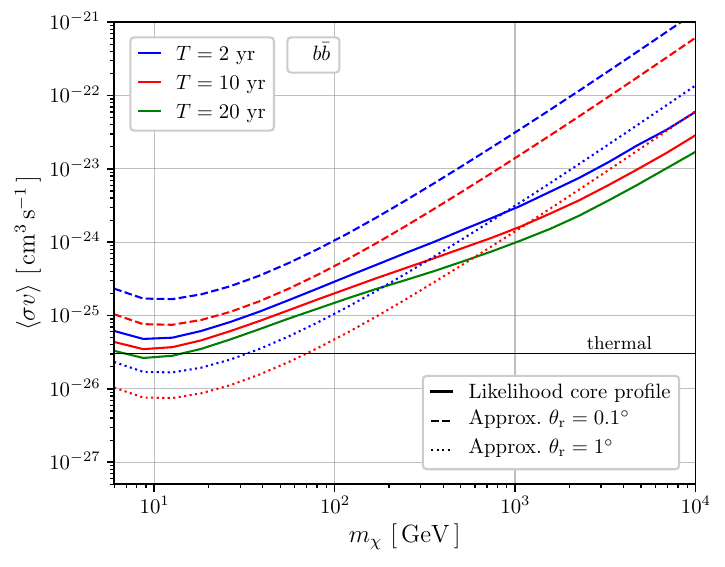}
\includegraphics[width = 0.485\textwidth]{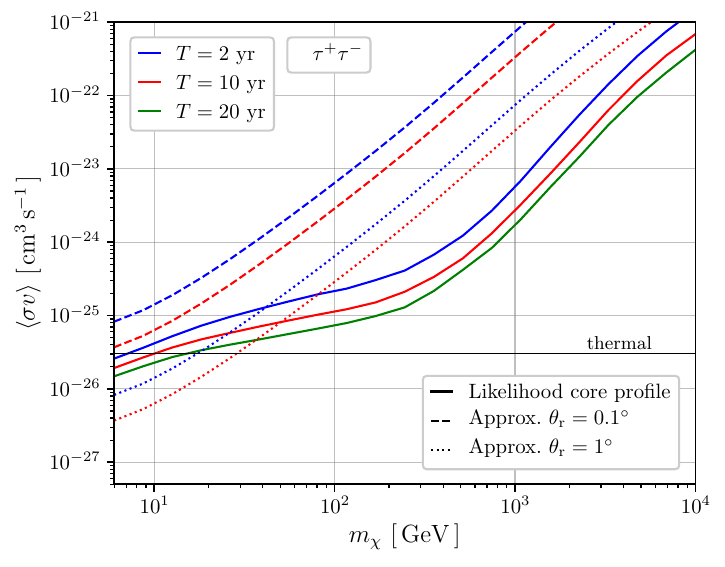}
\caption{\small Limits on \sigv, \ie~$\sigv_{\rm max}$, as a function of the WIMP mass \mchi\ for a Fermi-LAT-like telescope and for different observation times. Limits are set from: (i) the simplified statistical method presented in \citesec{par:dM_bar_bg}, with $\tilde{n}_\sigma=3$, an angular resolution $\theta_{\rm r}=0.1^\circ$ (dashed lines) or $\theta_{\rm r}=1^\circ$ (dotted lines), pointing to Galactic coordinates $(l_c,b_c)=(0^\circ,10^\circ)$; (ii) a full likelihood analysis performed on mock data, discussed in \citesec{sssec:likelihood}, and using background parameters $(\bar\alpha_{\rm b},\sigma_{\rm b}) =(1.3,0.1)$. The likelihood limits correspond to 3-$\sigma$ exclusion curves (solid curves). {\bf Top panels}: Limits for both our reference NFW halo and the halo shape used in the Fermi-LAT analysis (dubbed ``diffuse+12''---the dotted-dashed curve), together with the Fermi-LAT limits obtained from the diffuse Galactic emission \cite{FermiLATEtAl2012} (dark gray area), and from dwarf galaxies \cite{AckermannEtAl2015c,AlbertEtAl2017} (light gray area). {\bf Bottom panels}: Same for our reference cored halo profile. {\bf Left/right panels}: Full annihilation to $b\bar b$/$\tau^+\tau^-$ is assumed.}
\label{fig:sigvmax}
\end{figure*}

In \citefig{fig:sigvmax}, we show the results obtained using \citeeq{eq:sigvmax} for the determination of $\sigv_{\rm max}$ as a function of the WIMP mass \mchi, after integration of the gamma-ray fluxes in the $1-100$~GeV energy range and using typical efficiency parameters for \href{https://www.slac.stanford.edu/exp/glast/groups/canda/lat_Performance.htm}{Fermi}, recalled at the beginning of \citesec{sssec:diff_bgs}. We report the limits derived from the very simplified statistical analysis presented just above as dashed (for $\theta_{\rm r}= 0.1^\circ$) and dotted curves ($\theta_{\rm r}=1^\circ$, respectively), which have been obtained in a pointing direction $(l_{c},b_c)=(0^\circ,10^\circ)$---dubbed ``approx.'' in the legends. We assume DM annihilation into $b\bar{b}$ (left panels) and $\tau^+\tau^-$ pairs (right panels), use $\tilde n_\sigma=3$, and take two values for the observation time $\tilde{\cal T}$: 2 (blue), and 10~yr (red curves, respectively). We have considered both an NFW Galactic halo (top panels) and a cored halo (bottom panels). We compare our results with the limits obtained by the Fermi Collaboration from the analysis of the diffuse Galactic emission \cite{FermiLATEtAl2012} (dark gray area), using two years of data, and, for the sake of completeness, from satellite dwarf galaxies \cite{AckermannEtAl2015c,AlbertEtAl2017} (light gray area). We also report results from a more complete likelihood analysis that will be discussed later (solid and dotted-dashed curves). We see that the simplified approach underestimates the real experimental sensitivity by almost an order of magnitude. \revise{Notwithstanding, it has} a rather similar dependence in WIMP mass. The difference in sensitivity mostly comes from the fact that we use a single angular and energy bin, and therefore neglect a significant amount of available information. However, it is interesting to note that once we correctly rescale our effective sensitivity number $\tilde n_\sigma$, we can grossly match with the correct limit. This means that this simplified formalism may help capture the \revise{main dependencies and} asymptotic behavior of \revise{a more realistic sensitivity} to DM subhalos.

Assuming that the limit on \sigv\ reaches the upper bound $\sigv_{\rm max}$, \ie~the diffuse DM component is at the verge of being detected but is still not so, we can replace \sigv\ by $\sigv_{\rm max}$ in \citeeq{eq:def_jmin}. This provides us with a critical value for the pointlike subhalo detection threshold:
\ben
\label{eq:jmin_crit}
J_{\rm min}^{\rm crit}(l,b,\Delta E) &=& \eta_\sigma^{\rm eff}
\,\sqrt{\Delta E\,\left \langle\frac{\dd R^{\rm bg/cr}}{\dd E}\right \rangle} \\
&& \times \underset{(l_c,b_c)}{\rm max}
\Bigg\{
\frac{J_\psi^{\rm diff}}{\sqrt{\Delta E\,\langle  \dd R^{\rm bg/cr}/\dd E \rangle}}
\Bigg\}\nn\,,
\een
where
\ben
\eta_\sigma^{\rm eff}\equiv \frac{n_\sigma}{\tilde n_\sigma} \sqrt{\frac{\tilde{\cal T}}{\cal T}}
\approx \frac{n_\sigma}{\tilde n_\sigma} \,.
\een
Interestingly, this critical $J$~factor does not depend on the annihilation cross section anymore. Note that the background event rate $\langle \dd R^{\rm bg/cr}/\dd E\rangle$ is calculated at Galactic coordinates $(l_c,b_c)$ in the max term, while it is calculated at the target coordinates $(l,b)$ outside from the max term---all this is therefore fixed for a \revise{given} Galactic emission model. \revise{It turns convenient to combine the dependencies in the different observation times $\tilde{\cal T}$ (used to set the limit on \sigv) and ${\cal T}$ (on-subhalo-target time) and in the fluctuation thresholds $\tilde{n}_\sigma$ and $n_\sigma$ into a single effective sensitivity parameter $\eta_\sigma^{\rm eff}$. In pure Poisson statistics associated with an on-off method, and with ${\cal T}\sim \tilde{\cal T}$, we should have $\eta_{\sigma}^{\rm eff}\sim n_\sigma/\tilde{n}_\sigma\approx 5/2$ or 5/3. However, connecting with more advanced statistical analysis methods and different observational strategies allows for considering a much wider range of values, say ${\cal O}(1-10)$ per energy bin.}


\revise{For non-pointing experiments, like Fermi-LAT, $\tilde{\cal T} \approx {\cal T}$, and $J_{\rm min}^{\rm crit}$ further becomes a priori time-independent (this holds in the large-event-number limit). The fact that $J_{\rm min}^{\rm crit}$ is independent from both the annihilation cross section and the observation time (in the infinite-time limit) is, though derived from strongly simplifying assumtpions here, a very important result.} It is actually recovered when using a more sophisticated statistical analysis as we will see later. It means that we can rigorously answer the question of whether or not subhalos can be detected before the diffuse DM component, should DM self-annihilate and produce gamma-ray photons. Indeed, the derivation of $J_{\rm min}^{\rm crit}$ is based upon requiring the diffuse DM contribution to remain below the baryonic background. Therefore, irrespective of the annihilation cross section, one can simply infer the number of observable subhalos by integrating the probability distribution function of the $J$~factor shown in \citefig{fig:prob_jpsi_pl} above $J_{\rm min}^{\rm crit}$. If one finds the minimal $J$~factor needs to be lower than this critical value to get a sizable number of observable subhalos, then that means that subhalos could hardly be detected as individual sources before the smooth Galactic DM halo itself. 

That $J_{\rm min}^{\rm crit}$ does not explicitly depend on time needs further explanation. As said above, it is defined from the sensitivity $J_{\rm min}\propto (\sqrt{\cal T}\sigv)^{-1}$ (see \citeeqp{eq:def_jmin}), but evaluated at the maximal cross section $\sigv_{\rm max}\propto 1/\sqrt{\cal T}$ (see \citeeqp{eq:sigvmax}). This explains why the time dependence disappears in our simplified analysis. However, even though $J_{\rm min}^{\rm crit}$ is roughly expected to be time independent, it must still be associated with the time-dependent maximal annihilation cross section $\sigv_{\rm max}$.

\revise{In \citefig{fig:jmin_crit}, we trace $J_{\rm min}^{\rm crit}$ as a function of observation time from the simplified definition of \citeeq{eq:jmin_crit} on the one hand (with a conveniently rescaled $\eta_\sigma^{\rm eff}$---blue dashed curve), and from a more sophisticated likelihood analysis of mock data that will be discussed below (blue solid curve). When inferred from the simplified analysis, $J_{\rm min}^{\rm crit}$ is independent of time, as explained above. In slight contrast, it becomes flat only after a time of several years when inferred from a full likelihood analysis, because the latter correctly deals with the statistics of small numbers of events, but still asymptotically confirms the prediction obtained from the simplified method. The left and right panels differ only by the resolution angle (see caption). We also report the time-dependent sensitivity to pointlike subhalos $J_{\rm min}$ (in the direction where $J_{\rm min}$ is minimized) as a function of time, assuming an annihilation cross section set by a 3-$\sigma$ limit on the diffuse DM flux after 10~yr ($\sigv = \sigv_{\rm max}(10\,{\rm yrs})$, red curves) or 20~yr ($\sigv = \sigv_{\rm max}(20\,{\rm yrs})$, green curves)---the latter being $\sim\sqrt{2}$ smaller. The $J_{\rm min}$ curves cross the critical $J_{\rm min}^{\rm crit}$ ones at the corresponding times, as they should. Beyond these special crossing times, the decrease of $J_{\rm min}\propto 1/\sqrt{T}$ holds true assuming the diffuse DM-induced emission has truly been detected at these times. If not, then one should keep on following the critical blue lines until the detection of the diffuse emission (time from which $J_{\rm min}$ scales like $\propto 1/\sqrt{T}$ again). Therefore, if the values of $J_{\rm min}$ needed to detect a sizable number of subhalos lie below $J_{\rm min}^{\rm crit}$, that means that one should detect the diffuse DM-induced emission first.}

\begin{figure*}[!th]
\centering
\includegraphics[width = 0.495\textwidth]{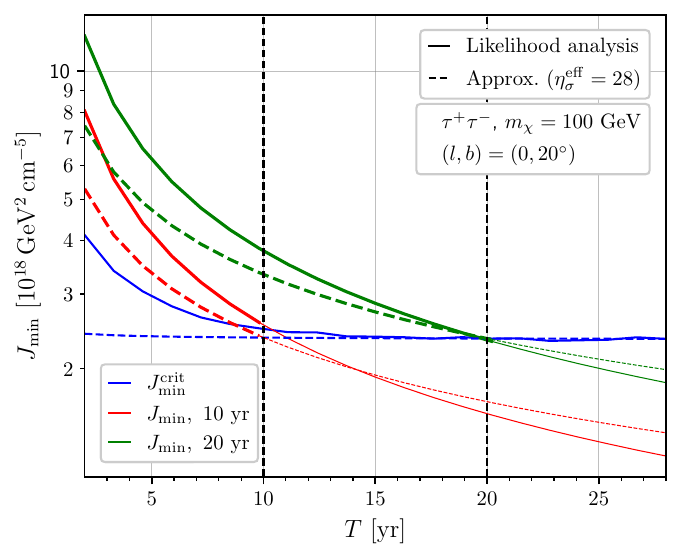}
\includegraphics[width = 0.495\textwidth]{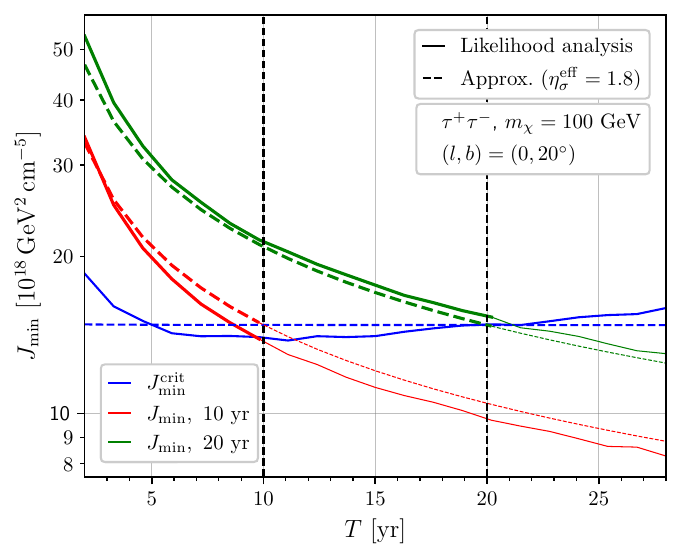}
\caption{\small Minimal $J$~factor (sensitivity to pointlike subhalos) as a function of time, assuming a subhalo population embedded in an NFW Galactic halo, and a WIMP of 100~GeV annihilating into $\tau^+\tau^-$. Solid curves are obtained from the full likelihood analysis of mock data, and dashed curves from the simplified statistical analysis, with a rescaled effective sensitivity parameter $\eta_\sigma^{\rm eff}$. \revise{Blue curves indicate the critical $J$~factor $J_{\rm min}^{\rm crit}$ at which the smooth DM contribution remains at its 3-$\sigma$ limit. Red curves show $J_{\rm min}(T)$ assuming an annihilation cross section set from the likelihood 3-$\sigma$ limit (non-detection of the smooth DM contribution) for 10~yr of observation, while green curves show $J_{\rm min}(T)$ assuming a lower annihilation cross section set from the likelihood limit for 20~yr (see \citefig{fig:sigvmax}). As expected, the red and green curves cross the critical blue ones at 10~yr and 20~yr, respectively.} {\bf Left panel}: $\theta_{\rm r}=0.1^\circ$. {\bf Right panel}: $\theta_{\rm r}=1^\circ$ (mimicking the sensitivity to extended sources).}
\label{fig:jmin_crit}
\end{figure*}

\subsubsection{A full likelihood analysis of mock data}
\label{sssec:likelihood}
In order to validate the previous results, we upgrade our statistical analysis method to get closer to the standards employed in the Fermi Collaboration for both the smooth Galactic DM searches \cite{FermiLATEtAl2012,FornasaEtAl2015,ChangEtAl2018} and the subhalo or pointlike source searches \cite{BelikovEtAl2012,BertoniEtAl2015,SchoonenbergEtAl2016,MirabalEtAl2016,HooperEtAl2017,CaloreEtAl2017,CaloreEtAl2019,GlawionEtAl2019,CoronadoBlazquezEtAl2019,CoronadoBlazquezEtAl2019a,FermiLATCollab2015,FermiLAT2019}. We therefore set up a full likelihood analysis.

\paragraph{\bf Mock data generation}:
We first generate mock data based on the signal and background configurations discussed above. However, here, we need to add a layer of subtlety. Indeed, to be as realistic as possible, we want to artificially reproduce the fact that like in the Fermi data analysis, our background model be not perfect, and that positive fluctuations arising from uncontrolled systematic effects degrade the sensitivity to DM searches. We also want to implement the fact that so far the smooth DM has not been convincingly detected. Therefore, our mock data will be based on a biased version of our baryonic diffuse emission model introduced in \citesec{ssec:background}, which will leave room for positive fluctuations possibly interpreted as DM annihilation in the absence of systematic uncertainties. To make it simple, the bias will simply amount to a systematic shift by 30\% of the Galactic baryonic foreground, \revise{inspired by the value of residuals found in the Fermi-LAT analysis \cite{FermiLATEtAl2012}}.

\revise{We divide the sky into $N_{\theta}$ angular bins labeled $i$ (also called pixels in the following) each divided into $N_{\rm E}$ energy bins labeled $j$. We denote $b_{ij}$ the averaged number of photons expected from our background emission model and instrumental specifications [see \citesec{ssec:background}, \citesec{sssec:specs}, \citeeqs{eq:nbg} and \eqref{eq:bg_evt_rate}] in a given two-dimensional (2D) bin. We then generate our mock data by drawing a corresponding number of gamma-ray photons $n_{ij}$ in that bin according to a Poisson distribution}
\ben
\revise{p(n_{ij} \,|\, b_{ij} ) =  \frac{b_{ij}^{n_{ij}}}{n_{ij}!} e^{-b_{ij}}\,.}
\label{eq:pmock_data}
\een
\revise{Note that since our goal is to set limits, we do not generate any signal event in our mock data.}

\revise{In \citefig{fig:mock_data}, we show an example of such mock data for a collection time of 2~yr in pixels of size $1^\circ \times 1^\circ$. We get $\sim 360,000$ photon events in the $1-100$~GeV energy range and in the selected ROI ($5^\circ<|b|<15^\circ$ and $|l|<80^\circ$), which is very close to the number count found in Ref.~\cite{FermiLATEtAl2012}  ($\lesssim 5\%$ larger). To further account for the point-source subtraction performed in the Fermi data analysis we remove $\sim 25\%$ of the bins randomly over the sky, which defines our initial sample of $\sim 270,000$ collected photons, still very close to the statistics used in Ref.~\cite{FermiLATEtAl2012}.} These mock data are further processed through a likelihood analysis discussed just below, which consists of two different steps: (i) setting the limit on \sigv\ from the diffuse emission; (ii) defining the sensitivity to pointlike subhalos.

\begin{figure*}[th!]
\centering
\includegraphics[width = 0.8\textwidth]{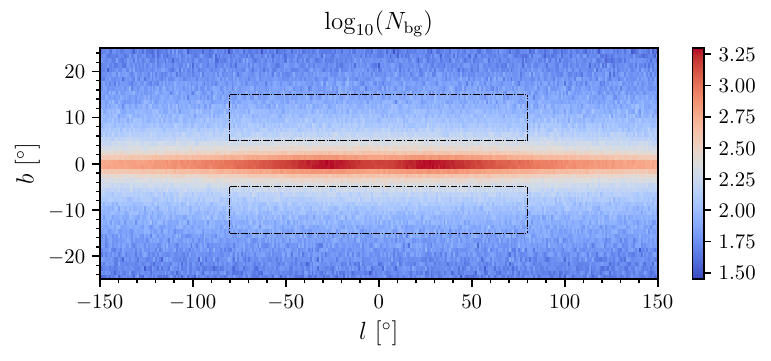}
\caption{Mock photon count map of biased background photons received in bins of size $1^\circ \times 1^\circ$ in the range $1-100$ GeV. The contour areas correspond to the ROI used to set constraints on \sigv.}
\label{fig:mock_data}
\end{figure*}

\paragraph{\bf Likelihood analysis of the diffuse emission: Limits on \sigv}:
In order to analyze our mock data, we set up a likelihood analysis similar to the one performed by the Fermi-LAT Collaboration to get limits on the diffuse Galactic DM-induced emission. \revise{We also want to account for the fact that significant fluctuations arise between the background model and the data due to an unperfect background modeling, which makes the likelihood possibly underestimate the background by $\sim 30\%$ \cite{FermiLATEtAl2012}. To this aim, we introduce a bias parameter $\alpha_{\rm b}$ that will be applied to the DGE background only. It is centered around a reconstruction efficiency $\varepsilon_{\rm rec}=0.7$, and with a Gaussian probability distribution such that an unbiased distribution would cost a $\sim 3$-$\sigma$ penalty, which is obtained from a Gaussian width $\sigma_{\rm b}=0.1$. This biasing procedure will mechanically degrade the limit derived on the DM annihilation signal, in the same vein as in the conservative analysis performed in Ref.~\cite{FermiLATEtAl2012}. These values for the bias parameters are inspired from the numbers quoted in Ref.~\cite{FermiLATEtAl2012}, and allow us to derive a limit on the annihilation cross section from our mock data analysis in reasonible agreement with the corresponding results. Changing the values of $\varepsilon_{\rm rec}$ and $\sigma_{\rm b}$ would not change our results qualitatively, keeping the final orders of magnitude unchanged.}


\revise{We can now construct a simple likelihood function to fit our signal and background models with a limited number of parameters (for a given WIMP mass \mchi): the annihilation cross section \sigv\ and the background bias parameter $\alpha_{\rm b}$. The chance of getting a number $n_{ij}$ of photons in bin $(i,j)$ can be estimated from the likelihood function}
\ben
{\cal L}_{ij}(n_{ij} \, | \, \sigv, \alpha_{\rm b})
&=& \frac{\left(a_{ij} \sigv + \alpha_{\rm b} \, b_{ij}^{\rm DGE} + b_{ij}^{\rm iso}\right)^{n_{ij}}}{n_{ij}!}\nn\\
\label{eq:Li_limit}
&\times &e^{-(a_{ij} \sigv + \alpha_{\rm b} b_{ij}^{\rm DGE}+ b_{ij}^{\rm iso})}\\
&\times&
\left\{  {\cal L}_{\rm sys}(\alpha_{\rm b}) \equiv
\frac{e^{-\frac{(\alpha_{\rm b} -\varepsilon_{\rm rec})^2}{2\sigma_{\rm b}^2}}}{\sqrt{2\pi \sigma_{\rm b}^2}} \right\}\,,\nn
\een
where $a_{ij}$ is defined such that the averaged number of photons expected from DM annihilation in bin $(i,j)$ be $s_{ij} = \sigv a_{ij}$, and $b_{ij}^{\rm DGE}$ and $b_{ij}^{\rm iso}$ stand, respectively, for the DGE and isotropic backgrounds given in \citeeq{eq:full_bckg}. ${\cal L}_{\rm sys}$ is our nuisance function that adds up a Gaussian penalty of $\sigma_{\rm b}$ if the bias parameter $\alpha_{\rm b}$ \revise{departs from the unperfect reconstruction efficiency $\varepsilon_{\rm rec}<1$}. This helps the model get closer to the mock data from below, while not too close to leave room for a possible DM contribution. This trick mimics a mismodeled background, which typically leads to 30\% fluctuations around the background-only hypothesis in the real Fermi data analysis \cite{FermiLATEtAl2012,FornasaEtAl2015,ChangEtAl2018}. This parametrizes our uncertainty in the background model, and allows us to calibrate our likelihood analysis to get results close enough to past or current data analyses, and then to more confidently extrapolate it to future times. Such a likelihood function is often called a profiled likelihood, because it is not normalized to unity with respect to the data. The total profiled likelihood associated with all bins is given by
\ben
{\cal L}( \sigv, \alpha)  = \prod_{ij} {\cal L}_{ij}(n_{ij} \, | \, \sigv, \alpha_{\rm b})\,. 
\label{eq:Ltot_limit}
\een

Equipped with this improved statistical setup, the first step is to find the best-fitting parameters of the model (including both the signal and the background), which we denote $(\widehat{\sigv}, \widehat{\alpha}_{\rm b})$ for a given WIMP mass and given annihilation channels. \revise{Whenever the isotropic background can be neglected (\eg~in the central Galactic regions at low energy), we could proceed semi-analytically, as explained in \citeapp{sapp:L_diff}. However, in the general case, we have to resort to the numerical method presented in \citeapp{sapp:newton}.}

Eventually, to set a conservative upper bound on $\sigv$ without directly comparing the background-only and the signal-and-background hypotheses, we standardly define our null hypothesis as our signal-and-background best-fitting model characterized by $(\widehat{\sigv},\widehat{\alpha}_{\rm b})$, and compute the likelihood ratio to that null hypothesis,
\ben
{\cal R}(\sigv) \equiv
\frac{{\cal L}(\sigv, \tilde{\alpha}_{\rm b}(\sigv))}{{\cal L}(\widehat{\sigv}, \widehat{\alpha}_{\rm b})}
\,.
\label{eq:L_ratio}
\een
Here, $\tilde{\alpha}_{\rm b}(\sigv)$ characterizes the best-fitting model for which $\sigv$ is now a fixed parameter.

Let us now present as clearly as possible the way we set a limit, and its precise statistical meaning. Wilks' theorem \cite{Wilks1938,Wilks1962} tells us that when the number of data points goes to infinity, under the condition that the null hypothesis holds true, the log-likelihood ratio defined as $- 2 \ln {\cal R}(\sigv)$ satisfies a $\chi^2(1)$ distribution \cite{CowanEtAl2011}, where the probability density of $\chi^2(k)$ is given by
\ben
f_{\chi^2(k)}(x) \equiv \frac{1}{2^{k/2}\Gamma(k/2)}x^{k/2-1} e^{-x/2} \, .
\een
If we denote $p_0$ the probability to have $-2 \ln {\cal R}(\sigv) > t$ under the null hypothesis, then $t$ is implicitly defined by
\ben
p_0 &=&  \int_t^\infty dy\, f_{\chi^2(1)}(y) \nn \\
&=& \int_t^\infty  dy \frac{1}{\sqrt{2\pi y}} e^{-y/2} \nn\\
&=& \sqrt{\frac{2}{\pi}} \int_{\sqrt{t}}^\infty  \dd x\, e^{-x^2/2}\,. 
\een
Therefore, if we demand a constraint at $\tilde{n}_\sigma \sigma$, then this translates into 
\ben
p_0 &=& 1 - \frac{1}{\sqrt{2\pi}}\int_{-\tilde{n}_\sigma}^{+ \tilde{n}_\sigma} \dd x\, e^{-x^2/2}\nn\\
&=& \sqrt{\frac{2}{\pi}}   \int_{\tilde{n}_\sigma}^{\infty} \dd x\, e^{-x^2/2} \,,
\label{eq:def_stat_limit}
\een
which implies from the previous equation that $t = \tilde{n}_\sigma^2$.

To summarize, a limit at $\tilde{n}_\sigma \sigma$ can be set by looking for the value of $\sigv$ such that $-2 \ln {\cal R}(\sigv) = t = \tilde{n}_\sigma^2$. If instead we want to define the limit from the probability itself, for example $p_0 = 0.05$ (equivalently a limit at 95\% confidence level), then we just have to solve
\ben
{\rm erfc}\left[\sqrt{\frac{t}{2}}\right] = 0.05\,,
\een 
which has solution $t \simeq  3.85$. Actually, parameter $t$ represents here what is generically called the test statistics (TS) \cite{MattoxEtAl1996} in Fermi-LAT data analyses.

We use this likelihood approach to derive limits on \sigv\ from the analysis of our mock data. This limit is important to assess whether pointlike subhalos can be detected before or after the DM-induced diffuse emission itself. It is the likelihood equivalent to $\sigv_{\rm max}$ defined in \citeeq{eq:sigvmax} and derived from our simplified statistical analysis. It fully determines $J_{\rm min}^{\rm crit}$ (see \citeeqp{eq:jmin_crit}), the critical threshold $J$~factor for subhalos, \revise{below which subhalos cannot be detected before the DM-induced diffuse emission itself}.

We first check whether the limit we get is consistent with the one derived by the Fermi-LAT Collaboration in Ref.~\cite{FermiLATEtAl2012}, calculated with two years of data. \revise{In fact, this comparison can help us check whether the unperfect reconstruction efficiency $\varepsilon_{\rm rec}$, which characterizes the bias between the background model and the mock data, and the Gaussian penalty $\sigma_{\rm b}$ paid by the reconstruction likelihood to catch up with the mock data, provide a realistic analysis framework.} Indeed, these parameters are meant to inject a tunable systematic error that degrades the limit on \sigv, in order to more correctly fake the results obtained from the real data analysis performed in Ref.~\cite{FermiLATEtAl2012}.

\revise{We analyze the mock data introduced before by selecting the same ROI as in Ref.~\cite{FermiLATEtAl2012}, \ie~$5^\circ<|b|<15^\circ$ and $|l|<80^\circ$, which we divide into $160\time 20$ angular bins of size $1^\circ\times 1^\circ$. We collect photons in an energy range of $1-100$~GeV further split into 5 logarithmic bins, using the experimental specifications listed in \citesec{sssec:specs} with a flat angular resolution of $\theta_{\rm r}=0.1^\circ$. After removal of virtual point sources randomly distributed in 25\% of the available pixels, we reach a total of $\sim 270,000$ collected photons in the ROI after two years, similar to the statistics found in Ref.~\cite{FermiLATEtAl2012}. Setting our systematic bias parameters to $\varepsilon_{\rm rec}=0.7$ and $\sigma_{\rm b}=0.1$ in the likelihood function, we derive the limits shown as solid curves in \citefig{fig:sigvmax} (using our Galactic halo model). We also report the likelihood limit inferred from the very same NFW halo parameters as in Ref.~\cite{FermiLATEtAl2012} as the dotted-dashed curve (top right panel, $\tau^+\tau^-$ channel), which can be more directly compared with the limit derived in Ref.~\cite{FermiLATEtAl2012} (dark gray shaded area). We see that in spite a slight and systematic underestimate of the genuine limit, the ``spectral'' agreement is quite reasonable up to WIMP masses of $\lesssim 1$~TeV for the $\tau^+\tau^-$ channel. This is a positive cross-check of our chain of mock data analysis (we cut the analysis above 100~GeV). The agreement is also reasonible for the $b\bar b$ channel (top left panel).}

These good qualitative matching and reasonably good quantitative agreement with a real data analysis validate the method, and make us confident to extrapolate our results to longer observation times. This is what we show also in \citefig{fig:sigvmax}, by extracting our limits for 10 and 20 years of observation (red and green curves, respectively). Since our mock data are generated without DM signal, we see that the limits improve as $\sim \sqrt{T}$, as expected. The next step is to study the sensitivity to individual subhalo detection.

\paragraph{\bf Likelihood analysis to set the sensitivity to pointlike subhalos}:

To determine the sensitivity to pointlike subhalos, we again implement a statistical method similar to the standards used in the Fermi collaboration \cite{FermiLATCollab2015,FermiLAT2019}, which are also based on a likelihood approach. In the following, the search for pointlike subhalos is performed over the full sky, except for for the disk region $|b|<5^\circ$ which is masked.

In the case of pointlike subhalo searches, the likelihood function should have the same form as the one used to set constraints on the diffuse emission model, \revise{but that diffuse-only model itself needs to be upgraded to allow for} the insertion of a pointlike subhalo in a pixel of angular resolution size.

Focusing on a specific direction in the sky and slightly around, \revise{and labeling our angular bins by $i$ (with a nominal resolution angle $\theta_{\rm r}=0.1^\circ$) and the energy bins by $j$}, we define the point-source search window as a region of $2.2^\circ\times 2.2^\circ$ about the pointing direction, divided in angular bins of $0.2^\circ\times 0.2^\circ$. \revise{When extending the nominal case to an increased resolution of $\theta_{\rm r}=1^\circ$, we shall increase the region to $6^\circ\times 6^\circ$ divided in bins of $2^\circ\times 2^\circ$}. We still use 5 logarithmic energy bins covering the $1-100$~GeV energy range.

The null hypothesis amounts to having no point source at all. We want to quantify the likelihood ratio change if we introduce a source in pixel $i_0$. To do so, we generate new mock data in the same way as for the diffuse emission for $i\neq i_0$, with the probability
\ben
p(n_{ij} \, | \, b_{ij}) = \frac{(b_{ij}  + a_{ij}\sigv)^{n_{ij}}}{n_{ij}!} e^{-\left(b_{ij} + a_{ij}\sigv \right)}\,,
\een
where $b_{ij}$ stands for both the DGE and the isotropic baryonic backgrounds, and where we impose $\sigv \leq \sigv_{\rm max}(T = 2 \,{\rm yr})$ since we consider cases for which we have not detected DM through the diffuse component at the time of observation (we could use $\sim 8$~yr \cite{ChangEtAl2018} instead, but this would not qualitatively change our results). In the central pixel $i_0$ we simply set
\ben
n_{ij} = b_{ij} + a_{ij}\sigv + \overline{J}  \sigv c_{ij}\,,
\een
where the product $\overline{J}  \sigv c_{ij}$ represents the number of photons received from a pointlike subhalo in pixel $i=i_0$ with a $J$~factor $\overline{J}$ and an annihilation cross section \sigv. The factor $c_{ij}$ satisfies $c_{ij} = c_{ij}^0 \delta_{i, i_0}$. We stress that here \sigv\ has to be considered as a fixed parameter of the model. Remind also that $\overline{J}$ is the true $J$~factor injected in the mock data.

The reconstruction likelihood function to consider should then be characterized by two free parameters (\sigv\ being fixed): $J$, \ie~the $J$~factor of the pointlike subhalo to estimate, and $\alpha_{\rm b}$, which represents the departure from central value of the background model. That likelihood function reads
\begin{align}
{\cal L}_{ij}(n_{ij} \, &| \,  J, \alpha \, ; \sigv ) =\nn\\
&\frac{\left(c_{ij}^0 \sigv J \delta_{i, i_0}+a_{ij}\sigv+\alpha_{\rm b} b_{ij}^{\rm DGE}+ b_{ij}^{\rm iso}\right)^{n_{ij}}}{n_{ij} !}\nn\\
\times& \displaystyle e^{-\left(c_{ij}^0 \sigv J  \delta_{i, i_0} + a_{ij} \sigv + \alpha_{\rm b} b_{ij}^{\rm DGE}+ b_{ij}^{\rm iso}\right)}\nn\\
\times& \frac{1}{\sqrt{2\pi \sigma_{\rm b}^2}}\, \displaystyle e^{-\frac{(\alpha_{\rm b}-\varepsilon_{\rm rec})^2}{2\sigma_{\rm b}^2}}\,.
\label{eq:lklhd_pt}
\end{align}
\revise{As before, departing from $\alpha_{\rm b} = \varepsilon_{\rm rec}$ to better match with the mock data costs a Gaussian penalty of $\sigma_{\rm b}$}, which again allows to artificially account for background mismodeling, as in the diffuse emission analysis.

The total likelihood function is then simply given by
\ben
{\cal L}(  J, \alpha \, ; \sigv ) = \prod_{ij}{\cal L}_{ij}(n_{ij} \, | \,  J, \alpha \, ; \sigv ) 
\,.
\een

We first want to determine the bias parameter $\tilde{\alpha}_{\rm b}$ that maximizes the likelihood function in the null hypothesis (no point source). This could be done semi-analytically if we neglected the isotropic background, as shown in \citeapp{sapp:L_pt}. However, contrary to the previous case, the signal hypothesis is now characterized by two maximizing parameters $(\widehat{\alpha}_{\rm b},\widehat{J})$, which are solutions to a system of equations hardly solvable by semi-analytical methods. Therefore, in the signal hypothesis, even in a simplified background modeling, we have to resort to the Newton-Ralphson algorithm, as explained in \citeapp{sapp:newton}.

We can eventually write down the likelihood ratio of the signal-to-null hypotheses 
\ben
{\cal R} \equiv  \frac{{\cal L}(\widehat{J}, \widehat{\alpha}_{\rm b}  \, ; \sigv )}{{\cal L}(0, \tilde{\alpha}_{\rm b}  \, ; \sigv )} \,,
\een
and unambiguously define a 5$\sigma$ detection threshold by demanding $2\ln {\cal R} > 25$. It is clear that the higher $\bar{J}$ in the generated mock data, the higher ${\cal R}$ in the analysis, as it drives the likelihood ratio further and further away from the null hypothesis. We denote $J_{\rm min}$ the value of $\bar{J}$ such that in average $2\ln {\cal R} = 25$, similarly to \citeeq{eq:def_jmin} in the simplified statistical analysis. More formally:
\ben
J_{\rm min}^{(l,b)} = \bar{J} \; |\; \ln {\cal R}(l,b) = \frac{25}{2}\,.
\label{eq:def_jmin_lklhd}
\een
This time, the sensitivity to pointlike subhalos $J_{\rm min}$, still a function of Galactic coordinates $(l,b)$, is determined from a much more rigorous statistical likelihood analysis of mock data, such as the ones currently used on real data.

Skymaps of $J_{\rm min}$ are shown in the right panels of \citefig{fig:nsub_baryons_and_dmbaryons} (baryonic background only in the top panel, and both baryonic and diffuse DM backgrounds in the bottom panel, setting \sigv\ to its 3$\sigma$ limit in the latter case, $\sim 5\times 10^{-26}\,{\rm cm^3s^{-1}}$, which can be read off \citefig{fig:sigvmax}). We see that the angular distribution strongly depends on the background, with a stronger contrast toward the central regions of the MW when the diffuse DM contribution is included. This obviously affects the angular distribution of detectable objects, as we will discuss later. We note that we get values of $J_{\rm min}\approx 10^{18}\,{\rm GeV^2/cm^{5}}$, which provide a rather generic order of magnitude for the subhalo detection threshold, which can be compared with the probability density function of subhalo $J$~factors in \citefig{fig:prob_jpsi_pl}.

The time dependence of $J_{\rm min}$ is further shown in \citefig{fig:jmin_crit} as the red and green solid curves (the corresponding dashed curves illustrate the simplified analysis). The former is obtained by setting the annihilation cross section to its limit after 10~yr of (virtual) observation without detection of the smooth halo, while the latter is based on the 20-yr limit (hence a value of \sigv\ smaller by a factor of $\sim \sqrt{2}$). The left (right) panel assumes an experimental angular resolution of $\theta_{\rm r}=0.1^\circ$ ($1^\circ$, respectively). We see that the prediction from the simplified analysis $J_{\rm min}\propto 1/\sqrt{T}$ is only recovered in the large $\theta_{\rm r}$ case, while for nominal angular resolution $J_{\rm min}$ decreases slightly
faster with time. This is a purely statistical effect which derives from the fact that some energy bins are empty or almost so in the latter case. This cannot be captured with our simplified analysis, while it is properly treated with the likelihood method. In particular, we see that the values obtained for $J_{\rm min}$ in that case are much more conservative at small observation time with the likelihood determination. However, the simplified analysis gets the qualitative trend of results correct, which shows its relevance to help understand the driving physical effects from analytical calculations.

By combining the sensitivity $J_{\rm min}$ with the 3$\sigma$ limit on \sigv\ obtained from the absence of DM signal in the diffuse emission in the mock data, we can determine the critical sensitivity $J_{\rm min}^{\rm crit}$, \ie~the threshold above which pointlike subhalos cannot be detected before the diffuse DM signal. As explained around \citeeq{eq:jmin_crit}, $J_{\rm min}^{\rm crit}$ is simply the time-dependent value of $J_{\rm min}$ obtained by setting $\sigv=\sigv_{\rm max}(T)$ in the likelihood function of \citeeq{eq:lklhd_pt} applied to the mock data generated for pointlike source searches. This can be formulated as
\ben
\label{eq:jmin_crit_L}
J_{\rm min}^{\rm crit}(T) = J_{\rm min}\left(T,\sigv_{\rm max}(T)\right)\,.
\een
Being the critical $J$~factor sensitivity below which the DM-induced diffuse emission should have already been detected, integrating the pdf of pointlike subhalo $J$~factors above $J_{\rm min}^{\rm crit}$ (see \citefig{fig:prob_jpsi_pl}) allows us to determine the number of subhalos that can be detected as pointlike objects before the smooth DM itself. With the rather involved statistical method described above, we can already check one of the main predictions of the earlier simplified statistical treatment: the fact that $J_{\rm min}^{\rm crit}$ should become asymptotically constant with time, and independent of annihilation cross section (as long as it is defined from the 3$\sigma$ limit on \sigv\ derived from the diffuse signal analysis, which does depend on observation time).

Values of $J_{\rm min}^{\rm crit}$ as functions of time and computed from the likelihood analysis are reported in \citefig{fig:jmin_crit} as the solid blue curves (the dashed blue curves show the results obtained with the simplified analysis). The left (right) panel assumes an angular resolution of $\theta_{\rm r}=0.1^\circ$ ($1^\circ$, respectively). \revise{Are also shown the evolutions of the subhalo detection threshold (or sensitivity) $J_{\rm min}$ derived assuming two different annihilation cross sections: one corresponding to the 10-yr limit in the diffuse signal (red curves), and the other one corresponding to the 20-yr limit (green curves)}. The $J_{\rm min}^{\rm crit}$ curves cross the $J_{\rm min}$ red (green) ones at 10~yr (20~yr, respectively), as expected, since they have been derived assuming $\sigv_{\rm max}(10/20\,{\rm yr})$.

These complete likelihood results for $J_{\rm min}^{\rm crit}$ do confirm the prediction obtained from the simplified analysis: $J_{\rm min}^{\rm crit}$ flattens and tends to a constant value at large observation time, which can be accurately determined from a likelihood analysis. It might look surprising that $J_{\rm min}^{\rm crit}$ is independent of time, but recall that it is calculated from $\sigv_{\rm max}(T)$ which does depend on time. The deep meaning of this time independence is that not detecting the diffuse component intrinsically limits the luminosity of subhalos, which is proportional to \sigv. Hence, this critical parameter self-consistently includes all the physical degeneracies of the problem.

\section{Summary results and conclusion}
\label{sec:summ}
After this pedestrian exploration of the issue of subhalo searches with Fermi-LAT-like gamma-ray experiments, it is worth summarizing our main results and drawing more quantitative conclusions.

First of all, the path we have followed in this study is complementary to many other similar works in that (i) it does not rely on a real data analysis, only on educated modeling; (ii) it is based on subhalo population models self-consistently embedded in full kinematically constrained Galactic mass models; (iii) it relies on semi-analytical calculations that allow us to integrate over the full available phase space that describes subhalos. Our subhalo population model accounts for tidal stripping induced by both the DM component and the baryonic disk, which are properly evaluated from the currently constrained distributions of DM and baryons. It is therefore not based on {\em ad hoc} rescaled formulations from cosmological simulations. This induces a tight dynamical correlation between the subhalo properties and the other Galactic components, which has to be treated self-consistently for a proper estimate of the detectability of subhalos. Indeed, this correlation strongly affects the angular distribution of the signal-to-noise ratio.

We have tried to address two different questions: {\bf (i) can have subhalos been plausibly detected and are they already present in the Fermi catalog as unidentified sources? (ii) how probable is it to detect subhalos without having detected the smooth halo first?} We have not fully answered these questions yet but shall do so just below. However, we have introduced or defined physical and statistical quantities appropriate to help us answer. As well known in the field, the physical quantity that best defines the gamma-ray flux of a dark matter object for an observer on Earth is the $J$~factor, first introduced in \cite{BergstroemEtAl1998a}.

The probability density function of subhalo $J$~factors, which is fully determined from the main subhalo characteristics (effective\footnote{Effective because they depend both on cosmological input functions (initial conditions) and on tidal stripping.} mass and concentration functions, and spatial distribution after tidal stripping), provides the most important piece of statistical information [see \citeeq{eq:prob_jpsi0} and \citefig{fig:prob_jpsi_pl}]. This was already noticed in \eg~\cite{HuettenEtAl2016}, but our probability function differs significantly from theirs because we account for tidal stripping, which modifies more naive scaling relations. This probability distribution of $J$~factors actually combines a complex mixture of different elements, each weighted by a specific though intricate probability: apparent size of a subhalo (fixed by angular resolution, position, mass and concentration), its intrinsic luminosity (mass and concentration), and its distance to the observer---all these are \revise{affected} by tidal effects.

This is of course not enough, since one also needs to figure out what the gamma-ray background is as precisely as possible, in particular its angular distribution. A rather sound model for the background allows us to define the sensitivity to pointlike subhalos, which can be expressed as a threshold $J$~factor. It is denoted $J_{\rm min}^{(l,b)}$ in this paper [see a simplified definition \citeeq{eq:def_jmin}, and a more statistically rigorous one in \citeeq{eq:def_jmin_lklhd}], and depends on Galactic coordinates $(l,b)$ via the background. It defines the $J$~factor necessary for a pointlike subhalo to fluctuate above the background emission significantly enough to be detected. That sensitivity to pointlike subhalos is closely related to the point-source flux sensitivity, more familiar to gamma-ray astronomers and defined in \citeeq{eq:def_phimin}. The accurate calculation of $J_{\rm min}^{(l,b)}$ is the key element to answer to question (i) above. Once it is calculated over the full sky (see the right panels of \citefig{fig:nsub_baryons_and_dmbaryons}), one can easily derive the expected number of visible subhalos by integrating the probability density of subhalo $J$~factors above $J_{\rm min}^{(l,b)}$ over the full sky (see \citefig{fig:prob_jpsi_pl}, where the green vertical thick line piles up the values of $J_{\rm min}$ in all directions).

We have explored the dependence of $J_{\rm min}$ on the main physical parameters with a simplified statistical method in \citesec{sssec:simple_stats}, and confirmed our results from a full likelihood analysis performed on mock data in \citesec{sssec:likelihood}. We can summarize the main dependencies as follows:
\bi
\item \sigv: The sensitivity to subhalos increases linearly with \sigv\ (\ie~$J_{\rm min}\overset{\sim}{\propto} 1/\sigv$) in a baryonic background domination, but only $\overset{\sim}{\propto}\sqrt{\sigv}$ when the DM-induced diffuse background becomes important as well. In contrast, the point-source flux sensitivity $\phi_{\rm min}$ is independent of \sigv\ in a baryonic background domination, and degrades like $\overset{\sim}{\propto}\sqrt{\sigv}$ when the DM-induced diffuse background takes over. These scaling relations assume that the Poissonian regime is reached.
\item $\alpha$: Interestingly enough, the sensitivity to subhalos slightly {\em degrades} if the initial mass function slope $\alpha>1.9$, because this increases the relative fraction of light (hence faint) subhalos with respect to heavier (hence brighter) ones, and thereby increases the contribution of unresolved subhalos to the diffuse emission (said differently, this increases the annihilation boost factor \revise{which contributes as an additional diffuse background}). See an illustration in \citefig{fig:nsub_dmbg}.
\item $m_{\rm min}$: The impact of the cutoff virial mass $m_{\rm min}$ is only important for $\alpha>1.9$. Then, decreasing $m_{\rm min}$ {\em degrades} the sensitivity to pointlike subhalos because this increases the background diffuse emission induced by unresolved subhalos, as explained just above.
\ei
Some other characteristics (most probable distances, masses, concentrations) are further illustrated in the appendix, see~\citeapp{app:sub_props}. They significantly depend on the angular resolution considered to define the pointlike character. By the way, extending the angular resolution beyond its nominal value of $\theta_{\rm r}=0.1^\circ$ in our calculations might be a way to address the sensitivity to extended objects.
\begin{figure*}[th!]
  \centering
  \begin{minipage}[t][0.5cm][c]{0.49\textwidth}\centering
    Global NFW Galactic halo $(\theta_{\rm r}=0.1^\circ)$
  \end{minipage}
  \begin{minipage}[t][0.5cm][c]{0.49\textwidth}\centering
    Global cored Galactic halo $(\theta_{\rm r}=0.1^\circ)$
  \end{minipage}
  \hrule
  \begin{minipage}[t][0.5cm][c]{0.9\textwidth}\centering
    Angular distribution of subhalos above a given $J_{\rm min}$ (colored), predicted $J_{\rm min}$
    (curves), and iso-$\log_{\rm 10}(N_{\rm vis})$
  \end{minipage}
  \includegraphics[width=0.4\textwidth]{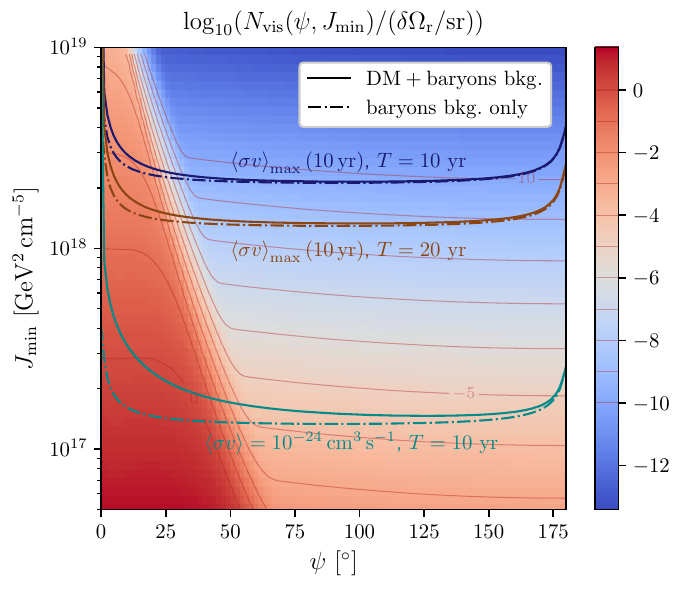}
  \includegraphics[width=0.4\textwidth]{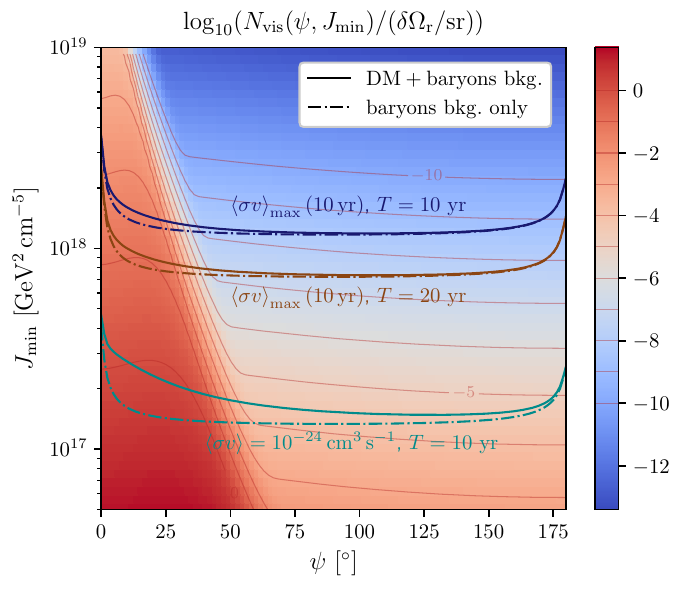}
  \hrule
  \begin{minipage}[t][0.5cm][c]{0.9\textwidth}\centering
  Same as above zoomed in the range $\psi\in[0^\circ,40^\circ]$
  \end{minipage}
  \includegraphics[width=0.4\textwidth]{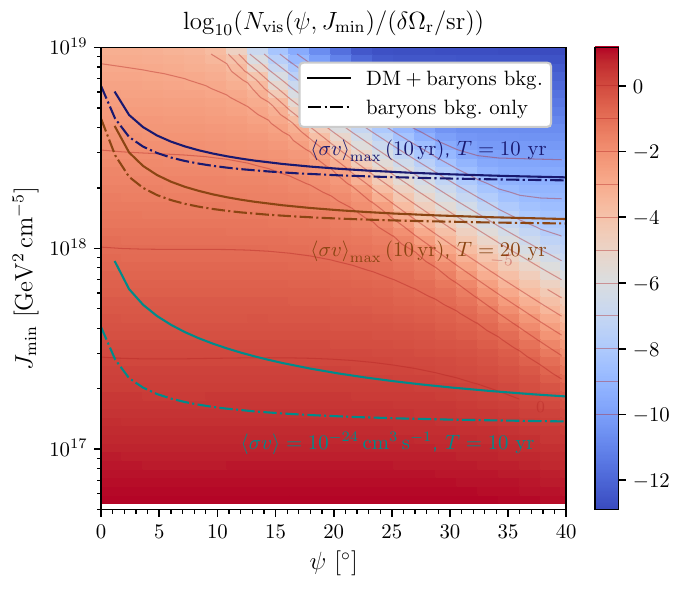}
  \includegraphics[width=0.4\textwidth]{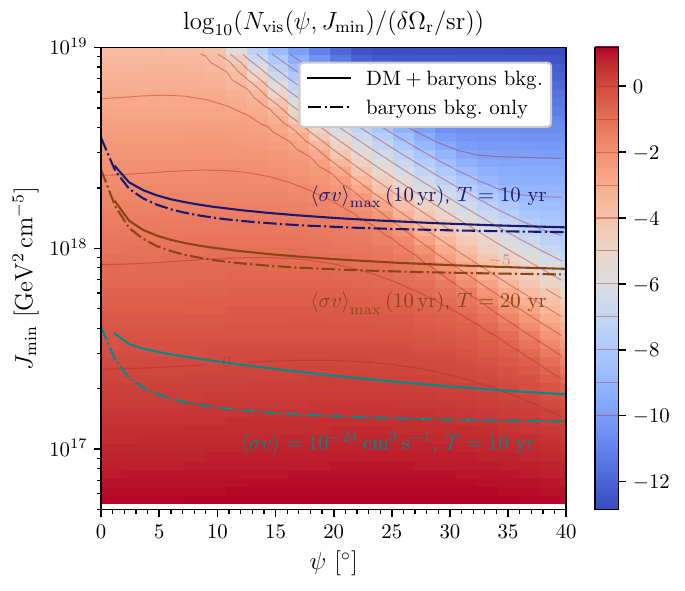}
  \hrule
  \begin{minipage}[t][0.5cm][c]{0.9\textwidth}\centering
  Corresponding angular distributions of visible subhalos
  \end{minipage}
  \includegraphics[width=0.4\textwidth]{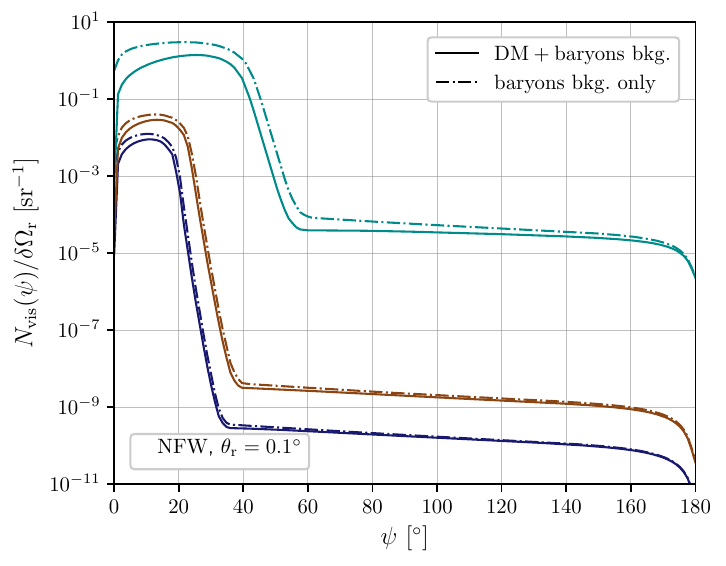}
  \includegraphics[width=0.4\textwidth]{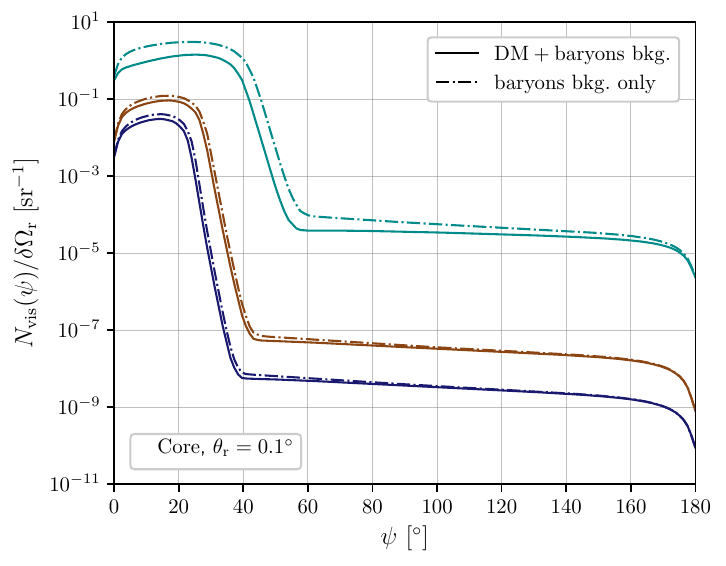}
  \caption{Angular profile of visible pointlike subhalos ($J>J_{\rm min}^{(l,b)}$) assuming $ \theta_{\rm r} = 0.1^\circ$ for a global NFW (left panels) or cored halo (right panels). Subhalo parameters are set to $(\alpha, m_{\rm min}/M_\odot,\epsilon_{\rm t}) = (1.9, 10^{-10},0.01)$. The $J_{\rm min}$ curves assume \sigv\ fixed to its 3$\sigma$ limit for 10~yr or to an already excluded value of $10^{24}\,{\rm cm^3/s}$ for a 100~GeV WIMP annihilating into $\tau^+\tau^-$. Observation times of 10 and 20~yr are considered. {\bf Top panels}: Angular distribution of subhalo $J$~factors (colored), $J_{\rm min}^{(l,b=\psi)}$ curves ($l=0^\circ,180^\circ$), and iso-$\log_{\rm 10}N_{\rm vis}$. {\bf Middle panels}: Zoom in the $\psi\in [0^\circ-40^\circ]$ range. {\bf Bottom panels}: Two-dimensional projection.} 
\label{fig:Nvis_vs_JminAndPsi_res01}
\end{figure*}

\begin{figure*}[th!]
  \centering
  \begin{minipage}[t][0.5cm][c]{0.49\textwidth}\centering
    Global NFW Galactic halo $(\theta_{\rm r}=1^\circ)$
  \end{minipage}
  \begin{minipage}[t][0.5cm][c]{0.49\textwidth}\centering
    Global cored Galactic halo $(\theta_{\rm r}=1^\circ)$
  \end{minipage}
  \hrule
  \begin{minipage}[t][0.5cm][c]{0.9\textwidth}\centering
    Angular distribution of subhalos above a given $J_{\rm min}$ (colored), predicted $J_{\rm min}$
    (curves), and iso-$\log_{\rm 10}(N_{\rm vis})$
  \end{minipage}
  \includegraphics[width=0.4\textwidth]{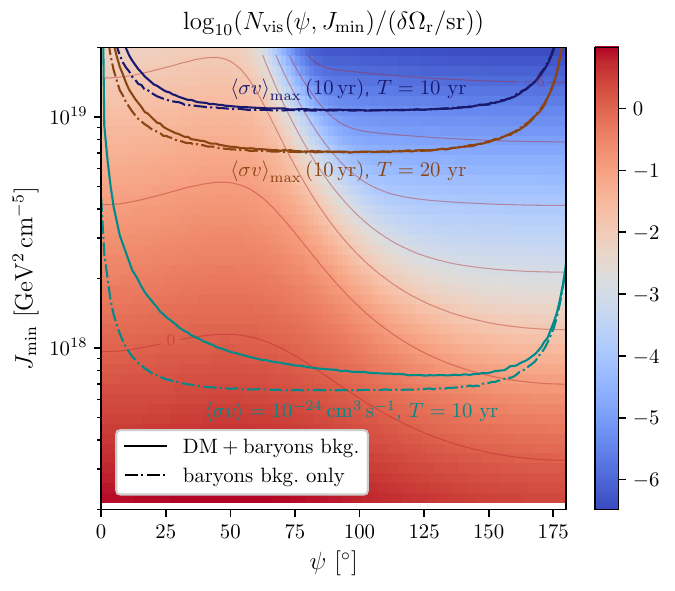}
  \includegraphics[width=0.4\textwidth]{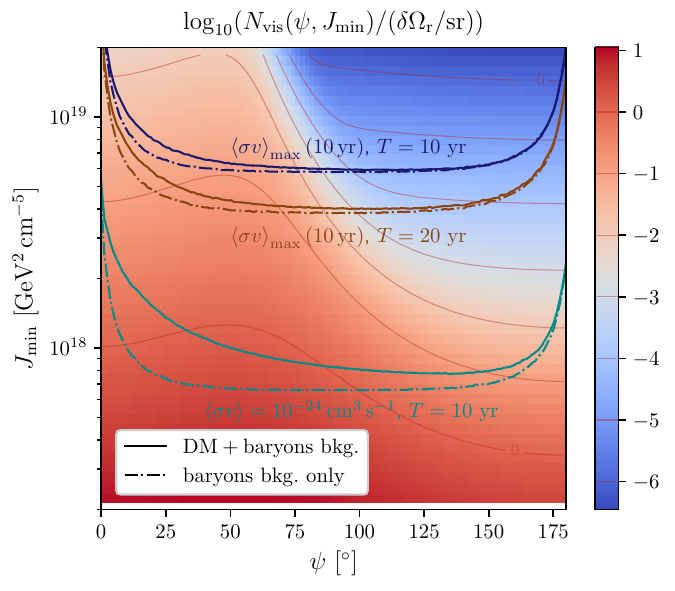}
  \hrule
  \begin{minipage}[t][0.5cm][c]{0.9\textwidth}\centering
  Same as above zoomed in the range $\psi\in[0^\circ,40^\circ]$
  \end{minipage}
  \includegraphics[width=0.4\textwidth]{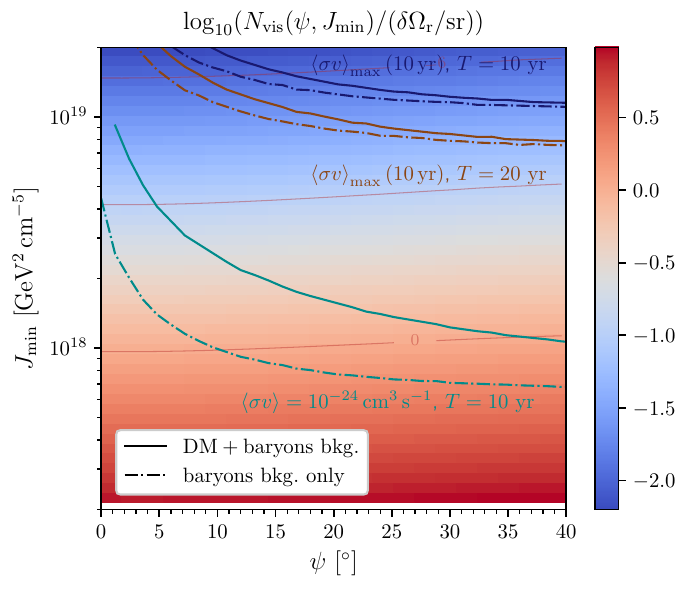}
  \includegraphics[width=0.4\textwidth]{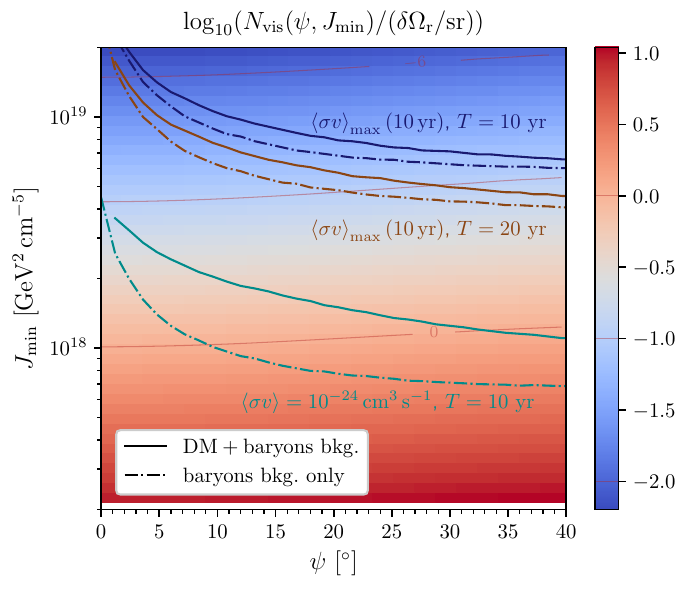}
  \hrule
  \begin{minipage}[t][0.5cm][c]{0.9\textwidth}\centering
  Corresponding angular distributions of visible subhalos
  \end{minipage}
  \includegraphics[width=0.4\textwidth]{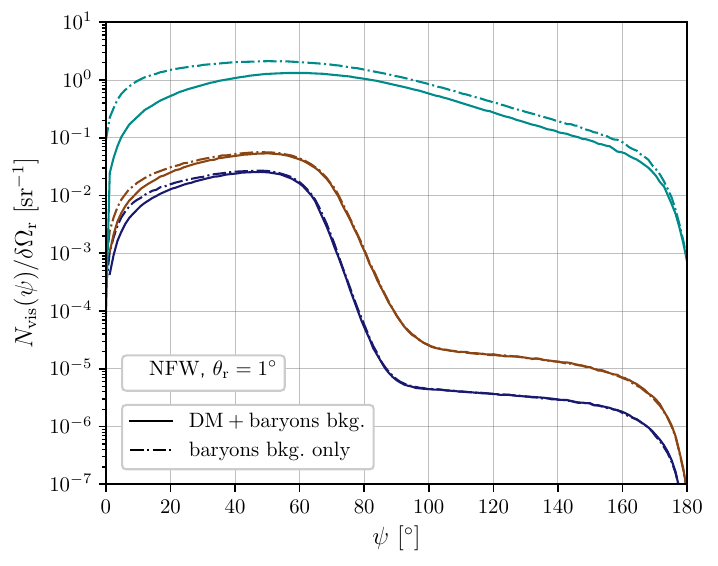}
  \includegraphics[width=0.4\textwidth]{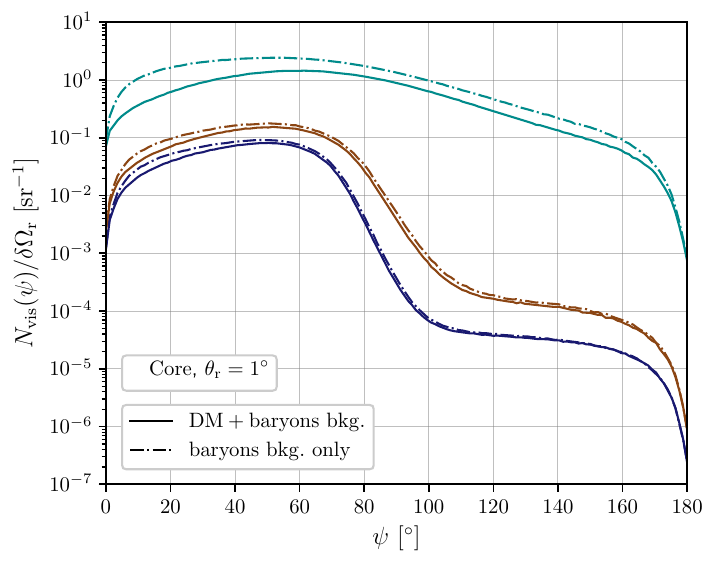}
  \caption{Same as \citefig{fig:Nvis_vs_JminAndPsi_res01} for an angular resolution of
    $\theta_{\rm r}=1^\circ$.} 
\label{fig:Nvis_vs_JminAndPsi_res1}
\end{figure*}

\begin{figure*}[ht!]
  \centering
  \begin{minipage}[t][0.5cm][c]{0.9\textwidth}\centering
    Predicted number of visible subhalos (10~yr @ $\sigv_{\rm max}$)
  \end{minipage}
  \begin{minipage}[t][0.5cm][c]{0.49\textwidth}\centering
    $\chi\bar\chi \longrightarrow \tau^+\tau^-$
  \end{minipage}
  \begin{minipage}[t][0.5cm][c]{0.49\textwidth}\centering
    $\chi\bar\chi \longrightarrow b\bar b$
  \end{minipage}
  \includegraphics[width=0.49\textwidth]{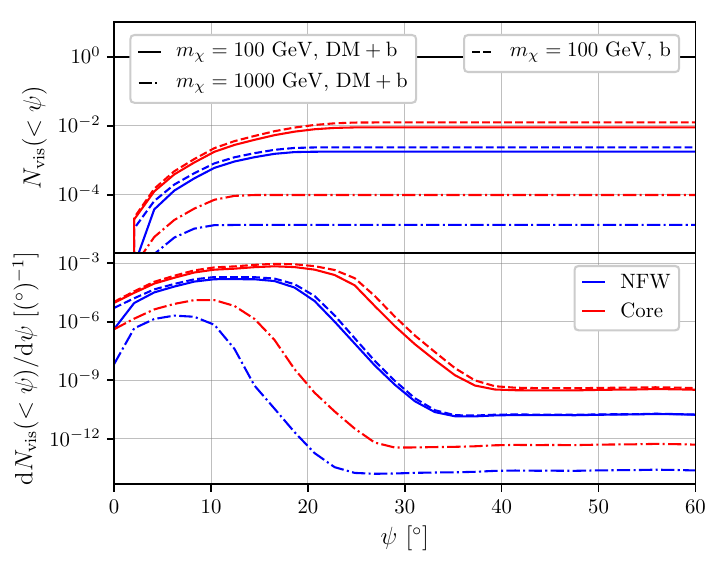}
  \includegraphics[width=0.49\textwidth]{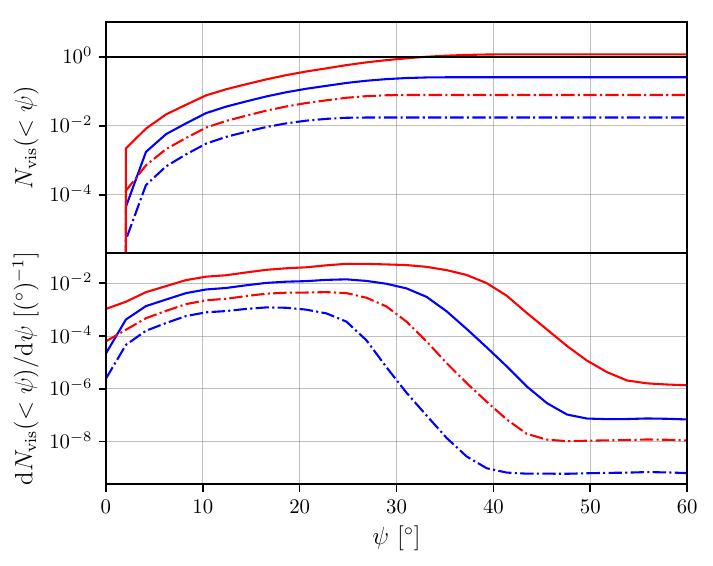}
  \caption{Predicted number of visible subhalos based on likelihood analyses on mock data generated for an observation time of 10~yr. Top (bottom) parts of the plots show the integrated (differential) number as a function of the line-of-sight angle $\psi$, for different WIMP benchmark models. The annihilation cross section is fixed to its 3$\sigma$ limit assuming the nondetection of the smooth halo (see \citefig{fig:sigvmax}). {\bf Left panel}: $\tau^+\tau^-$ annihilation channel. {\bf Right panel}: $b\bar b$ channel.} 
\label{fig:Nvis_summary}
\end{figure*}

We provide final summary results in \citefig{fig:Nvis_vs_JminAndPsi_res01}, in which the left (right) panels deal with a subhalo population model embedded within a global NFW (cored, respectively) Galactic halo. Top panels show sensitivity curves $J_{\rm min}^{(l,b)}$ [with $(l,b)=(0^\circ,\psi)||(180^\circ,\psi-180^\circ)$] as functions of the line-of-sight angle to the Galactic center $\psi$, in addition to the angular distribution of subhalos above a given threshold $J_{\rm min}$ (colored vertical scale and associated isolines). The $J_{\rm min}^{(l,b)}$ curves are calculated from different assumptions for the annihilation cross section and for the observation time --- $\sigv_{\rm max}(10\,{\rm yr})$ (dark blue and brownish curves), which corresponds to the 3$\sigma$ limit on \sigv\ derived from \citeeq{eq:def_stat_limit} ($\approx 6\times 10^{-26}{\rm cm^3/s}$, see \citefig{fig:sigvmax}), and an unrealistically large $\sigv=10^{-24}{\rm cm^3/s}$ (cyan curves); for $T=10$ (dark blue and cyan curves) or 20~yr (brownish curves). Two background configurations are assumed: baryonic background only (the DM contribution to the diffuse emission is unplugged--dot-dashed curves), and the complete background comprising both the baryonic and the DM-induced diffuse emissions (solid curves). All results assume WIMPs of 100~GeV annihilating into $\tau^+\tau^-$. These curves are inferred from the likelihood method introduced in \citesec{sssec:likelihood}, with the biased background parameters fixed to a degraded efficiency of $\varepsilon_{\rm rec}=0.7$ and a Gaussian penalty of $\sigma_{\rm b}=0.1$. These parameters artificially introduce a systematic mismodeling of the baryonic background and are tuned to match the limits obtained by the Fermi Collaboration on real data reasonably well, consistently with the analysis performed in Ref.~\cite{FermiLATEtAl2012}. The experimental angular resolution is fixed to $\theta_{\rm r}=0.1^\circ$---see the corresponding plots for $\theta_{\rm r}=1^\circ$ in \citefig{fig:Nvis_vs_JminAndPsi_res1}. Middle panels are just zoomed versions of the top panels in the range $\psi\in[0^\circ,40^\circ]$. Bottom panels show the corresponding averaged angular distributions of pointlike subhalos above $J_{\rm min}^{(l,b)}$, \ie~the visible subhalos (provided the integrated number exceeds 1). These angular distributions can be read off from the upper panels by looking at the background color gradient along the $J_{\rm min}^{(l,b)}$ curves.

Varying the background has almost no effect because the DM parameters are such that the baryonic background always dominate (sizable differences can only be seen in the case on the unrealistically large \sigv). For reasonable values of \sigv, we see that the global halo shape has no strong impact on the angular profile of detectable subhalos, with a peak found around $\sim 20^\circ$ falling sharply at larger angles, which strongly limits the angular search window. Still, the global halo shape has slightly more impact on the global distribution amplitude, making it slightly more probable to detect subhalos if they are embedded in cored Galactic halo. As seen in \citefig{fig:Nvis_vs_JminAndPsi_res1} though, increasing the angular resolution to $1^\circ$ has a more spectacular impact, since this strongly extends the angular distribution of visible pointlike subhalos, and also increases the associated amplitude in both the NFW and the cored Galactic halo cases. This might tend to indicate that searches for extended objects are a better strategy than searches for pointlike ones.

\begin{table*}[!ht]
\centering
\caption{\small Number of visible subhalos and 95\% confidence interval assuming angular resolutions of $\theta_{\rm r} = 0.1^\circ$ and $1^\circ$, and different WIMP models. Mock data are generated with the background model given in \citeeq{eq:full_bckg}, and the subhalo sensitivity is evaluated using the likelihood bias parameters $\varepsilon_{\rm rec}=0.7$ and $\sigma_{\rm b} = 0.1$ in the $1-100$~GeV energy range (5 logarithmic bins). The subhalo configuration is $(\alpha, m_{\rm min}/M_\odot,\epsilon_{\rm t}) = (1.9, 10^{-10},0.01)$, \ie~it describes a population of subhalos rather resilient to tidal stripping.}
\label{tab:Nvis} 
\begin{threeparttable}[t]
  \begin{tabular}{cccc||ccc|ccc||ccc|ccc}
    \hline
    \multirow{3}{*}{ $\frac{m_\chi}{\rm [GeV]}$ } &
    \multirow{3}{*}{channel} &
    \multirow{3}{*}{bkg.} &
    \multirow{3}{*}{$\frac{T}{\rm [yr]}$}  &
    \multicolumn{6}{c||}{$\theta_{\rm r}=0.1^\circ$}& \multicolumn{6}{c}{$\theta_{\rm r}=1^\circ$}\\
\cline{5-16}
  & & &  & \multicolumn{3}{c|}{NFW} &  \multicolumn{3}{c||}{Core}& \multicolumn{3}{c|}{NFW} &  \multicolumn{3}{c}{Core}\\
\cline{5-16}
 & & & &
    $N_{95\%}^-$ & $N_{\rm vis}$ & $N_{95\%}^+$ &  $N_{95\%}^-$ & $N_{\rm vis}$ & $N_{95\%}^+$ &
    $N_{95\%}^-$ & $N_{\rm vis}$ & $N_{95\%}^+$ &  $N_{95\%}^-$ & $N_{\rm vis}$ & $N_{95\%}^+$ \\
\hline
\hline
100  &  $\tau^+\tau^-$  &   DM+b  & 10\tnote{$\star$} &
0 & $1.8\times 10^{-3}$&  0.60 & 0  & $8.96\times 10^{-3}$ & 0.80 & 
0 & $4.97\times 10^{-2}$&  1.19 & 0  & 0.16 & 1.71\\ 
100  &  $\tau^+\tau^-$  &   b only  & 10\tnote{$\star$} &
0  & $2.4\times 10^{-3}$  &  0.63 & 0  & $1.25\times 10^{-2}$ & 0.85&  
0  & $5.44\times 10^{-2}$  &  1.22 & 0  & 0.19 & 1.81\\ 
100  &  $b\bar b$  &  DM+b & 10\tnote{$\star$}  &
0 & 0.26  &  2.04 & 0  & 1.18 & 4.2 & 
0 & 1.26  &  4.33 & 0  & 3.69 & 8.35 \\ 
1000  &  $b \bar b$  &   DM+b  & 10\tnote{$\star$} &
0 & $1.73\times 10^{-2}$  &  0.92 & 0  & $7.80\times 10^{-2}$ & 1.35 &  
0 & 0.20  &  1.84 & 0  & 0.57 & 2.88\\  
1000  &  $\tau^+ \tau^-$  &   DM+b  & 10\tnote{$\star$} &
0 & $1.3\times 10^{-5}$  &  0.34 & 0  & $9.8\times 10^{-5}$ & 0.41 &  
0 & $4.2\times 10^{-3}$  &  0.69 & 0  & $1.85\times 10^{-2}$ & 0.93\\  
\hline
100  &  $b \bar b$  &   DM+ b  & 20\tnote{$\dagger$} &
\dots & \dots  & \dots & 0  & 1.80 & 5.31 & 
\dots & \dots  &  \dots & 0.14  & 3.98 & 8.80\\ 
100  &  $b \bar b$  &   DM+ b  & 20\tnote{$\star$} &
\dots & \dots  &  \dots & 0  & 3.32 & 7.80 & 
\dots & \dots  &  \dots & 1.22 & 5.95 & 11.60\\ 
\hline
\end{tabular}
\begin{tablenotes}
  \tiny
  \item[$\star$] Using $\sigv_{\rm max}(10\,{\rm yr})$ for the corresponding channel.
  \item[$\dagger$] Using $\sigv_{\rm max}(20\,{\rm yr})$ for the corresponding channel.
\end{tablenotes}
\end{threeparttable}
\end{table*}

We further quantify our results in \citetab{tab:Nvis}, where we fully integrate over the statistical ensemble. We provide our predictions for the total number of visible subhalos and its 95\% C.L. range assuming several configurations for DM, the background, and the observation time. DM is taken in the form of WIMPs of 100~GeV or 1~TeV, distributed according to an NFW or a cored halo, annihilating into $b\bar b$ or $\tau^+\tau^-$, and with a cross section set to the 3$\sigma$ limit on the diffuse DM signal corresponding to 10 or 20~yr of unsuccessful observation (see \citefig{fig:sigvmax}). \revise{To derive the number of detectable subhalos, we have assumed an observation time of 10 or 20~yr. In the former case, we have fixed the annihilation cross section to the 10-yr limit for the diffuse signal, and in the later case, to either the 10- or 20-yr limit. A cross section set to the 10-yr limit together with a 20-yr observation time suppose that the diffuse DM signal has been detected for long a the time of subhalo searches.} We adopt nominal parameters for the resilient subhalo population model, and use not only the nominal angular resolution of $\theta_{\rm r}=0.1^\circ$ for pointlike subhalo searches, but also a more extended one of $\theta_{\rm r}=1^\circ$ to try to capture the potential reach of extended subhalo searches. Our main results, which are illustrated in \citefig{fig:Nvis_summary} in terms of angular distributions of visible subhalos for different model configurations, can be summarized as follows:
\bi
\item In most cases, the number of visible subhalos is presently $N_{\rm vis}<1$ at 95\% C.L.
\item The most optimistic case for a 10-yr search of pointlike subhalos (nominal resolution angle) is found for $\mchi = 100$~GeV annihilating into $b\bar b$, for which $N_{\rm vis}<5$ (3) at 95\% C.L. for a cored (NFW) Galactic halo. In that case $N_{\rm vis}=0$ is still part of the 95\% C.L. range.
\item Extending the analysis to 20~yr (same annihilation cross section), we find a minor improvement with $N_{\rm vis}<6$ (cored halo), though still consistent with 0 at 95\% C.L.
\item Increasing the angular resolution to $\theta_{\rm r}=1^\circ$ slightly increases the statitistics by adding bigger objects, which tends to show that there is a little bonus to be gained from extended source searches.
\item If to be hunted somewhere, subhalos should better be looked for in a latitude band extending from $\sim \pm 10^\circ$ to $\sim \pm 40^\circ$, and in a longitude band centered about $0^\circ$. With an angular resolution of $0.1^\circ$ ($1^\circ$), visible subhalos should have tidal masses of $\sim 10^4$-$10^5 M_\odot$ ( $\sim 10^6$-$10^7 M_\odot$) and be located at a distance of $\sim 10$~kpc ($\sim 10$-20~kpc) from Earth---see \citeapp{app:sub_props}. 
\ei
Based on these results, we conclude that it is unlikely that some of the unidentified sources of the Fermi catalog actually be Galactic subhalos; this might also hold for extended subhalo searches, if our large angular resolution example is confirmed to be a reasonable proxy for this complementary search window. The only configuration which may allow for subhalo detection is the cored halo case, owing to a reduced diffuse signal [detecting $\geq 1$ subhalo has a $p$ value of $\sim 0.7$ from \citeeq{eq:prob_geqn}]. \revise{Note that these statements are based upon a likelihood analysis of idealized mock data generated from a background model that underestimates the genuine DGE (see \citefig{fig:bckg_profile}), especially within the inner 10-40$^\circ$ from the GC, and that also leads to a slight underestimate of the current limits on \sigv\ (see \citefig{fig:sigvmax}). Therefore, despite the rather pessimistic prospects for subhalo detection, these can be still considered as lying on the optimistic side of possible predictions.}.

\revise{Another consequence of these limited detection perspectives is that further including subhalos to derive limits on the annihilation cross section, though necessary for self-consistency reasons, is not expected to significantly tighten those derived from the analysis of the diffuse Galactic emission only; neither from the absence of any individual detection, nor from their diffuse contribution which is lower than that of the smooth halo component at latitudes $\sim 10^\circ$-$15^\circ$ (which can otherwise be expressed as having a negligible subhalo boost factor in the central Galactic regions).} This answers to the question (i) raised above.

Finally, we have also defined a quantity, $J_{\rm min}^{\rm crit}$ [see \citeeq{eq:jmin_crit} for the definition in the simplified statistical analysis, and \citeeq{eq:jmin_crit_L} for the more rigorous one], which corresponds to the detection threshold $J_{\rm min}^{(l,b)}$ evaluated at the 3$\sigma$ limit cross section of the current (or future) observational time. That quantity formally allows us to answer to the question (ii) because it characterizes the critical $J$~factor threshold below which the diffuse signal should be detected before any pointlike subhalo. By comparing the flattish curves obtained for $J_{\rm min}^{\rm crit}$ in \citefig{fig:jmin_crit} with the probability density function of subhalo $J$~factors in \citefig{fig:prob_jpsi_pl}, we can readily claim that it is much more likely to detect the smooth halo before subhalos in the different configurations we have explored so far. Indeed, if the threshold $J_{\rm min}^{(l,b)}$ curves in \citefig{fig:jmin_crit} cross the $J_{\rm min}^{\rm crit}$ ones, that means that the smooth halo should have already been detected. We see from our results that $J_{\rm min}^{(l,b)}$ should definitely decrease below $J_{\rm min}^{\rm crit}$ in order to get a guaranteed sizable number of detectable subhalos.

What kind of physical effects could we think of to more optimistically change these conclusions? First of all, let us recall that our subhalo population model is on the optimistic side, since it is based on assuming a significant resilience to tidal effects (subhalo masses are still depleted by tides, but inner subhalo cusps survive). A systematic increase of the subhalo concentration could make them brighter without changing the more constrained smooth halo contribution. However, increasing the luminosity by a factor of $\sim 2$ would imply an aggressive change at the level of the width of the concentration distribution function (fully accounted for in our analysis), about 0.15~dex (log-normal distribution), which is not theoretically favored (\eg~\cite{MaccioEtAl2008,PradaEtAl2012,DuttonEtAl2014,Sanchez-CondeEtAl2011}). Moreover, this change would have to mostly affect the mass range of visible subhalos, otherwise it would increase the relative contribution of unresolved subhalos to the diffuse emission, and thereby temper the decrease of $J_{\rm min}^{(l,b)}$. Finally, one could also think about a distorted primordial spectrum that would inject additional power on the relevant subhalo mass scale, as is the case in the formation of primordial black holes or ultracompact mini-halos (\eg~\cite{BerezinskyEtAl2013,CarrEtAl2016}). However, even if possible, that would drive us in the study of a more fine-tuned model, which goes beyond the scope of this paper.



\begin{acknowledgments}
This work has been partly supported by the ANR project ANR-18-CE31-0006, the OCEVU Labex (ANR-11-LABX-0060), the national CNRS-INSU programs PNHE and PNCG, and European Union's Horizon 2020 research and innovation program under the Marie Sk\l{}odowska-Curie Grant Agreements No 690575 and No 674896 -- in addition to recurrent funding by CNRS-IN2P3 and the University of Montpellier.
\end{acknowledgments}

\clearpage

\appendix

\section{Subhalo model description}
\label{app:subs}

\begin{table*}[!ht]
\centering
\caption{\small Main characteristics of the subhalo population models used in this paper. Numbers are calculated using a minimal cutoff mass of $m_{\rm min}=10^{-10}M_\odot$, and for tidally resilient subhalos with $\epsilon_{\rm t}=0.01$. Are provided: $N_{\rm tot}$ the total number of surviving subhalos, and $f_{\rm tot}$, the total DM mass fraction they contain within the virial radius of the host halo.}\label{tab:subdetails}
\begin{ruledtabular}
\begin{tabular}{c||c c|c c|c c}
  & $\rho_\odot^{\rm tot}$ & $R_s^{\rm tot}$ & \multicolumn{2}{c}{$N_{\rm tot}$} & \multicolumn{2}{c}{$f_{\rm tot}$} \\
Galactic model& $[M_\odot/{\rm pc}^3]$& [kpc] & $\alpha=1.9$ & $\alpha=2$ & $\alpha=1.9$ & $\alpha=2$\\
\hline
 NFW $(\gamma=1)$   & 0.0101  & 18.6  &  $4.58\times 10^{18}$  &  $2.45\times 10^{20}$ & 0.16  & 0.52\\
 Cored $(\gamma=0)$ & 0.0103  & 7.7   &  $4.27\times 10^{18}$  &  $2.25\times 10^{20}$& 0.15 & 0.49 \\
\hline
\end{tabular}
\end{ruledtabular}
\end{table*}

Here we provide the details of the global galactic halos derived from fits on stellar kinematic data in Ref.~\cite{McMillan2017}. They are based on the following spherical profile:
\ben
\rho_{\rm tot}(R) = \rho_\odot^{\rm tot} \left\{\frac{R}{R_\odot}\right\}^{-\gamma}
\left\{\frac{1+X}{1+X_\odot}\right\}^{\gamma-3}\,,
\een
with $X=R/R_s^{\rm tot}$, $R_s^{\rm tot}$ the scale radius, $\rho_\odot^{\rm tot}$ the total average DM density in the solar system (including subhalos), and $R_\odot=8.2$~kpc the Sun's distance to the GC. We give additional details on the subhalo population models in \citetab{tab:subdetails}.

\section{Best-fitting solutions to the likelihood function}
\label{app:L_solutions}



\subsection{Semi-analytical solution (limit on \sigv\ with negligible isotropic background)}
\label{sapp:L_diff}
This method is suitable for quick analyses of diffuse photons and \revise{can formally be used when the rescaling or bias factor $\alpha_{\rm b}$ applies to the full background, which, for consistency, corresponds in our case to negligible isotropic background cases}. The best-fit couple of parameters $(\widetilde{\sigv},\widetilde{\alpha}_{\rm b})$ that maximizes the likelihood ${\cal L}(\sigv, \alpha_{\rm b})$ [see \citeeq{eq:Ltot_limit}] is given as a solution to the following system of equations:
\ben
\begin{cases}
  \frac{\partial {\cal L}(\sigv, \alpha_{\rm b})}{\partial \sigv}\Bigg|_{(\widehat{\sigv}, \widehat{\alpha}_{\rm b})}=0 \\
  \frac{\partial {\cal L}(\sigv, \alpha_{\rm b})}{\partial \alpha_{\rm b}}\Bigg|_{(\widehat{\sigv}, \widehat{\alpha}_{\rm b})}=0
\end{cases} \, .
\een
Since ${\cal L}(\sigv, \alpha_{\rm b}) > 0$, these equations are equivalent to much simpler ones involving the log-likelihood:
\ben
\begin{cases}
\frac{\partial \ln {\cal L}(\sigv, \alpha_{\rm b})}{\partial \sigv}\Bigg|_{(\widehat{\sigv},\widehat{\alpha}_{\rm b})}=0\\
\frac{\partial \ln {\cal L}(\sigv, \alpha_{\rm b})}{\partial \alpha_{\rm b}}\Bigg|_{(\widehat{\sigv},\widehat{\alpha}_{\rm b})}=0
\end{cases}\, .
\een
Inserting the expression of ${\cal L}$ given in \citeeqs{eq:Ltot_limit} and \eqref{eq:Li_limit},
we get
\ben
\begin{cases}
 \displaystyle \sum_{i,j} & a_{ij} \left(\frac{n_{ij}}{ \widehat{\sigv} a_{ij} + \widehat{\alpha}_{\rm b}b_{ij} } - 1\right) = 0\\
 \displaystyle \sum_{i,j} & b_{ij} \left(\frac{n_{ij}}{ \widehat{\sigv} a_{ij} + \widehat{\alpha}_{\rm b}b_{ij} } - 1\right) \\
 & - N_{\rm S} N_{\rm E} \frac{\widehat{\alpha}_{\rm b}-\varepsilon_{\rm rec}}{\sigma_{\rm b}^2} = 0
\end{cases}\,.
\label{eq:LogLikelihood}
\een
By a linear combination of these equations, we arrive to
\ben
\sum_{i,j} n_{ij} &-& \sum_{i,j} \left(\widehat{\sigv} a_{ij} + \widehat{\alpha}_{\rm b} b_{ij} \right) \nn\\
&-& N_{\rm S}N_{\rm E} \frac{\widehat{\alpha}_{\rm b}(\widehat{\alpha}_{\rm b}-\varepsilon_{\rm rec})}{\sigma_{\rm b}^2} = 0\,,
\een
which allows us to compute the value of $\widehat{\sigv}$ in terms of $\widehat{\alpha}_{\rm b}$ analytically from the following expression
\ben
\label{eq:BestFit}
\widehat{\sigv}  &=& \frac{1}{\sum_{i,j} a_{ij}}\\
&& \times \left[ \sum_{i,j} \left(n_{ij} - \widehat{\alpha}_{\rm b}b_{ij}\right)  - N_{\rm S} N_{\rm E}\frac{\widehat{\alpha}_{\rm b}(\widehat{\alpha}_{\rm b}-\varepsilon_{\rm rec})}{\sigma_{\rm b}^2} \right]
\,.\nn
\een
The best fit is then evaluated numerically by combining \citeeq{eq:BestFit} with one of the two expressions in \citeeq{eq:LogLikelihood}. \revise{This method is useful for quick analyses in the negligible isotropic background limit, or to test the correct implementation of the numerical algorithm presented in \citesec{sapp:newton} in the same configuration.}

\subsection{Solution to define the sensitivity to pointlike subhalos}
\label{sapp:L_pt}
Here, we derive the set of equations relevant to the case of pointlike subhalo searches, still when the isotropic background can be neglected (when the bias factor can be applied to the full background). The best-fit value of the null hypothesis (no point source) is obtained by solving
\ben
\frac{\partial \ln {\cal L}( 0, \alpha_{\rm b} \, ; \sigv)}{\partial \alpha_{\rm b}}
\Bigg|_{\tilde{\alpha}_{\rm b}}  = 0\,,
\een
which, in this case, corresponds to the solution to the equation
\ben
\sum_{ij}& b_{ij} \left(\frac{n_{ij}}{\sigv a_{ij} + \tilde{\alpha}_{\rm b}b_{ij}  } - 1\right) \nn\\
- & N_{\rm S} N_{\rm E} \frac{\tilde{\alpha}_{\rm b}-\varepsilon_{\rm rec}}{\sigma_{\rm b}^2} = 0 \, .
\een
Then we need to find the global best-fit model denoted $(\widehat{J}, \widehat{\alpha}_{\rm b})$ that is given as a solution of the two combined equations on the derivative of the log-likelihood, 
\ben
\begin{cases}
  \frac{\partial \ln {\cal L}(J, \alpha_{\rm b}  \, ; \sigv )}{\partial J} \Bigg|_{(\widehat{J}, \widehat{\alpha}_{\rm b})} = 0\\
\frac{\partial \ln  {\cal L}(J, \alpha_{\rm b}  \, ; \sigv )}{\partial \alpha_{\rm b}} \Bigg|_{(\widehat{J}, \widehat{\alpha}_{\rm b})}  = 0
\end{cases} \, .
\een
Inserting the expression of ${\cal L}$, we get
\ben
\begin{cases}
 \displaystyle \sum_{ij} & b_{ij} \left(\frac{n_{ij}}{\sigv a_{ij} + \widehat{\alpha}_{\rm b} b_{ij} + c_{ij}^0 \sigv \widehat{J} \delta_{i_0, i} } - 1\right)\\
  \displaystyle & -  N_{\rm S} N_{\rm E} \frac{\widehat{\alpha}_{\rm b}-\varepsilon_{\rm rec}}{\sigma_{\rm b}^2}  = 0  \\
  \displaystyle \sum_{ij} & c_{ij}^0 \sigv  \left(\frac{n_{ij}}{\sigv a_{ij} + \widehat{\alpha}_{\rm b} b_{ij} +
    c_{ij}^0 \sigv \widehat{J} \delta_{i_0, i} } - 1\right)  = 0
\end{cases}\,.\nn\\
\een
This system of coupled equations is actually very hard to solve. A way out is to use the
Newton-Ralphson algorithm (see below), which is well suited for this kind of problems. 

\subsection{The Newton-Ralphson algorithm}
\label{sapp:newton}
Here, we summarize our implementation of the Newton-Ralphson algorithm, which is a standard likelihood maximization procedure in gamma-ray astronomy \cite{MattoxEtAl1996}. Let us assume a likelihood function given by ${\cal L}(\Theta, \Xi)$, where $\Theta$ is a set of parameters, from which we are seeking the one, $\widehat\Theta$, that maximizes ${\cal L}$--- $\Xi$ is another set of fixed parameters. Let $\lambda (\Theta, \Xi) = \ln {\cal L}(\Theta, \Xi)$ be the corresponding log-likelihood function, and let us seek for the maximum of $\lambda$. To proceed, we introduce the gradient vector of $\lambda$ defined as ${\cal D}(\Theta, \Xi) =  \nabla_\Theta \lambda(\Theta, \Xi)$ such that, by definition, ${\cal D}(\widehat\Theta, \Xi) = 0$. We can now Taylor expand ${\cal D}$ around the best-fit point of coordinates $\widehat \Theta$ as follows:
\ben
{\cal D}(\Theta, \Xi) &=& {\cal D}(\widehat\Theta, \Xi) +  \left[(\Theta - \widehat\Theta). \nabla_\Theta\right] (\Theta, \Xi)  + \dots\nn\\
&=&  \left[(\Theta - \widehat\Theta). \nabla_\Theta\right] {\cal D}(\Theta, \Xi)  + \dots\,.
\een
By massaging this expression---and making explicit in the notation the dependence in
$(\Theta, \Xi)$---we find that
\ben
{\cal D} &=&  {\cal H}^T (\Theta - \widehat\Theta) + \dots \,, \\
{\rm with} \quad
{\cal H}_{k\ell} &\equiv&  \frac{\partial^2 \lambda(\Theta, \Xi)}{\partial\theta_k \partial
\theta_\ell} \nn
\een
the Hessian matrix defined using the elements $\Theta = (\theta_0, \theta_1, ...)$. Since the Hessian matrix is real symmetric by definition, by inverting the previous expression we get at first order
\ben
\widehat\Theta \simeq \Theta  - {\cal H}^{-1} {\cal D}\,.
\een
Like in the one-dimensional Newton algorithm, it is possible (provided ${\cal D}$ is well behaved) to find $\widehat \Theta$ simply by starting from an initial value $\Theta_0$ and defining an iterating procedure as follows:
\ben
\widehat\Theta_{n+1} &\simeq& \Theta_n  - {\cal H}^{-1}(\Theta_n, \Xi) {\cal D}(\Theta_n, \Xi)\nn\\
\text{such that} \quad \widehat \Theta  &=& \lim_{n\to\infty} \Theta_n\,. 
\een 
In practice, this converges very fast.

\section{Internal properties of visible subhalos}
\label{app:sub_props}
The most probable tidal masses, concentrations, and distances of visible subhalos are shown in \citefig{fig:NsubVis_vs_m_and_c200_th01}, \citefig{fig:NsubVis_vs_m_and_c200_th1}, and \citefig{fig:NsubVis_vs_s}.

\begin{figure*}[th!]
  \centering
\hrule
{NFW Galactic halo - Properties of subhalos visible with $\theta_{\rm r}=0.1^\circ$}\\
\includegraphics[width=0.485\linewidth]{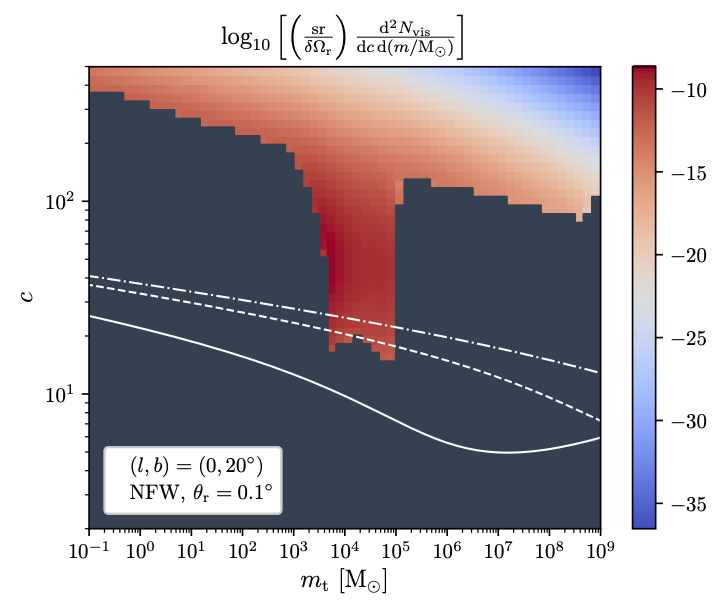}
\includegraphics[width=0.485\linewidth]{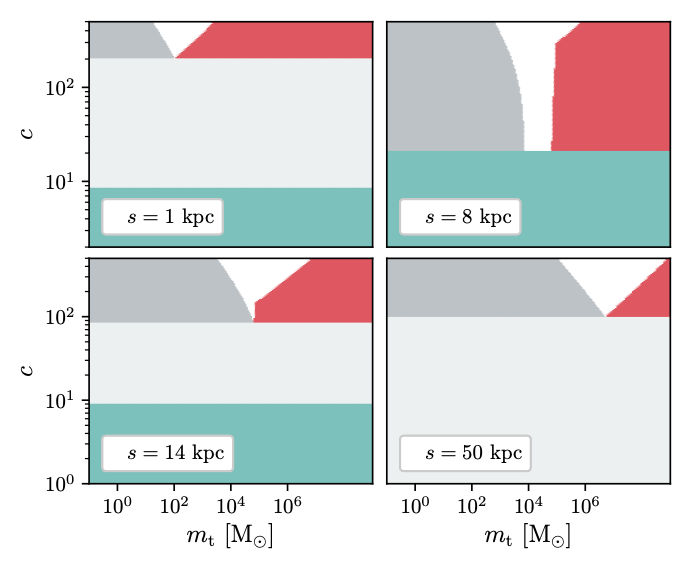}
\hrule
{Cored Galactic halo - Properties of subhalos visible with $\theta_{\rm r}=0.1^\circ$}\\
\includegraphics[width=0.485\linewidth]{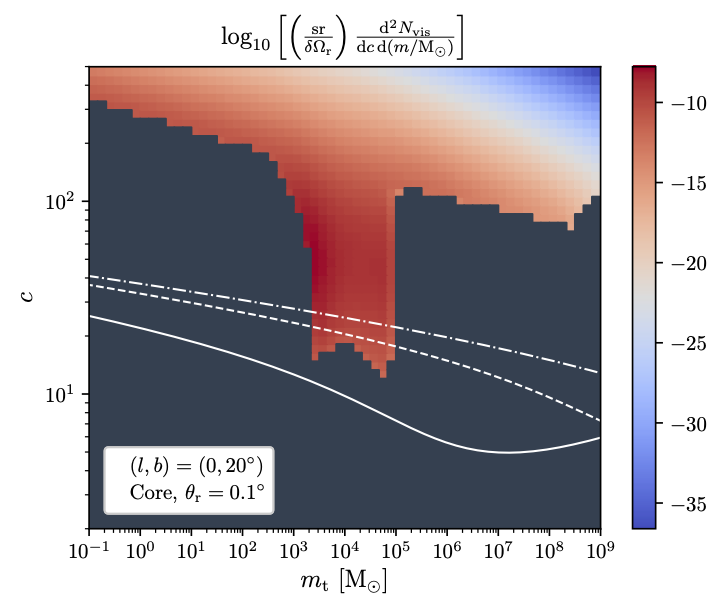}
\includegraphics[width=0.485\linewidth]{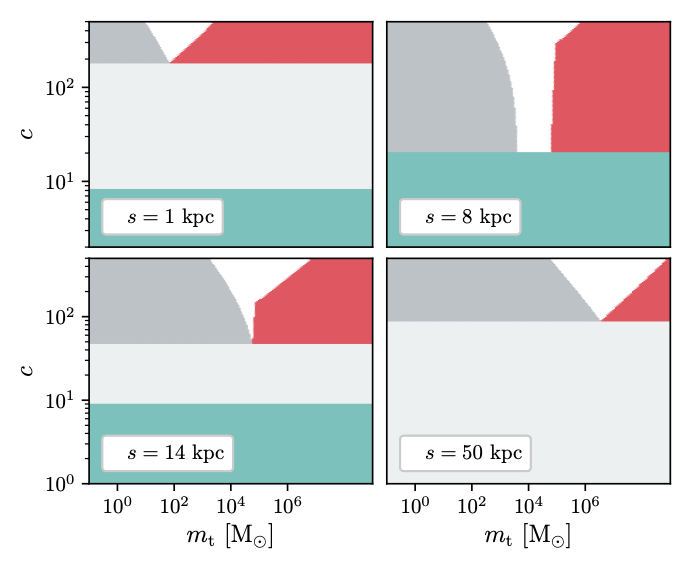}
   \caption{\small {\bf Left panels}: Concentrations and physical masses of the most visible subhalos in the direction of Galactic coordinates $(l,b) = (0^\circ, 20^\circ)$. The solid (dashed, dotted-dashed) white curve indicates the median concentration of a subhalo of virial mass $m_{200}$ that would be pruned off down to the tidal mass $m_{\rm t}$ in abscissa at a galactocentric distance of 1~kpc (10 and 100~kpc, respectively) if tidal disruption were unplugged (though not tidal stripping). This shows that subhalos with a given $m_{\rm t}$ originate from heavier and heavier objects as they are found closer and closer to the GC (\ie~tidal stripping is more and more efficient), should tidal stripping not be destructive---see in comparison the minimal concentration needed to survive tidal effects in the associated right panels. {\bf Right panels}: Exclusion areas for the computation of the probability and for different distances to the observer: subhalos that are not seen as points (red), subhalos that are below the critical/minimal allowed concentration and then tidally disrupted (turquoise---$\epsilon_{\rm t}=0.01$), subhalos that are too faint (dark gray on the left), subhalos that are either too faint or not point sources (light gray). Visible: those lying in the white area. {\bf Top panels}: NFW Galactic halo. {\bf Bottom panels}: Cored Galactic halo.}
\label{fig:NsubVis_vs_m_and_c200_th01}
\end{figure*}

\begin{figure*}[th!]
  \centering
\hrule
{NFW Galactic halo - Properties of subhalos visible with $\theta_{\rm r}=1^\circ$}\\
  \includegraphics[width=0.485\linewidth]{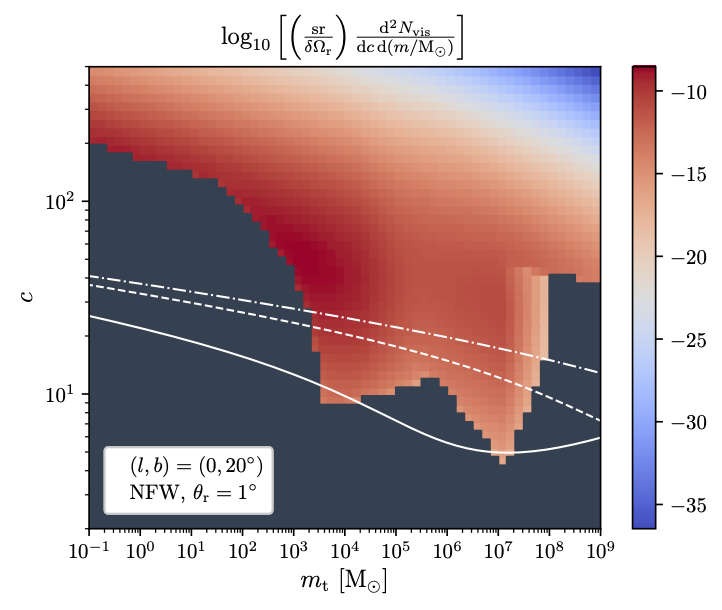}
  \includegraphics[width=0.485\linewidth]{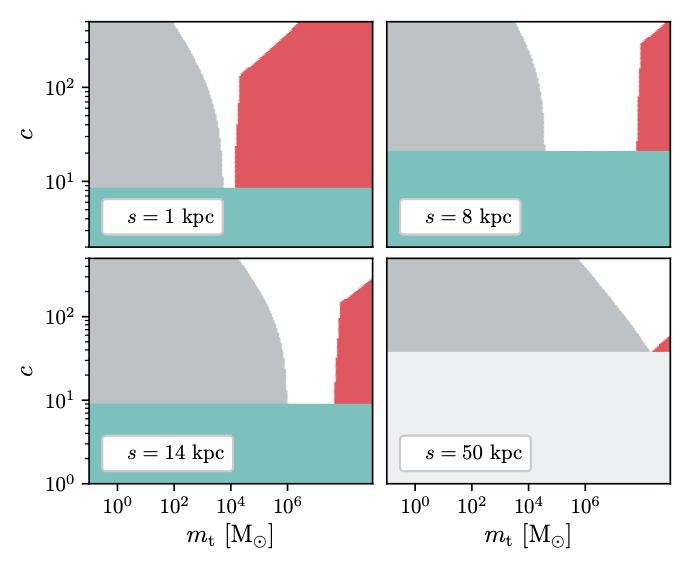}
\hrule
{Cored Galactic halo - Properties of subhalos visible with $\theta_{\rm r}=1^\circ$}\\
  \includegraphics[width=0.485\linewidth]{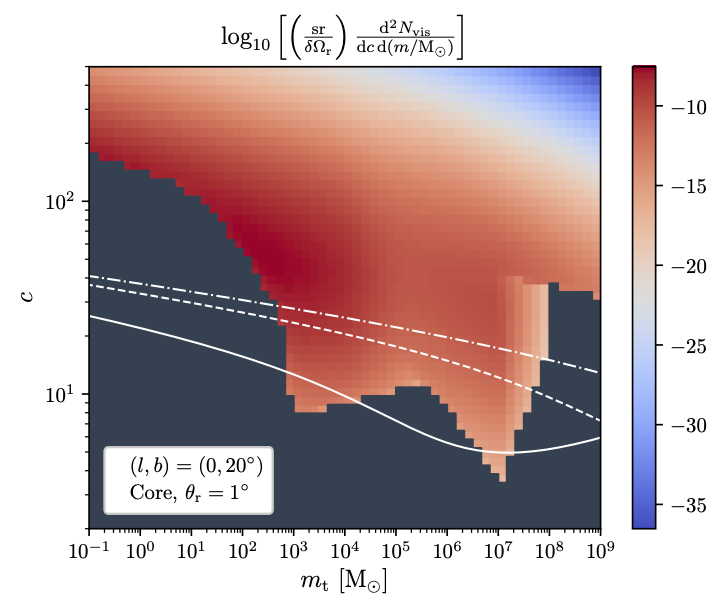}
  \includegraphics[width=0.485\linewidth]{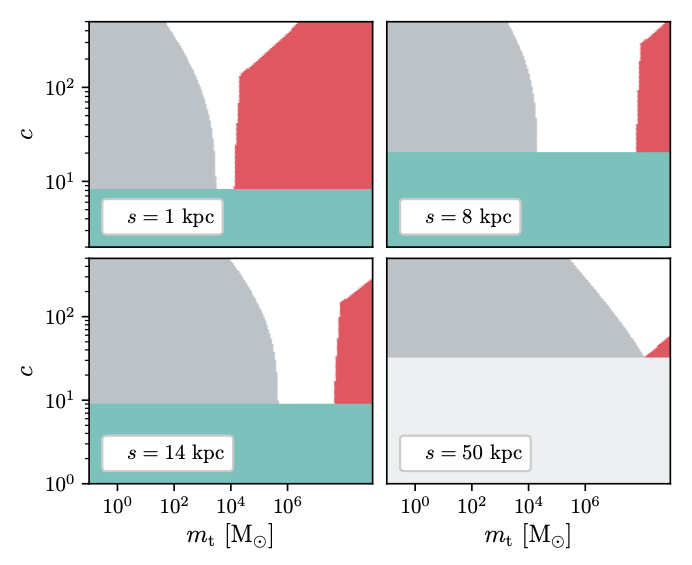}
   \caption{\small Same as \citefig{fig:NsubVis_vs_m_and_c200_th01} but with $\theta_{\rm r}=1^\circ$.}
   \label{fig:NsubVis_vs_m_and_c200_th1}
\end{figure*}

\begin{figure}[th!]
\centering
\includegraphics[width=0.95\linewidth]{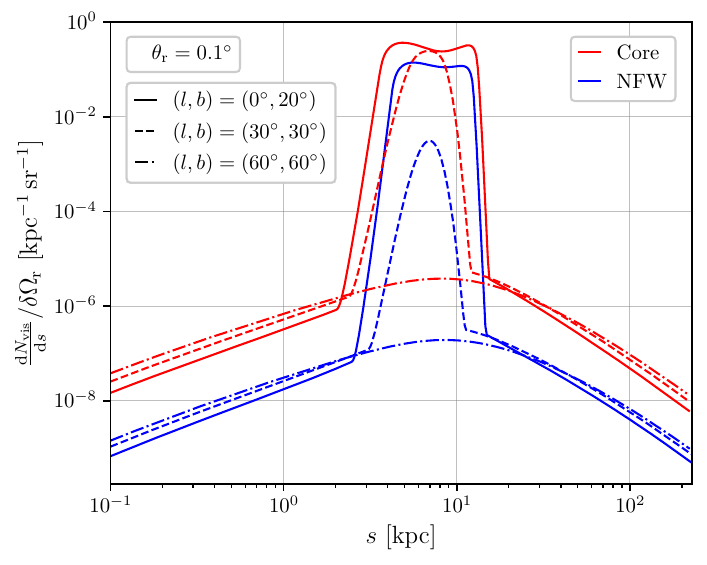}
\caption{\small Distance distribution of the visible subhalos (from the observer), for different pointing directions. This corresponds to the case in which $J_{\rm min}^{(l,b)}$ is computed assuming $\sigv_{\rm max}(10\,{\rm yr})$, $\chi\bar\chi\to b\bar b$, $\theta_{\rm r}=0.1$, and an observation time of 20~yr (the smooth halo should have already been detected).}
\label{fig:NsubVis_vs_s}
\end{figure}

\bibliographystyle{apsrev4-1}
\bibliography{biblio_jabref}

\begin{thebibliography}{142}%
\makeatletter
\providecommand \@ifxundefined [1]{%
 \@ifx{#1\undefined}
}%
\providecommand \@ifnum [1]{%
 \ifnum #1\expandafter \@firstoftwo
 \else \expandafter \@secondoftwo
 \fi
}%
\providecommand \@ifx [1]{%
 \ifx #1\expandafter \@firstoftwo
 \else \expandafter \@secondoftwo
 \fi
}%
\providecommand \natexlab [1]{#1}%
\providecommand \enquote  [1]{``#1''}%
\providecommand \bibnamefont  [1]{#1}%
\providecommand \bibfnamefont [1]{#1}%
\providecommand \citenamefont [1]{#1}%
\providecommand \href@noop [0]{\@secondoftwo}%
\providecommand \href [0]{\begingroup \@sanitize@url \@href}%
\providecommand \@href[1]{\@@startlink{#1}\@@href}%
\providecommand \@@href[1]{\endgroup#1\@@endlink}%
\providecommand \@sanitize@url [0]{\catcode `\\12\catcode `\$12\catcode
  `\&12\catcode `\#12\catcode `\^12\catcode `\_12\catcode `\%12\relax}%
\providecommand \@@startlink[1]{}%
\providecommand \@@endlink[0]{}%
\providecommand \url  [0]{\begingroup\@sanitize@url \@url }%
\providecommand \@url [1]{\endgroup\@href {#1}{\urlprefix }}%
\providecommand \urlprefix  [0]{URL }%
\providecommand \Eprint [0]{\href }%
\providecommand \doibase [0]{http://dx.doi.org/}%
\providecommand \selectlanguage [0]{\@gobble}%
\providecommand \bibinfo  [0]{\@secondoftwo}%
\providecommand \bibfield  [0]{\@secondoftwo}%
\providecommand \translation [1]{[#1]}%
\providecommand \BibitemOpen [0]{}%
\providecommand \bibitemStop [0]{}%
\providecommand \bibitemNoStop [0]{.\EOS\space}%
\providecommand \EOS [0]{\spacefactor3000\relax}%
\providecommand \BibitemShut  [1]{\csname bibitem#1\endcsname}%
\let\auto@bib@innerbib\@empty
\bibitem [{\citenamefont {{Peebles}}(1982)}]{Peebles1982}%
  \BibitemOpen
  \bibfield  {author} {\bibinfo {author} {\bibfnamefont {P.~J.~E.}\
  \bibnamefont {{Peebles}}},\ }\href {\doibase 10.1086/183911} {\bibfield
  {journal} {\bibinfo  {journal} {\apjl}\ }\textbf {\bibinfo {volume} {263}},\
  \bibinfo {pages} {L1} (\bibinfo {year} {1982})}\BibitemShut {NoStop}%
\bibitem [{\citenamefont {Blumenthal}\ \emph {et~al.}(1984)\citenamefont
  {Blumenthal}, \citenamefont {Faber}, \citenamefont {Primack},\ and\
  \citenamefont {Rees}}]{BlumenthalEtAl1984}%
  \BibitemOpen
  \bibfield  {author} {\bibinfo {author} {\bibfnamefont {G.~R.}\ \bibnamefont
  {Blumenthal}}, \bibinfo {author} {\bibfnamefont {S.~M.}\ \bibnamefont
  {Faber}}, \bibinfo {author} {\bibfnamefont {J.~R.}\ \bibnamefont {Primack}},
  \ and\ \bibinfo {author} {\bibfnamefont {M.~J.}\ \bibnamefont {Rees}},\
  }\href {\doibase 10.1038/311517a0} {\bibfield  {journal} {\bibinfo  {journal}
  {\nat}\ }\textbf {\bibinfo {volume} {311}},\ \bibinfo {pages} {517} (\bibinfo
  {year} {1984})}\BibitemShut {NoStop}%
\bibitem [{\citenamefont {{Bertone}}\ and\ \citenamefont
  {{Hooper}}(2018)}]{BertoneEtAl2018}%
  \BibitemOpen
  \bibfield  {author} {\bibinfo {author} {\bibfnamefont {G.}~\bibnamefont
  {{Bertone}}}\ and\ \bibinfo {author} {\bibfnamefont {D.}~\bibnamefont
  {{Hooper}}},\ }\href {\doibase 10.1103/RevModPhys.90.045002} {\bibfield
  {journal} {\bibinfo  {journal} {Reviews of Modern Physics}\ }\textbf
  {\bibinfo {volume} {90}},\ \bibinfo {eid} {045002} (\bibinfo {year}
  {2018})},\ \Eprint {http://arxiv.org/abs/1605.04909} {arXiv:1605.04909}
  \BibitemShut {NoStop}%
\bibitem [{\citenamefont {{Lee}}\ and\ \citenamefont
  {{Weinberg}}(1977)}]{LeeEtAl1977a}%
  \BibitemOpen
  \bibfield  {author} {\bibinfo {author} {\bibfnamefont {B.~W.}\ \bibnamefont
  {{Lee}}}\ and\ \bibinfo {author} {\bibfnamefont {S.}~\bibnamefont
  {{Weinberg}}},\ }\href {\doibase 10.1103/PhysRevLett.39.165} {\bibfield
  {journal} {\bibinfo  {journal} {\prl}\ }\textbf {\bibinfo {volume} {39}},\
  \bibinfo {pages} {165} (\bibinfo {year} {1977})}\BibitemShut {NoStop}%
\bibitem [{\citenamefont {{Bond}}\ \emph {et~al.}(1982)\citenamefont {{Bond}},
  \citenamefont {{Szalay}},\ and\ \citenamefont {{Turner}}}]{BondEtAl1982}%
  \BibitemOpen
  \bibfield  {author} {\bibinfo {author} {\bibfnamefont {J.~R.}\ \bibnamefont
  {{Bond}}}, \bibinfo {author} {\bibfnamefont {A.~S.}\ \bibnamefont
  {{Szalay}}}, \ and\ \bibinfo {author} {\bibfnamefont {M.~S.}\ \bibnamefont
  {{Turner}}},\ }\href {\doibase 10.1103/PhysRevLett.48.1636} {\bibfield
  {journal} {\bibinfo  {journal} {\prl}\ }\textbf {\bibinfo {volume} {48}},\
  \bibinfo {pages} {1636} (\bibinfo {year} {1982})}\BibitemShut {NoStop}%
\bibitem [{\citenamefont {{Bin{\'e}truy}}\ \emph {et~al.}(1984)\citenamefont
  {{Bin{\'e}truy}}, \citenamefont {{Girardi}},\ and\ \citenamefont
  {{Salati}}}]{BinetruyEtAl1984a}%
  \BibitemOpen
  \bibfield  {author} {\bibinfo {author} {\bibfnamefont {P.}~\bibnamefont
  {{Bin{\'e}truy}}}, \bibinfo {author} {\bibfnamefont {G.}~\bibnamefont
  {{Girardi}}}, \ and\ \bibinfo {author} {\bibfnamefont {P.}~\bibnamefont
  {{Salati}}},\ }\href {\doibase 10.1016/0550-3213(84)90161-5} {\bibfield
  {journal} {\bibinfo  {journal} {Nuclear Physics B}\ }\textbf {\bibinfo
  {volume} {237}},\ \bibinfo {pages} {285} (\bibinfo {year}
  {1984})}\BibitemShut {NoStop}%
\bibitem [{\citenamefont {{Srednicki}}\ \emph {et~al.}(1988)\citenamefont
  {{Srednicki}}, \citenamefont {{Watkins}},\ and\ \citenamefont
  {{Olive}}}]{SrednickiEtAl1988}%
  \BibitemOpen
  \bibfield  {author} {\bibinfo {author} {\bibfnamefont {M.}~\bibnamefont
  {{Srednicki}}}, \bibinfo {author} {\bibfnamefont {R.}~\bibnamefont
  {{Watkins}}}, \ and\ \bibinfo {author} {\bibfnamefont {K.~A.}\ \bibnamefont
  {{Olive}}},\ }\href {\doibase 10.1016/0550-3213(88)90099-5} {\bibfield
  {journal} {\bibinfo  {journal} {Nuclear Physics B}\ }\textbf {\bibinfo
  {volume} {310}},\ \bibinfo {pages} {693} (\bibinfo {year}
  {1988})}\BibitemShut {NoStop}%
\bibitem [{\citenamefont {Giudice}(2017)}]{Giudice2017}%
  \BibitemOpen
  \bibfield  {author} {\bibinfo {author} {\bibfnamefont {G.~F.}\ \bibnamefont
  {Giudice}},\ }\href {https://ui.adsabs.harvard.edu/abs/2017arXiv171007663G}
  {\bibfield  {journal} {\bibinfo  {journal} {arXiv e-prints}\ } (\bibinfo
  {year} {2017})},\ \Eprint {http://arxiv.org/abs/1710.07663}
  {arXiv:1710.07663} \BibitemShut {NoStop}%
\bibitem [{\citenamefont {Arcadi}\ \emph {et~al.}(2018)\citenamefont {Arcadi},
  \citenamefont {Dutra}, \citenamefont {Ghosh}, \citenamefont {Lindner},
  \citenamefont {Mambrini}, \citenamefont {Pierre}, \citenamefont {Profumo},\
  and\ \citenamefont {Queiroz}}]{ArcadiEtAl2017}%
  \BibitemOpen
  \bibfield  {author} {\bibinfo {author} {\bibfnamefont {G.}~\bibnamefont
  {Arcadi}}, \bibinfo {author} {\bibfnamefont {M.}~\bibnamefont {Dutra}},
  \bibinfo {author} {\bibfnamefont {P.}~\bibnamefont {Ghosh}}, \bibinfo
  {author} {\bibfnamefont {M.}~\bibnamefont {Lindner}}, \bibinfo {author}
  {\bibfnamefont {Y.}~\bibnamefont {Mambrini}}, \bibinfo {author}
  {\bibfnamefont {M.}~\bibnamefont {Pierre}}, \bibinfo {author} {\bibfnamefont
  {S.}~\bibnamefont {Profumo}}, \ and\ \bibinfo {author} {\bibfnamefont
  {F.~S.}\ \bibnamefont {Queiroz}},\ }\href {\doibase
  10.1140/epjc/s10052-018-5662-y} {\bibfield  {journal} {\bibinfo  {journal}
  {\epjc}\ }\textbf {\bibinfo {volume} {78}},\ \bibinfo {eid} {203} (\bibinfo
  {year} {2018})},\ \Eprint {http://arxiv.org/abs/1703.07364} {arXiv:1703.07364
  [hep-ph]} \BibitemShut {NoStop}%
\bibitem [{\citenamefont {{Leane}}\ \emph {et~al.}(2018)\citenamefont
  {{Leane}}, \citenamefont {{Slatyer}}, \citenamefont {{Beacom}},\ and\
  \citenamefont {{Ng}}}]{LeaneEtAl2018}%
  \BibitemOpen
  \bibfield  {author} {\bibinfo {author} {\bibfnamefont {R.~K.}\ \bibnamefont
  {{Leane}}}, \bibinfo {author} {\bibfnamefont {T.~R.}\ \bibnamefont
  {{Slatyer}}}, \bibinfo {author} {\bibfnamefont {J.~F.}\ \bibnamefont
  {{Beacom}}}, \ and\ \bibinfo {author} {\bibfnamefont {K.~C.~Y.}\ \bibnamefont
  {{Ng}}},\ }\href {\doibase 10.1103/PhysRevD.98.023016} {\bibfield  {journal}
  {\bibinfo  {journal} {\prd}\ }\textbf {\bibinfo {volume} {98}},\ \bibinfo
  {eid} {023016} (\bibinfo {year} {2018})},\ \Eprint
  {http://arxiv.org/abs/1805.10305} {arXiv:1805.10305 [hep-ph]} \BibitemShut
  {NoStop}%
\bibitem [{\citenamefont {{Gunn}}\ \emph {et~al.}(1978)\citenamefont {{Gunn}},
  \citenamefont {{Lee}}, \citenamefont {{Lerche}}, \citenamefont {{Schramm}},\
  and\ \citenamefont {{Steigman}}}]{GunnEtAl1978}%
  \BibitemOpen
  \bibfield  {author} {\bibinfo {author} {\bibfnamefont {J.~E.}\ \bibnamefont
  {{Gunn}}}, \bibinfo {author} {\bibfnamefont {B.~W.}\ \bibnamefont {{Lee}}},
  \bibinfo {author} {\bibfnamefont {I.}~\bibnamefont {{Lerche}}}, \bibinfo
  {author} {\bibfnamefont {D.~N.}\ \bibnamefont {{Schramm}}}, \ and\ \bibinfo
  {author} {\bibfnamefont {G.}~\bibnamefont {{Steigman}}},\ }\href {\doibase
  10.1086/156335} {\bibfield  {journal} {\bibinfo  {journal} {\apj}\ }\textbf
  {\bibinfo {volume} {223}},\ \bibinfo {pages} {1015} (\bibinfo {year}
  {1978})}\BibitemShut {NoStop}%
\bibitem [{\citenamefont {{Silk}}\ and\ \citenamefont
  {{Srednicki}}(1984)}]{SilkEtAl1984}%
  \BibitemOpen
  \bibfield  {author} {\bibinfo {author} {\bibfnamefont {J.}~\bibnamefont
  {{Silk}}}\ and\ \bibinfo {author} {\bibfnamefont {M.}~\bibnamefont
  {{Srednicki}}},\ }\href {\doibase 10.1103/PhysRevLett.53.624} {\bibfield
  {journal} {\bibinfo  {journal} {\prl}\ }\textbf {\bibinfo {volume} {53}},\
  \bibinfo {pages} {624} (\bibinfo {year} {1984})}\BibitemShut {NoStop}%
\bibitem [{\citenamefont {{Goodman}}\ and\ \citenamefont
  {{Witten}}(1985)}]{GoodmanEtAl1985}%
  \BibitemOpen
  \bibfield  {author} {\bibinfo {author} {\bibfnamefont {M.~W.}\ \bibnamefont
  {{Goodman}}}\ and\ \bibinfo {author} {\bibfnamefont {E.}~\bibnamefont
  {{Witten}}},\ }\href {\doibase 10.1103/PhysRevD.31.3059} {\bibfield
  {journal} {\bibinfo  {journal} {\prd}\ }\textbf {\bibinfo {volume} {31}},\
  \bibinfo {pages} {3059} (\bibinfo {year} {1985})}\BibitemShut {NoStop}%
\bibitem [{\citenamefont {{Primack}}\ \emph {et~al.}(1988)\citenamefont
  {{Primack}}, \citenamefont {{Seckel}},\ and\ \citenamefont
  {{Sadoulet}}}]{PrimackEtAl1988}%
  \BibitemOpen
  \bibfield  {author} {\bibinfo {author} {\bibfnamefont {J.~R.}\ \bibnamefont
  {{Primack}}}, \bibinfo {author} {\bibfnamefont {D.}~\bibnamefont {{Seckel}}},
  \ and\ \bibinfo {author} {\bibfnamefont {B.}~\bibnamefont {{Sadoulet}}},\
  }\href {\doibase 10.1146/annurev.ns.38.120188.003535} {\bibfield  {journal}
  {\bibinfo  {journal} {Annual Review of Nuclear and Particle Science}\
  }\textbf {\bibinfo {volume} {38}},\ \bibinfo {pages} {751} (\bibinfo {year}
  {1988})}\BibitemShut {NoStop}%
\bibitem [{\citenamefont {{Jungman}}\ \emph {et~al.}(1996)\citenamefont
  {{Jungman}}, \citenamefont {{Kamionkowski}},\ and\ \citenamefont
  {{Griest}}}]{JungmanEtAl1996}%
  \BibitemOpen
  \bibfield  {author} {\bibinfo {author} {\bibfnamefont {G.}~\bibnamefont
  {{Jungman}}}, \bibinfo {author} {\bibfnamefont {M.}~\bibnamefont
  {{Kamionkowski}}}, \ and\ \bibinfo {author} {\bibfnamefont {K.}~\bibnamefont
  {{Griest}}},\ }\href {\doibase 10.1016/0370-1573(95)00058-5} {\bibfield
  {journal} {\bibinfo  {journal} {\physrep}\ }\textbf {\bibinfo {volume}
  {267}},\ \bibinfo {pages} {195} (\bibinfo {year} {1996})},\ \Eprint
  {http://arxiv.org/abs/hep-ph/9506380} {hep-ph/9506380} \BibitemShut {NoStop}%
\bibitem [{\citenamefont {{Feng}}(2010)}]{Feng2010}%
  \BibitemOpen
  \bibfield  {author} {\bibinfo {author} {\bibfnamefont {J.~L.}\ \bibnamefont
  {{Feng}}},\ }\href {\doibase 10.1146/annurev-astro-082708-101659} {\bibfield
  {journal} {\bibinfo  {journal} {\araa}\ }\textbf {\bibinfo {volume} {48}},\
  \bibinfo {pages} {495} (\bibinfo {year} {2010})},\ \Eprint
  {http://arxiv.org/abs/1003.0904} {arXiv:1003.0904 [astro-ph.CO]} \BibitemShut
  {NoStop}%
\bibitem [{\citenamefont {{Bergstr{\"o}m}}(2000)}]{Bergstroem2000}%
  \BibitemOpen
  \bibfield  {author} {\bibinfo {author} {\bibfnamefont {L.}~\bibnamefont
  {{Bergstr{\"o}m}}},\ }\href {\doibase 10.1088/0034-4885/63/5/2r3} {\bibfield
  {journal} {\bibinfo  {journal} {Reports on Progress in Physics}\ }\textbf
  {\bibinfo {volume} {63}},\ \bibinfo {pages} {793} (\bibinfo {year} {2000})},\
  \Eprint {http://arxiv.org/abs/hep-ph/0002126} {hep-ph/0002126} \BibitemShut
  {NoStop}%
\bibitem [{\citenamefont {{Lavalle}}\ and\ \citenamefont
  {{Salati}}(2012)}]{LavalleEtAl2012}%
  \BibitemOpen
  \bibfield  {author} {\bibinfo {author} {\bibfnamefont {J.}~\bibnamefont
  {{Lavalle}}}\ and\ \bibinfo {author} {\bibfnamefont {P.}~\bibnamefont
  {{Salati}}},\ }\href {\doibase 10.1016/j.crhy.2012.05.001} {\bibfield
  {journal} {\bibinfo  {journal} {Comptes Rendus Physique}\ }\textbf {\bibinfo
  {volume} {13}},\ \bibinfo {pages} {740} (\bibinfo {year} {2012})},\ \Eprint
  {http://arxiv.org/abs/1205.1004} {arXiv:1205.1004 [astro-ph.HE]} \BibitemShut
  {NoStop}%
\bibitem [{\citenamefont {{Bringmann}}\ and\ \citenamefont
  {{Weniger}}(2012)}]{BringmannEtAl2012c}%
  \BibitemOpen
  \bibfield  {author} {\bibinfo {author} {\bibfnamefont {T.}~\bibnamefont
  {{Bringmann}}}\ and\ \bibinfo {author} {\bibfnamefont {C.}~\bibnamefont
  {{Weniger}}},\ }\href {\doibase 10.1016/j.dark.2012.10.005} {\bibfield
  {journal} {\bibinfo  {journal} {Physics of the Dark Universe}\ }\textbf
  {\bibinfo {volume} {1}},\ \bibinfo {pages} {194} (\bibinfo {year} {2012})},\
  \Eprint {http://arxiv.org/abs/1208.5481} {arXiv:1208.5481 [hep-ph]}
  \BibitemShut {NoStop}%
\bibitem [{\citenamefont {{Liu}}\ \emph {et~al.}(2016)\citenamefont {{Liu}},
  \citenamefont {{Slatyer}},\ and\ \citenamefont {{Zavala}}}]{LiuEtAl2016}%
  \BibitemOpen
  \bibfield  {author} {\bibinfo {author} {\bibfnamefont {H.}~\bibnamefont
  {{Liu}}}, \bibinfo {author} {\bibfnamefont {T.~R.}\ \bibnamefont
  {{Slatyer}}}, \ and\ \bibinfo {author} {\bibfnamefont {J.}~\bibnamefont
  {{Zavala}}},\ }\href {\doibase 10.1103/PhysRevD.94.063507} {\bibfield
  {journal} {\bibinfo  {journal} {\prd}\ }\textbf {\bibinfo {volume} {94}},\
  \bibinfo {eid} {063507} (\bibinfo {year} {2016})},\ \Eprint
  {http://arxiv.org/abs/1604.02457} {arXiv:1604.02457} \BibitemShut {NoStop}%
\bibitem [{\citenamefont {Boudaud}\ \emph {et~al.}(2019)\citenamefont
  {Boudaud}, \citenamefont {Lacroix}, \citenamefont {Stref},\ and\
  \citenamefont {Lavalle}}]{BoudaudEtAl2019a}%
  \BibitemOpen
  \bibfield  {author} {\bibinfo {author} {\bibfnamefont {M.}~\bibnamefont
  {Boudaud}}, \bibinfo {author} {\bibfnamefont {T.}~\bibnamefont {Lacroix}},
  \bibinfo {author} {\bibfnamefont {M.}~\bibnamefont {Stref}}, \ and\ \bibinfo
  {author} {\bibfnamefont {J.}~\bibnamefont {Lavalle}},\ }\href {\doibase
  10.1103/PhysRevD.99.061302} {\bibfield  {journal} {\bibinfo  {journal}
  {\prd}\ }\textbf {\bibinfo {volume} {99}},\ \bibinfo {eid} {061302} (\bibinfo
  {year} {2019})},\ \Eprint {http://arxiv.org/abs/1810.01680} {arXiv:1810.01680
  [astro-ph.HE]} \BibitemShut {NoStop}%
\bibitem [{\citenamefont {{Lewin}}\ and\ \citenamefont
  {{Smith}}(1996)}]{LewinEtAl1996}%
  \BibitemOpen
  \bibfield  {author} {\bibinfo {author} {\bibfnamefont {J.~D.}\ \bibnamefont
  {{Lewin}}}\ and\ \bibinfo {author} {\bibfnamefont {P.~F.}\ \bibnamefont
  {{Smith}}},\ }\href {\doibase 10.1016/S0927-6505(96)00047-3} {\bibfield
  {journal} {\bibinfo  {journal} {Astroparticle Physics}\ }\textbf {\bibinfo
  {volume} {6}},\ \bibinfo {pages} {87} (\bibinfo {year} {1996})}\BibitemShut
  {NoStop}%
\bibitem [{\citenamefont {{Freese}}\ \emph {et~al.}(2013)\citenamefont
  {{Freese}}, \citenamefont {{Lisanti}},\ and\ \citenamefont
  {{Savage}}}]{FreeseEtAl2013}%
  \BibitemOpen
  \bibfield  {author} {\bibinfo {author} {\bibfnamefont {K.}~\bibnamefont
  {{Freese}}}, \bibinfo {author} {\bibfnamefont {M.}~\bibnamefont {{Lisanti}}},
  \ and\ \bibinfo {author} {\bibfnamefont {C.}~\bibnamefont {{Savage}}},\
  }\href {\doibase 10.1103/RevModPhys.85.1561} {\bibfield  {journal} {\bibinfo
  {journal} {Reviews of Modern Physics}\ }\textbf {\bibinfo {volume} {85}},\
  \bibinfo {pages} {1561} (\bibinfo {year} {2013})},\ \Eprint
  {http://arxiv.org/abs/1209.3339} {arXiv:1209.3339} \BibitemShut {NoStop}%
\bibitem [{\citenamefont {{Fairbairn}}\ \emph {et~al.}(2007)\citenamefont
  {{Fairbairn}}, \citenamefont {{Kraan}}, \citenamefont {{Milstead}},
  \citenamefont {{Sj{\"o}strand}}, \citenamefont {{Skands}},\ and\
  \citenamefont {{Sloan}}}]{FairbairnEtAl2007}%
  \BibitemOpen
  \bibfield  {author} {\bibinfo {author} {\bibfnamefont {M.}~\bibnamefont
  {{Fairbairn}}}, \bibinfo {author} {\bibfnamefont {A.~C.}\ \bibnamefont
  {{Kraan}}}, \bibinfo {author} {\bibfnamefont {D.~A.}\ \bibnamefont
  {{Milstead}}}, \bibinfo {author} {\bibfnamefont {T.}~\bibnamefont
  {{Sj{\"o}strand}}}, \bibinfo {author} {\bibfnamefont {P.}~\bibnamefont
  {{Skands}}}, \ and\ \bibinfo {author} {\bibfnamefont {T.}~\bibnamefont
  {{Sloan}}},\ }\href {\doibase 10.1016/j.physrep.2006.10.002} {\bibfield
  {journal} {\bibinfo  {journal} {\physrep}\ }\textbf {\bibinfo {volume}
  {438}},\ \bibinfo {pages} {1} (\bibinfo {year} {2007})},\ \Eprint
  {http://arxiv.org/abs/hep-ph/0611040} {hep-ph/0611040} \BibitemShut {NoStop}%
\bibitem [{\citenamefont {{Abdallah}}\ \emph {et~al.}}(2015)]{AbdallahEtAl2015}%
  \BibitemOpen
  \bibfield  {author} {\bibinfo {author} {\bibfnamefont {J.}~\bibnamefont
  {{Abdallah}~\emph{et al.}} },\ }\href {\doibase 10.1016/j.dark.2015.08.001}
  {\bibfield  {journal} {\bibinfo  {journal} {\pdu}\ }\textbf {\bibinfo
  {volume} {9}},\ \bibinfo {pages} {8} (\bibinfo {year} {2015})},\ \Eprint
  {http://arxiv.org/abs/1506.03116} {arXiv:1506.03116 [hep-ph]} \BibitemShut
  {NoStop}%
\bibitem [{\citenamefont {{Strigari}}(2013)}]{Strigari2013}%
  \BibitemOpen
  \bibfield  {author} {\bibinfo {author} {\bibfnamefont {L.~E.}\ \bibnamefont
  {{Strigari}}},\ }\href {\doibase 10.1016/j.physrep.2013.05.004} {\bibfield
  {journal} {\bibinfo  {journal} {\physrep}\ }\textbf {\bibinfo {volume}
  {531}},\ \bibinfo {pages} {1} (\bibinfo {year} {2013})},\ \Eprint
  {http://arxiv.org/abs/1211.7090} {arXiv:1211.7090 [astro-ph.CO]} \BibitemShut
  {NoStop}%
\bibitem [{\citenamefont {Schmid}\ \emph {et~al.}(1999)\citenamefont {Schmid},
  \citenamefont {Schwarz},\ and\ \citenamefont {Widerin}}]{SchmidEtAl1999}%
  \BibitemOpen
  \bibfield  {author} {\bibinfo {author} {\bibfnamefont {C.}~\bibnamefont
  {Schmid}}, \bibinfo {author} {\bibfnamefont {D.~J.}\ \bibnamefont {Schwarz}},
  \ and\ \bibinfo {author} {\bibfnamefont {P.}~\bibnamefont {Widerin}},\ }\href
  {\doibase 10.1103/PhysRevD.59.043517} {\bibfield  {journal} {\bibinfo
  {journal} {\prd}\ }\textbf {\bibinfo {volume} {59}},\ \bibinfo {eid} {043517}
  (\bibinfo {year} {1999})},\ \Eprint {http://arxiv.org/abs/astro-ph/9807257}
  {astro-ph/9807257} \BibitemShut {NoStop}%
\bibitem [{\citenamefont {{B{\oe}hm}}\ \emph {et~al.}(2001)\citenamefont
  {{B{\oe}hm}}, \citenamefont {{Fayet}},\ and\ \citenamefont
  {{Schaeffer}}}]{BoehmEtAl2001}%
  \BibitemOpen
  \bibfield  {author} {\bibinfo {author} {\bibfnamefont {C.}~\bibnamefont
  {{B{\oe}hm}}}, \bibinfo {author} {\bibfnamefont {P.}~\bibnamefont {{Fayet}}},
  \ and\ \bibinfo {author} {\bibfnamefont {R.}~\bibnamefont {{Schaeffer}}},\
  }\href {\doibase 10.1016/S0370-2693(01)01060-7} {\bibfield  {journal}
  {\bibinfo  {journal} {Physics Letters B}\ }\textbf {\bibinfo {volume}
  {518}},\ \bibinfo {pages} {8} (\bibinfo {year} {2001})},\ \Eprint
  {http://arxiv.org/abs/astro-ph/0012504} {astro-ph/0012504} \BibitemShut
  {NoStop}%
\bibitem [{\citenamefont {Chen}\ \emph {et~al.}(2001)\citenamefont {Chen},
  \citenamefont {Kamionkowski},\ and\ \citenamefont {Zhang}}]{ChenEtAl2001}%
  \BibitemOpen
  \bibfield  {author} {\bibinfo {author} {\bibfnamefont {X.}~\bibnamefont
  {Chen}}, \bibinfo {author} {\bibfnamefont {M.}~\bibnamefont {Kamionkowski}},
  \ and\ \bibinfo {author} {\bibfnamefont {X.}~\bibnamefont {Zhang}},\ }\href
  {\doibase 10.1103/PhysRevD.64.021302} {\bibfield  {journal} {\bibinfo
  {journal} {\prd}\ }\textbf {\bibinfo {volume} {64}},\ \bibinfo {eid} {021302}
  (\bibinfo {year} {2001})},\ \Eprint {http://arxiv.org/abs/astro-ph/0103452}
  {astro-ph/0103452} \BibitemShut {NoStop}%
\bibitem [{\citenamefont {Hofmann}\ \emph {et~al.}(2001)\citenamefont
  {Hofmann}, \citenamefont {Schwarz},\ and\ \citenamefont
  {St{\"o}cker}}]{HofmannEtAl2001}%
  \BibitemOpen
  \bibfield  {author} {\bibinfo {author} {\bibfnamefont {S.}~\bibnamefont
  {Hofmann}}, \bibinfo {author} {\bibfnamefont {D.~J.}\ \bibnamefont
  {Schwarz}}, \ and\ \bibinfo {author} {\bibfnamefont {H.}~\bibnamefont
  {St{\"o}cker}},\ }\href {\doibase 10.1103/PhysRevD.64.083507} {\bibfield
  {journal} {\bibinfo  {journal} {\prd}\ }\textbf {\bibinfo {volume} {64}},\
  \bibinfo {pages} {083507} (\bibinfo {year} {2001})},\ \Eprint
  {http://arxiv.org/abs/astro-ph/0104173} {arXiv:astro-ph/0104173 [astro-ph]}
  \BibitemShut {NoStop}%
\bibitem [{\citenamefont {{Berezinsky}}\ \emph {et~al.}(2003)\citenamefont
  {{Berezinsky}}, \citenamefont {{Dokuchaev}},\ and\ \citenamefont
  {{Eroshenko}}}]{BerezinskyEtAl2003}%
  \BibitemOpen
  \bibfield  {author} {\bibinfo {author} {\bibfnamefont {V.}~\bibnamefont
  {{Berezinsky}}}, \bibinfo {author} {\bibfnamefont {V.}~\bibnamefont
  {{Dokuchaev}}}, \ and\ \bibinfo {author} {\bibfnamefont {Y.}~\bibnamefont
  {{Eroshenko}}},\ }\href {\doibase 10.1103/PhysRevD.68.103003} {\bibfield
  {journal} {\bibinfo  {journal} {\prd}\ }\textbf {\bibinfo {volume} {68}},\
  \bibinfo {eid} {103003} (\bibinfo {year} {2003})},\ \Eprint
  {http://arxiv.org/abs/astro-ph/0301551} {astro-ph/0301551} \BibitemShut
  {NoStop}%
\bibitem [{\citenamefont {{Green}}\ \emph {et~al.}(2004)\citenamefont
  {{Green}}, \citenamefont {{Hofmann}},\ and\ \citenamefont
  {{Schwarz}}}]{GreenEtAl2004}%
  \BibitemOpen
  \bibfield  {author} {\bibinfo {author} {\bibfnamefont {A.~M.}\ \bibnamefont
  {{Green}}}, \bibinfo {author} {\bibfnamefont {S.}~\bibnamefont {{Hofmann}}},
  \ and\ \bibinfo {author} {\bibfnamefont {D.~J.}\ \bibnamefont {{Schwarz}}},\
  }\href {\doibase 10.1111/j.1365-2966.2004.08232.x} {\bibfield  {journal}
  {\bibinfo  {journal} {\mnras}\ }\textbf {\bibinfo {volume} {353}},\ \bibinfo
  {pages} {L23} (\bibinfo {year} {2004})},\ \Eprint
  {http://arxiv.org/abs/astro-ph/0309621} {astro-ph/0309621} \BibitemShut
  {NoStop}%
\bibitem [{\citenamefont {{Bertschinger}}(2006)}]{Bertschinger2006a}%
  \BibitemOpen
  \bibfield  {author} {\bibinfo {author} {\bibfnamefont {E.}~\bibnamefont
  {{Bertschinger}}},\ }\href {\doibase 10.1103/PhysRevD.74.063509} {\bibfield
  {journal} {\bibinfo  {journal} {\prd}\ }\textbf {\bibinfo {volume} {74}},\
  \bibinfo {eid} {063509} (\bibinfo {year} {2006})},\ \Eprint
  {http://arxiv.org/abs/astro-ph/0607319} {astro-ph/0607319} \BibitemShut
  {NoStop}%
\bibitem [{\citenamefont {{Bringmann}}\ and\ \citenamefont
  {{Hofmann}}(2007)}]{BringmannEtAl2007a}%
  \BibitemOpen
  \bibfield  {author} {\bibinfo {author} {\bibfnamefont {T.}~\bibnamefont
  {{Bringmann}}}\ and\ \bibinfo {author} {\bibfnamefont {S.}~\bibnamefont
  {{Hofmann}}},\ }\href {\doibase 10.1088/1475-7516/2007/04/016} {\bibfield
  {journal} {\bibinfo  {journal} {\jcap}\ }\textbf {\bibinfo {volume} {4}},\
  \bibinfo {eid} {016} (\bibinfo {year} {2007})},\ \Eprint
  {http://arxiv.org/abs/hep-ph/0612238} {hep-ph/0612238} \BibitemShut {NoStop}%
\bibitem [{\citenamefont {{Press}}\ and\ \citenamefont
  {{Schechter}}(1974)}]{PressEtAl1974}%
  \BibitemOpen
  \bibfield  {author} {\bibinfo {author} {\bibfnamefont {W.~H.}\ \bibnamefont
  {{Press}}}\ and\ \bibinfo {author} {\bibfnamefont {P.}~\bibnamefont
  {{Schechter}}},\ }\href {\doibase 10.1086/152650} {\bibfield  {journal}
  {\bibinfo  {journal} {\apj}\ }\textbf {\bibinfo {volume} {187}},\ \bibinfo
  {pages} {425} (\bibinfo {year} {1974})}\BibitemShut {NoStop}%
\bibitem [{\citenamefont {{Bond}}\ \emph {et~al.}(1991)\citenamefont {{Bond}},
  \citenamefont {{Cole}}, \citenamefont {{Efstathiou}},\ and\ \citenamefont
  {{Kaiser}}}]{BondEtAl1991a}%
  \BibitemOpen
  \bibfield  {author} {\bibinfo {author} {\bibfnamefont {J.~R.}\ \bibnamefont
  {{Bond}}}, \bibinfo {author} {\bibfnamefont {S.}~\bibnamefont {{Cole}}},
  \bibinfo {author} {\bibfnamefont {G.}~\bibnamefont {{Efstathiou}}}, \ and\
  \bibinfo {author} {\bibfnamefont {N.}~\bibnamefont {{Kaiser}}},\ }\href
  {\doibase 10.1086/170520} {\bibfield  {journal} {\bibinfo  {journal} {\apj}\
  }\textbf {\bibinfo {volume} {379}},\ \bibinfo {pages} {440} (\bibinfo {year}
  {1991})}\BibitemShut {NoStop}%
\bibitem [{\citenamefont {{Lacey}}\ and\ \citenamefont
  {{Cole}}(1993)}]{LaceyEtAl1993}%
  \BibitemOpen
  \bibfield  {author} {\bibinfo {author} {\bibfnamefont {C.}~\bibnamefont
  {{Lacey}}}\ and\ \bibinfo {author} {\bibfnamefont {S.}~\bibnamefont
  {{Cole}}},\ }\href {https://ui.adsabs.harvard.edu/abs/1993MNRAS.262..627L}
  {\bibfield  {journal} {\bibinfo  {journal} {\mnras}\ }\textbf {\bibinfo
  {volume} {262}},\ \bibinfo {pages} {627} (\bibinfo {year}
  {1993})}\BibitemShut {NoStop}%
\bibitem [{\citenamefont {{Diemand}}\ \emph {et~al.}(2005)\citenamefont
  {{Diemand}}, \citenamefont {{Moore}},\ and\ \citenamefont
  {{Stadel}}}]{DiemandEtAl2005a}%
  \BibitemOpen
  \bibfield  {author} {\bibinfo {author} {\bibfnamefont {J.}~\bibnamefont
  {{Diemand}}}, \bibinfo {author} {\bibfnamefont {B.}~\bibnamefont {{Moore}}},
  \ and\ \bibinfo {author} {\bibfnamefont {J.}~\bibnamefont {{Stadel}}},\
  }\href {\doibase 10.1038/nature03270} {\bibfield  {journal} {\bibinfo
  {journal} {\nat}\ }\textbf {\bibinfo {volume} {433}},\ \bibinfo {pages} {389}
  (\bibinfo {year} {2005})},\ \Eprint {http://arxiv.org/abs/astro-ph/0501589}
  {astro-ph/0501589} \BibitemShut {NoStop}%
\bibitem [{\citenamefont {{Ishiyama}}\ \emph {et~al.}(2010)\citenamefont
  {{Ishiyama}}, \citenamefont {{Makino}},\ and\ \citenamefont
  {{Ebisuzaki}}}]{IshiyamaEtAl2010}%
  \BibitemOpen
  \bibfield  {author} {\bibinfo {author} {\bibfnamefont {T.}~\bibnamefont
  {{Ishiyama}}}, \bibinfo {author} {\bibfnamefont {J.}~\bibnamefont
  {{Makino}}}, \ and\ \bibinfo {author} {\bibfnamefont {T.}~\bibnamefont
  {{Ebisuzaki}}},\ }\href {\doibase 10.1088/2041-8205/723/2/L195} {\bibfield
  {journal} {\bibinfo  {journal} {\apjl}\ }\textbf {\bibinfo {volume} {723}},\
  \bibinfo {pages} {L195} (\bibinfo {year} {2010})},\ \Eprint
  {http://arxiv.org/abs/1006.3392} {arXiv:1006.3392 [astro-ph.CO]} \BibitemShut
  {NoStop}%
\bibitem [{\citenamefont {{Berezinsky}}\ \emph {et~al.}(2014)\citenamefont
  {{Berezinsky}}, \citenamefont {{Dokuchaev}},\ and\ \citenamefont
  {{Eroshenko}}}]{BerezinskyEtAl2014}%
  \BibitemOpen
  \bibfield  {author} {\bibinfo {author} {\bibfnamefont {V.~S.}\ \bibnamefont
  {{Berezinsky}}}, \bibinfo {author} {\bibfnamefont {V.~I.}\ \bibnamefont
  {{Dokuchaev}}}, \ and\ \bibinfo {author} {\bibfnamefont {Y.~N.}\ \bibnamefont
  {{Eroshenko}}},\ }\href {\doibase 10.3367/UFNe.0184.201401a.0003} {\bibfield
  {journal} {\bibinfo  {journal} {Physics Uspekhi}\ }\textbf {\bibinfo {volume}
  {57}},\ \bibinfo {eid} {1} (\bibinfo {year} {2014})},\ \Eprint
  {http://arxiv.org/abs/1405.2204} {arXiv:1405.2204 [astro-ph.HE]} \BibitemShut
  {NoStop}%
\bibitem [{\citenamefont {{Stref}}\ and\ \citenamefont
  {{Lavalle}}(2017)}]{StrefEtAl2017}%
  \BibitemOpen
  \bibfield  {author} {\bibinfo {author} {\bibfnamefont {M.}~\bibnamefont
  {{Stref}}}\ and\ \bibinfo {author} {\bibfnamefont {J.}~\bibnamefont
  {{Lavalle}}},\ }\href {\doibase 10.1103/PhysRevD.95.063003} {\bibfield
  {journal} {\bibinfo  {journal} {\prd}\ }\textbf {\bibinfo {volume} {95}},\
  \bibinfo {pages} {063003} (\bibinfo {year} {2017})},\ \Eprint
  {http://arxiv.org/abs/1610.02233} {arXiv:1610.02233} \BibitemShut {NoStop}%
\bibitem [{\citenamefont {Ishiyama}\ and\ \citenamefont
  {Ando}(2020)}]{IshiyamaEtAl2020a}%
  \BibitemOpen
  \bibfield  {author} {\bibinfo {author} {\bibfnamefont {T.}~\bibnamefont
  {Ishiyama}}\ and\ \bibinfo {author} {\bibfnamefont {S.}~\bibnamefont
  {Ando}},\ }\href {\doibase 10.1093/mnras/staa069} {\bibfield  {journal}
  {\bibinfo  {journal} {\mnras}\ }\textbf {\bibinfo {volume} {492}},\ \bibinfo
  {pages} {3662} (\bibinfo {year} {2020})},\ \Eprint
  {http://arxiv.org/abs/1907.03642} {arXiv:1907.03642 [astro-ph.CO]}
  \BibitemShut {NoStop}%
\bibitem [{\citenamefont {{Silk}}\ and\ \citenamefont
  {{Stebbins}}(1993)}]{SilkEtAl1993}%
  \BibitemOpen
  \bibfield  {author} {\bibinfo {author} {\bibfnamefont {J.}~\bibnamefont
  {{Silk}}}\ and\ \bibinfo {author} {\bibfnamefont {A.}~\bibnamefont
  {{Stebbins}}},\ }\href {\doibase 10.1086/172846} {\bibfield  {journal}
  {\bibinfo  {journal} {\apj}\ }\textbf {\bibinfo {volume} {411}},\ \bibinfo
  {pages} {439} (\bibinfo {year} {1993})}\BibitemShut {NoStop}%
\bibitem [{\citenamefont {{Bergstr{\"o}m}}\ \emph {et~al.}(1999)\citenamefont
  {{Bergstr{\"o}m}}, \citenamefont {{Edsj{\"o}}}, \citenamefont {{Gondolo}},\
  and\ \citenamefont {{Ullio}}}]{BergstroemEtAl1999a}%
  \BibitemOpen
  \bibfield  {author} {\bibinfo {author} {\bibfnamefont {L.}~\bibnamefont
  {{Bergstr{\"o}m}}}, \bibinfo {author} {\bibfnamefont {J.}~\bibnamefont
  {{Edsj{\"o}}}}, \bibinfo {author} {\bibfnamefont {P.}~\bibnamefont
  {{Gondolo}}}, \ and\ \bibinfo {author} {\bibfnamefont {P.}~\bibnamefont
  {{Ullio}}},\ }\href {\doibase 10.1103/PhysRevD.59.043506} {\bibfield
  {journal} {\bibinfo  {journal} {\prd}\ }\textbf {\bibinfo {volume} {59}},\
  \bibinfo {eid} {043506} (\bibinfo {year} {1999})},\ \Eprint
  {http://arxiv.org/abs/astro-ph/9806072} {astro-ph/9806072} \BibitemShut
  {NoStop}%
\bibitem [{\citenamefont {{Lavalle}}\ \emph {et~al.}(2007)\citenamefont
  {{Lavalle}}, \citenamefont {{Pochon}}, \citenamefont {{Salati}},\ and\
  \citenamefont {{Taillet}}}]{LavalleEtAl2007}%
  \BibitemOpen
  \bibfield  {author} {\bibinfo {author} {\bibfnamefont {J.}~\bibnamefont
  {{Lavalle}}}, \bibinfo {author} {\bibfnamefont {J.}~\bibnamefont {{Pochon}}},
  \bibinfo {author} {\bibfnamefont {P.}~\bibnamefont {{Salati}}}, \ and\
  \bibinfo {author} {\bibfnamefont {R.}~\bibnamefont {{Taillet}}},\ }\href
  {\doibase 10.1051/0004-6361:20065312} {\bibfield  {journal} {\bibinfo
  {journal} {\aap}\ }\textbf {\bibinfo {volume} {462}},\ \bibinfo {pages} {827}
  (\bibinfo {year} {2007})},\ \Eprint
  {http://arxiv.org/abs/arXiv:astro-ph/0603796} {arXiv:astro-ph/0603796}
  \BibitemShut {NoStop}%
\bibitem [{\citenamefont {{Lavalle}}\ \emph {et~al.}(2008)\citenamefont
  {{Lavalle}}, \citenamefont {{Yuan}}, \citenamefont {{Maurin}},\ and\
  \citenamefont {{Bi}}}]{LavalleEtAl2008}%
  \BibitemOpen
  \bibfield  {author} {\bibinfo {author} {\bibfnamefont {J.}~\bibnamefont
  {{Lavalle}}}, \bibinfo {author} {\bibfnamefont {Q.}~\bibnamefont {{Yuan}}},
  \bibinfo {author} {\bibfnamefont {D.}~\bibnamefont {{Maurin}}}, \ and\
  \bibinfo {author} {\bibfnamefont {X.-J.}\ \bibnamefont {{Bi}}},\ }\href
  {\doibase 10.1051/0004-6361:20078723} {\bibfield  {journal} {\bibinfo
  {journal} {\aap}\ }\textbf {\bibinfo {volume} {479}},\ \bibinfo {pages} {427}
  (\bibinfo {year} {2008})},\ \Eprint {http://arxiv.org/abs/0709.3634}
  {arXiv:0709.3634} \BibitemShut {NoStop}%
\bibitem [{\citenamefont {Pieri}\ \emph {et~al.}(2011)\citenamefont {Pieri},
  \citenamefont {Lavalle}, \citenamefont {Bertone},\ and\ \citenamefont
  {Branchini}}]{PieriEtAl2011}%
  \BibitemOpen
  \bibfield  {author} {\bibinfo {author} {\bibfnamefont {L.}~\bibnamefont
  {Pieri}}, \bibinfo {author} {\bibfnamefont {J.}~\bibnamefont {Lavalle}},
  \bibinfo {author} {\bibfnamefont {G.}~\bibnamefont {Bertone}}, \ and\
  \bibinfo {author} {\bibfnamefont {E.}~\bibnamefont {Branchini}},\ }\href
  {\doibase 10.1103/PhysRevD.83.023518} {\bibfield  {journal} {\bibinfo
  {journal} {\prd}\ }\textbf {\bibinfo {volume} {83}},\ \bibinfo {eid} {023518}
  (\bibinfo {year} {2011})},\ \Eprint {http://arxiv.org/abs/0908.0195}
  {arXiv:0908.0195 [astro-ph.HE]} \BibitemShut {NoStop}%
\bibitem [{\citenamefont {{Ando}}\ and\ \citenamefont
  {{Komatsu}}(2006)}]{AndoEtAl2006}%
  \BibitemOpen
  \bibfield  {author} {\bibinfo {author} {\bibfnamefont {S.}~\bibnamefont
  {{Ando}}}\ and\ \bibinfo {author} {\bibfnamefont {E.}~\bibnamefont
  {{Komatsu}}},\ }\href {\doibase 10.1103/PhysRevD.73.023521} {\bibfield
  {journal} {\bibinfo  {journal} {\prd}\ }\textbf {\bibinfo {volume} {73}},\
  \bibinfo {eid} {023521} (\bibinfo {year} {2006})},\ \Eprint
  {http://arxiv.org/abs/astro-ph/0512217} {astro-ph/0512217} \BibitemShut
  {NoStop}%
\bibitem [{\citenamefont {{Fornasa}}\ \emph {et~al.}(2016)\citenamefont
  {{Fornasa}}, \citenamefont {{Cuoco}}, \citenamefont {{Zavala}}, \citenamefont
  {{Gaskins}}, \citenamefont {{S{\'a}nchez-Conde}}, \citenamefont
  {{Gomez-Vargas}}, \citenamefont {{Komatsu}}, \citenamefont {{Linden}},
  \citenamefont {{Prada}}, \citenamefont {{Zandanel}},\ and\ \citenamefont
  {{Morselli}}}]{FornasaEtAl2016}%
  \BibitemOpen
  \bibfield  {author} {\bibinfo {author} {\bibfnamefont {M.}~\bibnamefont
  {{Fornasa}}}, \bibinfo {author} {\bibfnamefont {A.}~\bibnamefont {{Cuoco}}},
  \bibinfo {author} {\bibfnamefont {J.}~\bibnamefont {{Zavala}}}, \bibinfo
  {author} {\bibfnamefont {J.~M.}\ \bibnamefont {{Gaskins}}}, \bibinfo {author}
  {\bibfnamefont {M.~A.}\ \bibnamefont {{S{\'a}nchez-Conde}}}, \bibinfo
  {author} {\bibfnamefont {G.}~\bibnamefont {{Gomez-Vargas}}}, \bibinfo
  {author} {\bibfnamefont {E.}~\bibnamefont {{Komatsu}}}, \bibinfo {author}
  {\bibfnamefont {T.}~\bibnamefont {{Linden}}}, \bibinfo {author}
  {\bibfnamefont {F.}~\bibnamefont {{Prada}}}, \bibinfo {author} {\bibfnamefont
  {F.}~\bibnamefont {{Zandanel}}}, \ and\ \bibinfo {author} {\bibfnamefont
  {A.}~\bibnamefont {{Morselli}}},\ }\href {\doibase
  10.1103/PhysRevD.94.123005} {\bibfield  {journal} {\bibinfo  {journal}
  {\prd}\ }\textbf {\bibinfo {volume} {94}},\ \bibinfo {eid} {123005} (\bibinfo
  {year} {2016})},\ \Eprint {http://arxiv.org/abs/1608.07289} {arXiv:1608.07289
  [astro-ph.HE]} \BibitemShut {NoStop}%
\bibitem [{\citenamefont {Tasitsiomi}\ and\ \citenamefont
  {Olinto}(2002)}]{TasitsiomiEtAl2002}%
  \BibitemOpen
  \bibfield  {author} {\bibinfo {author} {\bibfnamefont {A.}~\bibnamefont
  {Tasitsiomi}}\ and\ \bibinfo {author} {\bibfnamefont {A.~V.}\ \bibnamefont
  {Olinto}},\ }\href {\doibase 10.1103/PhysRevD.66.083006} {\bibfield
  {journal} {\bibinfo  {journal} {\prd}\ }\textbf {\bibinfo {volume} {66}},\
  \bibinfo {eid} {083006} (\bibinfo {year} {2002})},\ \Eprint
  {http://arxiv.org/abs/astro-ph/0206040} {astro-ph/0206040} \BibitemShut
  {NoStop}%
\bibitem [{\citenamefont {{Stoehr}}\ \emph {et~al.}(2003)\citenamefont
  {{Stoehr}}, \citenamefont {{White}}, \citenamefont {{Springel}},
  \citenamefont {{Tormen}},\ and\ \citenamefont {{Yoshida}}}]{StoehrEtAl2003}%
  \BibitemOpen
  \bibfield  {author} {\bibinfo {author} {\bibfnamefont {F.}~\bibnamefont
  {{Stoehr}}}, \bibinfo {author} {\bibfnamefont {S.~D.~M.}\ \bibnamefont
  {{White}}}, \bibinfo {author} {\bibfnamefont {V.}~\bibnamefont {{Springel}}},
  \bibinfo {author} {\bibfnamefont {G.}~\bibnamefont {{Tormen}}}, \ and\
  \bibinfo {author} {\bibfnamefont {N.}~\bibnamefont {{Yoshida}}},\ }\href
  {\doibase 10.1046/j.1365-2966.2003.07052.x} {\bibfield  {journal} {\bibinfo
  {journal} {\mnras}\ }\textbf {\bibinfo {volume} {345}},\ \bibinfo {pages}
  {1313} (\bibinfo {year} {2003})},\ \Eprint
  {http://arxiv.org/abs/astro-ph/0307026} {astro-ph/0307026} \BibitemShut
  {NoStop}%
\bibitem [{\citenamefont {Aloisio}\ \emph {et~al.}(2004)\citenamefont
  {Aloisio}, \citenamefont {Blasi},\ and\ \citenamefont
  {Olinto}}]{AloisioEtAl2004a}%
  \BibitemOpen
  \bibfield  {author} {\bibinfo {author} {\bibfnamefont {R.}~\bibnamefont
  {Aloisio}}, \bibinfo {author} {\bibfnamefont {P.}~\bibnamefont {Blasi}}, \
  and\ \bibinfo {author} {\bibfnamefont {A.~V.}\ \bibnamefont {Olinto}},\
  }\href {\doibase 10.1086/380425} {\bibfield  {journal} {\bibinfo  {journal}
  {\apj}\ }\textbf {\bibinfo {volume} {601}},\ \bibinfo {pages} {47} (\bibinfo
  {year} {2004})},\ \Eprint {http://arxiv.org/abs/astro-ph/0206036}
  {astro-ph/0206036} \BibitemShut {NoStop}%
\bibitem [{\citenamefont {{Pieri}}\ \emph {et~al.}(2005)\citenamefont
  {{Pieri}}, \citenamefont {{Branchini}},\ and\ \citenamefont
  {{Hofmann}}}]{PieriEtAl2005}%
  \BibitemOpen
  \bibfield  {author} {\bibinfo {author} {\bibfnamefont {L.}~\bibnamefont
  {{Pieri}}}, \bibinfo {author} {\bibfnamefont {E.}~\bibnamefont
  {{Branchini}}}, \ and\ \bibinfo {author} {\bibfnamefont {S.}~\bibnamefont
  {{Hofmann}}},\ }\href {\doibase 10.1103/PhysRevLett.95.211301} {\bibfield
  {journal} {\bibinfo  {journal} {\prl}\ }\textbf {\bibinfo {volume} {95}},\
  \bibinfo {eid} {211301} (\bibinfo {year} {2005})},\ \Eprint
  {http://arxiv.org/abs/astro-ph/0505356} {astro-ph/0505356} \BibitemShut
  {NoStop}%
\bibitem [{\citenamefont {{Kuhlen}}\ \emph {et~al.}(2008)\citenamefont
  {{Kuhlen}}, \citenamefont {{Diemand}},\ and\ \citenamefont
  {{Madau}}}]{KuhlenEtAl2008}%
  \BibitemOpen
  \bibfield  {author} {\bibinfo {author} {\bibfnamefont {M.}~\bibnamefont
  {{Kuhlen}}}, \bibinfo {author} {\bibfnamefont {J.}~\bibnamefont {{Diemand}}},
  \ and\ \bibinfo {author} {\bibfnamefont {P.}~\bibnamefont {{Madau}}},\ }\href
  {\doibase 10.1086/590337} {\bibfield  {journal} {\bibinfo  {journal} {\apj}\
  }\textbf {\bibinfo {volume} {686}},\ \bibinfo {pages} {262} (\bibinfo {year}
  {2008})},\ \Eprint {http://arxiv.org/abs/0805.4416} {arXiv:0805.4416}
  \BibitemShut {NoStop}%
\bibitem [{\citenamefont {{Anderson}}\ \emph {et~al.}(2010)\citenamefont
  {{Anderson}}, \citenamefont {{Kuhlen}}, \citenamefont {{Diemand}},
  \citenamefont {{Johnson}},\ and\ \citenamefont {{Madau}}}]{AndersonEtAl2010}%
  \BibitemOpen
  \bibfield  {author} {\bibinfo {author} {\bibfnamefont {B.}~\bibnamefont
  {{Anderson}}}, \bibinfo {author} {\bibfnamefont {M.}~\bibnamefont
  {{Kuhlen}}}, \bibinfo {author} {\bibfnamefont {J.}~\bibnamefont {{Diemand}}},
  \bibinfo {author} {\bibfnamefont {R.~P.}\ \bibnamefont {{Johnson}}}, \ and\
  \bibinfo {author} {\bibfnamefont {P.}~\bibnamefont {{Madau}}},\ }\href
  {\doibase 10.1088/0004-637X/718/2/899} {\bibfield  {journal} {\bibinfo
  {journal} {\apj}\ }\textbf {\bibinfo {volume} {718}},\ \bibinfo {pages} {899}
  (\bibinfo {year} {2010})},\ \Eprint {http://arxiv.org/abs/1006.1628}
  {arXiv:1006.1628 [astro-ph.HE]} \BibitemShut {NoStop}%
\bibitem [{\citenamefont {H{\"u}tten}\ \emph {et~al.}(2016)\citenamefont
  {H{\"u}tten}, \citenamefont {Combet}, \citenamefont {Maier},\ and\
  \citenamefont {Maurin}}]{HuettenEtAl2016}%
  \BibitemOpen
  \bibfield  {author} {\bibinfo {author} {\bibfnamefont {M.}~\bibnamefont
  {H{\"u}tten}}, \bibinfo {author} {\bibfnamefont {C.}~\bibnamefont {Combet}},
  \bibinfo {author} {\bibfnamefont {G.}~\bibnamefont {Maier}}, \ and\ \bibinfo
  {author} {\bibfnamefont {D.}~\bibnamefont {Maurin}},\ }\href {\doibase
  10.1088/1475-7516/2016/09/047} {\bibfield  {journal} {\bibinfo  {journal}
  {\jcap}\ }\textbf {\bibinfo {volume} {9}},\ \bibinfo {eid} {047} (\bibinfo
  {year} {2016})},\ \Eprint {http://arxiv.org/abs/1606.04898} {arXiv:1606.04898
  [astro-ph.HE]} \BibitemShut {NoStop}%
\bibitem [{\citenamefont {Baltz}\ and\ \citenamefont
  {Wai}(2004)}]{BaltzEtAl2004}%
  \BibitemOpen
  \bibfield  {author} {\bibinfo {author} {\bibfnamefont {E.~A.}\ \bibnamefont
  {Baltz}}\ and\ \bibinfo {author} {\bibfnamefont {L.}~\bibnamefont {Wai}},\
  }\href {\doibase 10.1103/PhysRevD.70.023512} {\bibfield  {journal} {\bibinfo
  {journal} {\prd}\ }\textbf {\bibinfo {volume} {70}},\ \bibinfo {eid} {023512}
  (\bibinfo {year} {2004})},\ \Eprint {http://arxiv.org/abs/astro-ph/0403528}
  {astro-ph/0403528} \BibitemShut {NoStop}%
\bibitem [{\citenamefont {{Blanchet}}\ and\ \citenamefont {{Lavalle}}(2012)}]{FermiLATCollab2015}%
  \BibitemOpen
  \bibfield  {author} {\bibinfo {author} {\bibnamefont {{The Fermi-LAT
  Collaboration}}},\ }\href {\doibase 10.1088/0067-0049/218/2/23}
  {\bibfield  {journal} {\bibinfo  {journal} {\apjs}\ }\textbf {\bibinfo
  {volume} {218}},\ \bibinfo {eid} {23} (\bibinfo {year}
  {2015}{\natexlab{a}})},\ \Eprint {http://arxiv.org/abs/1501.02003}
  {arXiv:1501.02003 [astro-ph.HE]} \BibitemShut {NoStop}%
\bibitem [{\citenamefont {{The Fermi-LAT collaboration}}(2019)}]{FermiLAT2019}%
  \BibitemOpen
  \bibfield  {author} {\bibinfo {author} {\bibnamefont {{The Fermi-LAT
  collaboration}}},\ }\href
  {https://ui.adsabs.harvard.edu/abs/2019arXiv190210045T} {\bibfield  {journal}
  {\bibinfo  {journal} {arXiv e-prints}\ } (\bibinfo {year} {2019})},\ \Eprint
  {http://arxiv.org/abs/1902.10045} {arXiv:1902.10045 [astro-ph.HE]}
  \BibitemShut {NoStop}%
\bibitem [{\citenamefont {{Belikov}}\ \emph {et~al.}(2012)\citenamefont
  {{Belikov}}, \citenamefont {{Buckley}},\ and\ \citenamefont
  {{Hooper}}}]{BelikovEtAl2012}%
  \BibitemOpen
  \bibfield  {author} {\bibinfo {author} {\bibfnamefont {A.~V.}\ \bibnamefont
  {{Belikov}}}, \bibinfo {author} {\bibfnamefont {M.~R.}\ \bibnamefont
  {{Buckley}}}, \ and\ \bibinfo {author} {\bibfnamefont {D.}~\bibnamefont
  {{Hooper}}},\ }\href {\doibase 10.1103/PhysRevD.86.043504} {\bibfield
  {journal} {\bibinfo  {journal} {\prd}\ }\textbf {\bibinfo {volume} {86}},\
  \bibinfo {eid} {043504} (\bibinfo {year} {2012})},\ \Eprint
  {http://arxiv.org/abs/1111.2613} {arXiv:1111.2613 [hep-ph]} \BibitemShut
  {NoStop}%
\bibitem [{\citenamefont {Bertoni}\ \emph {et~al.}(2015)\citenamefont
  {Bertoni}, \citenamefont {Hooper},\ and\ \citenamefont
  {Linden}}]{BertoniEtAl2015}%
  \BibitemOpen
  \bibfield  {author} {\bibinfo {author} {\bibfnamefont {B.}~\bibnamefont
  {Bertoni}}, \bibinfo {author} {\bibfnamefont {D.}~\bibnamefont {Hooper}}, \
  and\ \bibinfo {author} {\bibfnamefont {T.}~\bibnamefont {Linden}},\ }\href
  {\doibase 10.1088/1475-7516/2015/12/035} {\bibfield  {journal} {\bibinfo
  {journal} {\jcap}\ }\textbf {\bibinfo {volume} {12}},\ \bibinfo {eid} {035}
  (\bibinfo {year} {2015})},\ \Eprint {http://arxiv.org/abs/1504.02087}
  {arXiv:1504.02087 [astro-ph.HE]} \BibitemShut {NoStop}%
\bibitem [{\citenamefont {Schoonenberg}\ \emph {et~al.}(2016)\citenamefont
  {Schoonenberg}, \citenamefont {Gaskins}, \citenamefont {Bertone},\ and\
  \citenamefont {Diemand}}]{SchoonenbergEtAl2016}%
  \BibitemOpen
  \bibfield  {author} {\bibinfo {author} {\bibfnamefont {D.}~\bibnamefont
  {Schoonenberg}}, \bibinfo {author} {\bibfnamefont {J.}~\bibnamefont
  {Gaskins}}, \bibinfo {author} {\bibfnamefont {G.}~\bibnamefont {Bertone}}, \
  and\ \bibinfo {author} {\bibfnamefont {J.}~\bibnamefont {Diemand}},\ }\href
  {\doibase 10.1088/1475-7516/2016/05/028} {\bibfield  {journal} {\bibinfo
  {journal} {\jcap}\ }\textbf {\bibinfo {volume} {2016}},\ \bibinfo {eid} {028}
  (\bibinfo {year} {2016})},\ \Eprint {http://arxiv.org/abs/1601.06781}
  {arXiv:1601.06781 [astro-ph.HE]} \BibitemShut {NoStop}%
\bibitem [{\citenamefont {Mirabal}\ \emph {et~al.}(2016)\citenamefont
  {Mirabal}, \citenamefont {Charles}, \citenamefont {Ferrara}, \citenamefont
  {Gonthier}, \citenamefont {Harding}, \citenamefont {S{\'a}nchez-Conde},\ and\
  \citenamefont {Thompson}}]{MirabalEtAl2016}%
  \BibitemOpen
  \bibfield  {author} {\bibinfo {author} {\bibfnamefont {N.}~\bibnamefont
  {Mirabal}}, \bibinfo {author} {\bibfnamefont {E.}~\bibnamefont {Charles}},
  \bibinfo {author} {\bibfnamefont {E.~C.}\ \bibnamefont {Ferrara}}, \bibinfo
  {author} {\bibfnamefont {P.~L.}\ \bibnamefont {Gonthier}}, \bibinfo {author}
  {\bibfnamefont {A.~K.}\ \bibnamefont {Harding}}, \bibinfo {author}
  {\bibfnamefont {M.~A.}\ \bibnamefont {S{\'a}nchez-Conde}}, \ and\ \bibinfo
  {author} {\bibfnamefont {D.~J.}\ \bibnamefont {Thompson}},\ }\href {\doibase
  10.3847/0004-637X/825/1/69} {\bibfield  {journal} {\bibinfo  {journal}
  {\apj}\ }\textbf {\bibinfo {volume} {825}},\ \bibinfo {eid} {69} (\bibinfo
  {year} {2016})},\ \Eprint {http://arxiv.org/abs/1605.00711} {arXiv:1605.00711
  [astro-ph.HE]} \BibitemShut {NoStop}%
\bibitem [{\citenamefont {Hooper}\ and\ \citenamefont
  {Witte}(2017)}]{HooperEtAl2017}%
  \BibitemOpen
  \bibfield  {author} {\bibinfo {author} {\bibfnamefont {D.}~\bibnamefont
  {Hooper}}\ and\ \bibinfo {author} {\bibfnamefont {S.~J.}\ \bibnamefont
  {Witte}},\ }\href {\doibase 10.1088/1475-7516/2017/04/018} {\bibfield
  {journal} {\bibinfo  {journal} {\jcap}\ }\textbf {\bibinfo {volume} {4}},\
  \bibinfo {eid} {018} (\bibinfo {year} {2017})},\ \Eprint
  {http://arxiv.org/abs/1610.07587} {arXiv:1610.07587 [astro-ph.HE]}
  \BibitemShut {NoStop}%
\bibitem [{\citenamefont {Calore}\ \emph {et~al.}(2017)\citenamefont {Calore},
  \citenamefont {De~Romeri}, \citenamefont {Di~Mauro}, \citenamefont {Donato},\
  and\ \citenamefont {Marinacci}}]{CaloreEtAl2017}%
  \BibitemOpen
  \bibfield  {author} {\bibinfo {author} {\bibfnamefont {F.}~\bibnamefont
  {Calore}}, \bibinfo {author} {\bibfnamefont {V.}~\bibnamefont {De~Romeri}},
  \bibinfo {author} {\bibfnamefont {M.}~\bibnamefont {Di~Mauro}}, \bibinfo
  {author} {\bibfnamefont {F.}~\bibnamefont {Donato}}, \ and\ \bibinfo {author}
  {\bibfnamefont {F.}~\bibnamefont {Marinacci}},\ }\href {\doibase
  10.1103/PhysRevD.96.063009} {\bibfield  {journal} {\bibinfo  {journal}
  {\prd}\ }\textbf {\bibinfo {volume} {96}},\ \bibinfo {eid} {063009} (\bibinfo
  {year} {2017})},\ \Eprint {http://arxiv.org/abs/1611.03503} {arXiv:1611.03503
  [astro-ph.HE]} \BibitemShut {NoStop}%
\bibitem [{\citenamefont {Calore}\ \emph {et~al.}(2019)\citenamefont {Calore},
  \citenamefont {H{\"u}tten},\ and\ \citenamefont {Stref}}]{CaloreEtAl2019}%
  \BibitemOpen
  \bibfield  {author} {\bibinfo {author} {\bibfnamefont {F.}~\bibnamefont
  {Calore}}, \bibinfo {author} {\bibfnamefont {M.}~\bibnamefont {H{\"u}tten}},
  \ and\ \bibinfo {author} {\bibfnamefont {M.}~\bibnamefont {Stref}},\ }\href
  {\doibase 10.3390/galaxies7040090} {\bibfield  {journal} {\bibinfo  {journal}
  {Galaxies}\ }\textbf {\bibinfo {volume} {7}},\ \bibinfo {pages} {90}
  (\bibinfo {year} {2019})},\ \Eprint {http://arxiv.org/abs/1910.13722}
  {arXiv:1910.13722 [astro-ph.HE]} \BibitemShut {NoStop}%
\bibitem [{\citenamefont {Glawion}\ \emph {et~al.}(2019)\citenamefont
  {Glawion}, \citenamefont {Malyshev}, \citenamefont {Moulin}, \citenamefont
  {Oakes}, \citenamefont {Rinchiuso},\ and\ \citenamefont
  {Viana}}]{GlawionEtAl2019}%
  \BibitemOpen
  \bibfield  {author} {\bibinfo {author} {\bibfnamefont {D.}~\bibnamefont
  {Glawion}}, \bibinfo {author} {\bibfnamefont {D.}~\bibnamefont {Malyshev}},
  \bibinfo {author} {\bibfnamefont {E.}~\bibnamefont {Moulin}}, \bibinfo
  {author} {\bibfnamefont {L.}~\bibnamefont {Oakes}}, \bibinfo {author}
  {\bibfnamefont {L.}~\bibnamefont {Rinchiuso}}, \ and\ \bibinfo {author}
  {\bibfnamefont {A.}~\bibnamefont {Viana}},\ }\href
  {https://ui.adsabs.harvard.edu/abs/2019arXiv190901072G} {\bibfield  {journal}
  {\bibinfo  {journal} {arXiv e-prints}\ } (\bibinfo {year} {2019})},\ \Eprint
  {http://arxiv.org/abs/1909.01072} {arXiv:1909.01072 [astro-ph.HE]}
  \BibitemShut {NoStop}%
\bibitem [{\citenamefont {Coronado-Bl{\'a}zquez}\ \emph
  {et~al.}(2019{\natexlab{a}})\citenamefont {Coronado-Bl{\'a}zquez},
  \citenamefont {S{\'a}nchez-Conde}, \citenamefont {Dom{\'\i}nguez},
  \citenamefont {Aguirre-Santaella}, \citenamefont {Di~Mauro}, \citenamefont
  {Mirabal}, \citenamefont {Nieto},\ and\ \citenamefont
  {Charles}}]{CoronadoBlazquezEtAl2019}%
  \BibitemOpen
  \bibfield  {author} {\bibinfo {author} {\bibfnamefont {J.}~\bibnamefont
  {Coronado-Bl{\'a}zquez}}, \bibinfo {author} {\bibfnamefont {M.~A.}\
  \bibnamefont {S{\'a}nchez-Conde}}, \bibinfo {author} {\bibfnamefont
  {A.}~\bibnamefont {Dom{\'\i}nguez}}, \bibinfo {author} {\bibfnamefont
  {A.}~\bibnamefont {Aguirre-Santaella}}, \bibinfo {author} {\bibfnamefont
  {M.}~\bibnamefont {Di~Mauro}}, \bibinfo {author} {\bibfnamefont
  {N.}~\bibnamefont {Mirabal}}, \bibinfo {author} {\bibfnamefont
  {D.}~\bibnamefont {Nieto}}, \ and\ \bibinfo {author} {\bibfnamefont
  {E.}~\bibnamefont {Charles}},\ }\href {\doibase
  10.1088/1475-7516/2019/07/020} {\bibfield  {journal} {\bibinfo  {journal}
  {\jcap}\ }\textbf {\bibinfo {volume} {2019}},\ \bibinfo {eid} {020} (\bibinfo
  {year} {2019}{\natexlab{a}})},\ \Eprint {http://arxiv.org/abs/1906.11896}
  {arXiv:1906.11896 [astro-ph.HE]} \BibitemShut {NoStop}%
\bibitem [{\citenamefont {Coronado-Bl{\'a}zquez}\ \emph
  {et~al.}(2019{\natexlab{b}})\citenamefont {Coronado-Bl{\'a}zquez},
  \citenamefont {S{\'a}nchez-Conde}, \citenamefont {Di~Mauro}, \citenamefont
  {Aguirre-Santaella}, \citenamefont {Ciuc{\u{a}}}, \citenamefont
  {Dom{\'\i}nguez}, \citenamefont {Kawata},\ and\ \citenamefont
  {Mirabal}}]{CoronadoBlazquezEtAl2019a}%
  \BibitemOpen
  \bibfield  {author} {\bibinfo {author} {\bibfnamefont {J.}~\bibnamefont
  {Coronado-Bl{\'a}zquez}}, \bibinfo {author} {\bibfnamefont {M.~A.}\
  \bibnamefont {S{\'a}nchez-Conde}}, \bibinfo {author} {\bibfnamefont
  {M.}~\bibnamefont {Di~Mauro}}, \bibinfo {author} {\bibfnamefont
  {A.}~\bibnamefont {Aguirre-Santaella}}, \bibinfo {author} {\bibfnamefont
  {I.}~\bibnamefont {Ciuc{\u{a}}}}, \bibinfo {author} {\bibfnamefont
  {A.}~\bibnamefont {Dom{\'\i}nguez}}, \bibinfo {author} {\bibfnamefont
  {D.}~\bibnamefont {Kawata}}, \ and\ \bibinfo {author} {\bibfnamefont
  {N.}~\bibnamefont {Mirabal}},\ }\href {\doibase
  10.1088/1475-7516/2019/11/045} {\bibfield  {journal} {\bibinfo  {journal}
  {\jcap}\ }\textbf {\bibinfo {volume} {2019}},\ \bibinfo {eid} {045} (\bibinfo
  {year} {2019}{\natexlab{b}})},\ \Eprint {http://arxiv.org/abs/1910.14429}
  {arXiv:1910.14429 [astro-ph.HE]} \BibitemShut {NoStop}%
\bibitem [{\citenamefont {Hiroshima}\ \emph {et~al.}(2018)\citenamefont
  {Hiroshima}, \citenamefont {Ando},\ and\ \citenamefont
  {Ishiyama}}]{HiroshimaEtAl2018}%
  \BibitemOpen
  \bibfield  {author} {\bibinfo {author} {\bibfnamefont {N.}~\bibnamefont
  {Hiroshima}}, \bibinfo {author} {\bibfnamefont {S.}~\bibnamefont {Ando}}, \
  and\ \bibinfo {author} {\bibfnamefont {T.}~\bibnamefont {Ishiyama}},\ }\href
  {\doibase 10.1103/PhysRevD.97.123002} {\bibfield  {journal} {\bibinfo
  {journal} {\prd}\ }\textbf {\bibinfo {volume} {97}},\ \bibinfo {eid} {123002}
  (\bibinfo {year} {2018})},\ \Eprint {http://arxiv.org/abs/1803.07691}
  {arXiv:1803.07691} \BibitemShut {NoStop}%
\bibitem [{\citenamefont {{Bartels}}\ and\ \citenamefont
  {{Ando}}(2015)}]{BartelsEtAl2015}%
  \BibitemOpen
  \bibfield  {author} {\bibinfo {author} {\bibfnamefont {R.}~\bibnamefont
  {{Bartels}}}\ and\ \bibinfo {author} {\bibfnamefont {S.}~\bibnamefont
  {{Ando}}},\ }\href {\doibase 10.1103/PhysRevD.92.123508} {\bibfield
  {journal} {\bibinfo  {journal} {\prd}\ }\textbf {\bibinfo {volume} {92}},\
  \bibinfo {eid} {123508} (\bibinfo {year} {2015})},\ \Eprint
  {http://arxiv.org/abs/1507.08656} {arXiv:1507.08656} \BibitemShut {NoStop}%
\bibitem [{\citenamefont {{Zavala}}\ and\ \citenamefont
  {{Afshordi}}(2014)}]{ZavalaEtAl2014a}%
  \BibitemOpen
  \bibfield  {author} {\bibinfo {author} {\bibfnamefont {J.}~\bibnamefont
  {{Zavala}}}\ and\ \bibinfo {author} {\bibfnamefont {N.}~\bibnamefont
  {{Afshordi}}},\ }\href {\doibase 10.1093/mnras/stu506} {\bibfield  {journal}
  {\bibinfo  {journal} {\mnras}\ }\textbf {\bibinfo {volume} {441}},\ \bibinfo
  {pages} {1329} (\bibinfo {year} {2014})},\ \Eprint
  {http://arxiv.org/abs/1311.3296} {arXiv:1311.3296} \BibitemShut {NoStop}%
\bibitem [{\citenamefont {Benson}(2012)}]{Benson2012}%
  \BibitemOpen
  \bibfield  {author} {\bibinfo {author} {\bibfnamefont {A.~J.}\ \bibnamefont
  {Benson}},\ }\href {\doibase 10.1016/j.newast.2011.07.004} {\bibfield
  {journal} {\bibinfo  {journal} {\na}\ }\textbf {\bibinfo {volume} {17}},\
  \bibinfo {pages} {175} (\bibinfo {year} {2012})},\ \Eprint
  {http://arxiv.org/abs/1008.1786} {arXiv:1008.1786} \BibitemShut {NoStop}%
\bibitem [{\citenamefont {{van den Bosch}}\ \emph {et~al.}(2005)\citenamefont
  {{van den Bosch}}, \citenamefont {{Tormen}},\ and\ \citenamefont
  {{Giocoli}}}]{vandenBoschEtAl2005a}%
  \BibitemOpen
  \bibfield  {author} {\bibinfo {author} {\bibfnamefont {F.~C.}\ \bibnamefont
  {{van den Bosch}}}, \bibinfo {author} {\bibfnamefont {G.}~\bibnamefont
  {{Tormen}}}, \ and\ \bibinfo {author} {\bibfnamefont {C.}~\bibnamefont
  {{Giocoli}}},\ }\href {\doibase 10.1111/j.1365-2966.2005.08964.x} {\bibfield
  {journal} {\bibinfo  {journal} {\mnras}\ }\textbf {\bibinfo {volume} {359}},\
  \bibinfo {pages} {1029} (\bibinfo {year} {2005})},\ \Eprint
  {http://arxiv.org/abs/astro-ph/0409201} {astro-ph/0409201} \BibitemShut
  {NoStop}%
\bibitem [{\citenamefont {Zavala}\ and\ \citenamefont
  {Frenk}(2019)}]{ZavalaEtAl2019a}%
  \BibitemOpen
  \bibfield  {author} {\bibinfo {author} {\bibfnamefont {J.}~\bibnamefont
  {Zavala}}\ and\ \bibinfo {author} {\bibfnamefont {C.~S.}\ \bibnamefont
  {Frenk}},\ }\href {\doibase 10.3390/galaxies7040081} {\bibfield  {journal}
  {\bibinfo  {journal} {Galaxies}\ }\textbf {\bibinfo {volume} {7}},\ \bibinfo
  {pages} {81} (\bibinfo {year} {2019})},\ \Eprint
  {http://arxiv.org/abs/1907.11775} {arXiv:1907.11775 [astro-ph.CO]}
  \BibitemShut {NoStop}%
\bibitem [{\citenamefont {{McMillan}}(2017)}]{McMillan2017}%
  \BibitemOpen
  \bibfield  {author} {\bibinfo {author} {\bibfnamefont {P.~J.}\ \bibnamefont
  {{McMillan}}},\ }\href {\doibase 10.1093/mnras/stw2759} {\bibfield  {journal}
  {\bibinfo  {journal} {\mnras}\ }\textbf {\bibinfo {volume} {465}},\ \bibinfo
  {pages} {76} (\bibinfo {year} {2017})},\ \Eprint
  {http://arxiv.org/abs/1608.00971} {arXiv:1608.00971} \BibitemShut {NoStop}%
\bibitem [{\citenamefont {H{\"u}tten}\ \emph
  {et~al.}(2019{\natexlab{a}})\citenamefont {H{\"u}tten}, \citenamefont
  {Stref}, \citenamefont {Combet}, \citenamefont {Lavalle},\ and\ \citenamefont
  {Maurin}}]{HuettenEtAl2019}%
  \BibitemOpen
  \bibfield  {author} {\bibinfo {author} {\bibfnamefont {M.}~\bibnamefont
  {H{\"u}tten}}, \bibinfo {author} {\bibfnamefont {M.}~\bibnamefont {Stref}},
  \bibinfo {author} {\bibfnamefont {C.}~\bibnamefont {Combet}}, \bibinfo
  {author} {\bibfnamefont {J.}~\bibnamefont {Lavalle}}, \ and\ \bibinfo
  {author} {\bibfnamefont {D.}~\bibnamefont {Maurin}},\ }\href {\doibase
  10.3390/galaxies7020060} {\bibfield  {journal} {\bibinfo  {journal}
  {Galaxies}\ }\textbf {\bibinfo {volume} {7}},\ \bibinfo {pages} {60}
  (\bibinfo {year} {2019}{\natexlab{a}})},\ \Eprint
  {http://arxiv.org/abs/1904.10935} {arXiv:1904.10935 [astro-ph.HE]}
  \BibitemShut {NoStop}%
\bibitem [{\citenamefont {{Charbonnier}}\ \emph {et~al.}(2012)\citenamefont
  {{Charbonnier}}, \citenamefont {{Combet}},\ and\ \citenamefont
  {{Maurin}}}]{CharbonnierEtAl2012}%
  \BibitemOpen
  \bibfield  {author} {\bibinfo {author} {\bibfnamefont {A.}~\bibnamefont
  {{Charbonnier}}}, \bibinfo {author} {\bibfnamefont {C.}~\bibnamefont
  {{Combet}}}, \ and\ \bibinfo {author} {\bibfnamefont {D.}~\bibnamefont
  {{Maurin}}},\ }\href {\doibase 10.1016/j.cpc.2011.10.017} {\bibfield
  {journal} {\bibinfo  {journal} {Computer Physics Communications}\ }\textbf
  {\bibinfo {volume} {183}},\ \bibinfo {pages} {656} (\bibinfo {year}
  {2012})},\ \Eprint {http://arxiv.org/abs/1201.4728} {arXiv:1201.4728
  [astro-ph.HE]} \BibitemShut {NoStop}%
\bibitem [{\citenamefont {H{\"u}tten}\ \emph
  {et~al.}(2019{\natexlab{b}})\citenamefont {H{\"u}tten}, \citenamefont
  {Combet},\ and\ \citenamefont {Maurin}}]{HuettenEtAl2019a}%
  \BibitemOpen
  \bibfield  {author} {\bibinfo {author} {\bibfnamefont {M.}~\bibnamefont
  {H{\"u}tten}}, \bibinfo {author} {\bibfnamefont {C.}~\bibnamefont {Combet}},
  \ and\ \bibinfo {author} {\bibfnamefont {D.}~\bibnamefont {Maurin}},\ }\href
  {\doibase 10.1016/j.cpc.2018.10.001} {\bibfield  {journal} {\bibinfo
  {journal} {Computer Physics Communications}\ }\textbf {\bibinfo {volume}
  {235}},\ \bibinfo {pages} {336} (\bibinfo {year} {2019}{\natexlab{b}})},\
  \Eprint {http://arxiv.org/abs/1806.08639} {arXiv:1806.08639} \BibitemShut
  {NoStop}%
\bibitem [{\citenamefont {Kelley}\ \emph {et~al.}(2019)\citenamefont {Kelley},
  \citenamefont {Bullock}, \citenamefont {Garrison-Kimmel}, \citenamefont
  {Boylan-Kolchin}, \citenamefont {Pawlowski},\ and\ \citenamefont
  {Graus}}]{KelleyEtAl2019}%
  \BibitemOpen
  \bibfield  {author} {\bibinfo {author} {\bibfnamefont {T.}~\bibnamefont
  {Kelley}}, \bibinfo {author} {\bibfnamefont {J.~S.}\ \bibnamefont {Bullock}},
  \bibinfo {author} {\bibfnamefont {S.}~\bibnamefont {Garrison-Kimmel}},
  \bibinfo {author} {\bibfnamefont {M.}~\bibnamefont {Boylan-Kolchin}},
  \bibinfo {author} {\bibfnamefont {M.~S.}\ \bibnamefont {Pawlowski}}, \ and\
  \bibinfo {author} {\bibfnamefont {A.~S.}\ \bibnamefont {Graus}},\ }\href
  {\doibase 10.1093/mnras/stz1553} {\bibfield  {journal} {\bibinfo  {journal}
  {\mnras}\ }\textbf {\bibinfo {volume} {487}},\ \bibinfo {pages} {4409}
  (\bibinfo {year} {2019})},\ \Eprint {http://arxiv.org/abs/1811.12413}
  {arXiv:1811.12413} \BibitemShut {NoStop}%
\bibitem [{\citenamefont {{Abdo}}\ \emph {et~al.}(2010)}]{AbdoEtAl2010d}%
  \BibitemOpen
  \bibfield  {author} {\bibinfo {author} {\bibfnamefont {{The Fermi-LAT Collaboration}}},\ }\href {\doibase 10.1088/1475-7516/2010/04/014} {\bibfield
  {journal} {\bibinfo  {journal} {\jcap}\ }\textbf {\bibinfo {volume} {4}},\
  \bibinfo {eid} {014} (\bibinfo {year} {2010})},\ \Eprint
  {http://arxiv.org/abs/1002.4415} {arXiv:1002.4415} \BibitemShut {NoStop}%
\bibitem [{\citenamefont {Cirelli}\ \emph {et~al.}(2010)\citenamefont
  {Cirelli}, \citenamefont {Panci},\ and\ \citenamefont
  {Serpico}}]{CirelliEtAl2010}%
  \BibitemOpen
  \bibfield  {author} {\bibinfo {author} {\bibfnamefont {M.}~\bibnamefont
  {Cirelli}}, \bibinfo {author} {\bibfnamefont {P.}~\bibnamefont {Panci}}, \
  and\ \bibinfo {author} {\bibfnamefont {P.~D.}\ \bibnamefont {Serpico}},\
  }\href {\doibase 10.1016/j.nuclphysb.2010.07.010} {\bibfield  {journal}
  {\bibinfo  {journal} {Nucl.Phys.}\ }\textbf {\bibinfo {volume} {B840}},\
  \bibinfo {pages} {284} (\bibinfo {year} {2010})},\ \Eprint
  {http://arxiv.org/abs/0912.0663} {arXiv:0912.0663 [astro-ph.CO]} \BibitemShut
  {NoStop}%
\bibitem [{\citenamefont {{Blanchet}}\ and\ \citenamefont
  {{Lavalle}}(2012)}]{BlanchetEtAl2012}%
  \BibitemOpen
  \bibfield  {author} {\bibinfo {author} {\bibfnamefont {S.}~\bibnamefont
  {{Blanchet}}}\ and\ \bibinfo {author} {\bibfnamefont {J.}~\bibnamefont
  {{Lavalle}}},\ }\href {\doibase 10.1088/1475-7516/2012/11/021} {\bibfield
  {journal} {\bibinfo  {journal} {\jcap}\ }\textbf {\bibinfo {volume} {11}},\
  \bibinfo {eid} {021} (\bibinfo {year} {2012})},\ \Eprint
  {http://arxiv.org/abs/1207.2476} {arXiv:1207.2476 [astro-ph.HE]} \BibitemShut
  {NoStop}%
\bibitem [{\citenamefont {{Blanchet}}\ and\ \citenamefont {{Lavalle}}(2012)}]{FermiLATEtAl2012}%
  \BibitemOpen
  \bibfield  {author} {\bibinfo {author} {\bibnamefont {{The Fermi-LAT
  Collaboration}}},\ }\href {\doibase 10.1088/0004-637X/761/2/91}
  {\bibfield  {journal} {\bibinfo  {journal} {\apj}\ }\textbf {\bibinfo
  {volume} {761}},\ \bibinfo {eid} {91} (\bibinfo {year} {2012})},\ \Eprint
  {http://arxiv.org/abs/1205.6474} {arXiv:1205.6474 [astro-ph.CO]} \BibitemShut
  {NoStop}%
\bibitem [{\citenamefont {{Essig}}\ \emph {et~al.}(2013)\citenamefont
  {{Essig}}, \citenamefont {{Kuflik}}, \citenamefont {{McDermott}},
  \citenamefont {{Volansky}},\ and\ \citenamefont {{Zurek}}}]{EssigEtAl2013a}%
  \BibitemOpen
  \bibfield  {author} {\bibinfo {author} {\bibfnamefont {R.}~\bibnamefont
  {{Essig}}}, \bibinfo {author} {\bibfnamefont {E.}~\bibnamefont {{Kuflik}}},
  \bibinfo {author} {\bibfnamefont {S.~D.}\ \bibnamefont {{McDermott}}},
  \bibinfo {author} {\bibfnamefont {T.}~\bibnamefont {{Volansky}}}, \ and\
  \bibinfo {author} {\bibfnamefont {K.~M.}\ \bibnamefont {{Zurek}}},\ }\href
  {\doibase 10.1007/JHEP11(2013)193} {\bibfield  {journal} {\bibinfo  {journal}
  {Journal of High Energy Physics}\ }\textbf {\bibinfo {volume} {11}},\
  \bibinfo {eid} {193} (\bibinfo {year} {2013})},\ \Eprint
  {http://arxiv.org/abs/1309.4091} {arXiv:1309.4091 [hep-ph]} \BibitemShut
  {NoStop}%
\bibitem [{\citenamefont {{Cirelli}}\ \emph {et~al.}(2015)\citenamefont
  {{Cirelli}}, \citenamefont {{Hambye}}, \citenamefont {{Panci}}, \citenamefont
  {{Sala}},\ and\ \citenamefont {{Taoso}}}]{CirelliEtAl2015}%
  \BibitemOpen
  \bibfield  {author} {\bibinfo {author} {\bibfnamefont {M.}~\bibnamefont
  {{Cirelli}}}, \bibinfo {author} {\bibfnamefont {T.}~\bibnamefont {{Hambye}}},
  \bibinfo {author} {\bibfnamefont {P.}~\bibnamefont {{Panci}}}, \bibinfo
  {author} {\bibfnamefont {F.}~\bibnamefont {{Sala}}}, \ and\ \bibinfo {author}
  {\bibfnamefont {M.}~\bibnamefont {{Taoso}}},\ }\href {\doibase
  10.1088/1475-7516/2015/10/026} {\bibfield  {journal} {\bibinfo  {journal}
  {\jcap}\ }\textbf {\bibinfo {volume} {10}},\ \bibinfo {eid} {026} (\bibinfo
  {year} {2015})},\ \Eprint {http://arxiv.org/abs/1507.05519} {arXiv:1507.05519
  [hep-ph]} \BibitemShut {NoStop}%
\bibitem [{\citenamefont {{Fornasa}}\ and\ \citenamefont
  {{S{\'a}nchez-Conde}}(2015)}]{FornasaEtAl2015}%
  \BibitemOpen
  \bibfield  {author} {\bibinfo {author} {\bibfnamefont {M.}~\bibnamefont
  {{Fornasa}}}\ and\ \bibinfo {author} {\bibfnamefont {M.~A.}\ \bibnamefont
  {{S{\'a}nchez-Conde}}},\ }\href {\doibase 10.1016/j.physrep.2015.09.002}
  {\bibfield  {journal} {\bibinfo  {journal} {\physrep}\ }\textbf {\bibinfo
  {volume} {598}},\ \bibinfo {pages} {1} (\bibinfo {year} {2015})},\ \Eprint
  {http://arxiv.org/abs/1502.02866} {arXiv:1502.02866} \BibitemShut {NoStop}%
\bibitem [{\citenamefont {Chang}\ \emph {et~al.}(2018)\citenamefont {Chang},
  \citenamefont {Lisanti},\ and\ \citenamefont
  {Mishra-Sharma}}]{ChangEtAl2018}%
  \BibitemOpen
  \bibfield  {author} {\bibinfo {author} {\bibfnamefont {L.~J.}\ \bibnamefont
  {Chang}}, \bibinfo {author} {\bibfnamefont {M.}~\bibnamefont {Lisanti}}, \
  and\ \bibinfo {author} {\bibfnamefont {S.}~\bibnamefont {Mishra-Sharma}},\
  }\href {\doibase 10.1103/PhysRevD.98.123004} {\bibfield  {journal} {\bibinfo
  {journal} {\prd}\ }\textbf {\bibinfo {volume} {98}},\ \bibinfo {eid} {123004}
  (\bibinfo {year} {2018})},\ \Eprint {http://arxiv.org/abs/1804.04132}
  {arXiv:1804.04132 [astro-ph.CO]} \BibitemShut {NoStop}%
\bibitem [{\citenamefont {{Zhao}}(1996)}]{Zhao1996}%
  \BibitemOpen
  \bibfield  {author} {\bibinfo {author} {\bibfnamefont {H.}~\bibnamefont
  {{Zhao}}},\ }\href {https://ui.adsabs.harvard.edu/abs/1996MNRAS.278..488Z}
  {\bibfield  {journal} {\bibinfo  {journal} {\mnras}\ }\textbf {\bibinfo
  {volume} {278}},\ \bibinfo {pages} {488} (\bibinfo {year} {1996})},\ \Eprint
  {http://arxiv.org/abs/astro-ph/9509122} {astro-ph/9509122} \BibitemShut
  {NoStop}%
\bibitem [{\citenamefont {{Navarro}}\ \emph {et~al.}(1996)\citenamefont
  {{Navarro}}, \citenamefont {{Frenk}},\ and\ \citenamefont
  {{White}}}]{NavarroEtAl1996a}%
  \BibitemOpen
  \bibfield  {author} {\bibinfo {author} {\bibfnamefont {J.~F.}\ \bibnamefont
  {{Navarro}}}, \bibinfo {author} {\bibfnamefont {C.~S.}\ \bibnamefont
  {{Frenk}}}, \ and\ \bibinfo {author} {\bibfnamefont {S.~D.~M.}\ \bibnamefont
  {{White}}},\ }\href {\doibase 10.1086/177173} {\bibfield  {journal} {\bibinfo
   {journal} {\apj}\ }\textbf {\bibinfo {volume} {462}},\ \bibinfo {pages}
  {563} (\bibinfo {year} {1996})},\ \Eprint
  {http://arxiv.org/abs/astro-ph/9508025} {astro-ph/9508025} \BibitemShut
  {NoStop}%
\bibitem [{\citenamefont {{Bullock}}\ \emph {et~al.}(2001)\citenamefont
  {{Bullock}}, \citenamefont {{Kolatt}}, \citenamefont {{Sigad}}, \citenamefont
  {{Somerville}}, \citenamefont {{Kravtsov}}, \citenamefont {{Klypin}},
  \citenamefont {{Primack}},\ and\ \citenamefont {{Dekel}}}]{BullockEtAl2001b}%
  \BibitemOpen
  \bibfield  {author} {\bibinfo {author} {\bibfnamefont {J.~S.}\ \bibnamefont
  {{Bullock}}}, \bibinfo {author} {\bibfnamefont {T.~S.}\ \bibnamefont
  {{Kolatt}}}, \bibinfo {author} {\bibfnamefont {Y.}~\bibnamefont {{Sigad}}},
  \bibinfo {author} {\bibfnamefont {R.~S.}\ \bibnamefont {{Somerville}}},
  \bibinfo {author} {\bibfnamefont {A.~V.}\ \bibnamefont {{Kravtsov}}},
  \bibinfo {author} {\bibfnamefont {A.~A.}\ \bibnamefont {{Klypin}}}, \bibinfo
  {author} {\bibfnamefont {J.~R.}\ \bibnamefont {{Primack}}}, \ and\ \bibinfo
  {author} {\bibfnamefont {A.}~\bibnamefont {{Dekel}}},\ }\href {\doibase
  10.1046/j.1365-8711.2001.04068.x} {\bibfield  {journal} {\bibinfo  {journal}
  {\mnras}\ }\textbf {\bibinfo {volume} {321}},\ \bibinfo {pages} {559}
  (\bibinfo {year} {2001})},\ \Eprint {http://arxiv.org/abs/astro-ph/9908159}
  {astro-ph/9908159} \BibitemShut {NoStop}%
\bibitem [{\citenamefont {{Diemand}}\ \emph {et~al.}(2008)\citenamefont
  {{Diemand}}, \citenamefont {{Kuhlen}}, \citenamefont {{Madau}}, \citenamefont
  {{Zemp}}, \citenamefont {{Moore}}, \citenamefont {{Potter}},\ and\
  \citenamefont {{Stadel}}}]{DiemandEtAl2008b}%
  \BibitemOpen
  \bibfield  {author} {\bibinfo {author} {\bibfnamefont {J.}~\bibnamefont
  {{Diemand}}}, \bibinfo {author} {\bibfnamefont {M.}~\bibnamefont {{Kuhlen}}},
  \bibinfo {author} {\bibfnamefont {P.}~\bibnamefont {{Madau}}}, \bibinfo
  {author} {\bibfnamefont {M.}~\bibnamefont {{Zemp}}}, \bibinfo {author}
  {\bibfnamefont {B.}~\bibnamefont {{Moore}}}, \bibinfo {author} {\bibfnamefont
  {D.}~\bibnamefont {{Potter}}}, \ and\ \bibinfo {author} {\bibfnamefont
  {J.}~\bibnamefont {{Stadel}}},\ }\href {\doibase 10.1038/nature07153}
  {\bibfield  {journal} {\bibinfo  {journal} {\nat}\ }\textbf {\bibinfo
  {volume} {454}},\ \bibinfo {pages} {735} (\bibinfo {year} {2008})},\ \Eprint
  {http://arxiv.org/abs/0805.1244} {arXiv:0805.1244} \BibitemShut {NoStop}%
\bibitem [{\citenamefont {{Springel}}\ \emph {et~al.}(2008)\citenamefont
  {{Springel}}, \citenamefont {{Wang}}, \citenamefont {{Vogelsberger}},
  \citenamefont {{Ludlow}}, \citenamefont {{Jenkins}}, \citenamefont {{Helmi}},
  \citenamefont {{Navarro}}, \citenamefont {{Frenk}},\ and\ \citenamefont
  {{White}}}]{SpringelEtAl2008}%
  \BibitemOpen
  \bibfield  {author} {\bibinfo {author} {\bibfnamefont {V.}~\bibnamefont
  {{Springel}}}, \bibinfo {author} {\bibfnamefont {J.}~\bibnamefont {{Wang}}},
  \bibinfo {author} {\bibfnamefont {M.}~\bibnamefont {{Vogelsberger}}},
  \bibinfo {author} {\bibfnamefont {A.}~\bibnamefont {{Ludlow}}}, \bibinfo
  {author} {\bibfnamefont {A.}~\bibnamefont {{Jenkins}}}, \bibinfo {author}
  {\bibfnamefont {A.}~\bibnamefont {{Helmi}}}, \bibinfo {author} {\bibfnamefont
  {J.~F.}\ \bibnamefont {{Navarro}}}, \bibinfo {author} {\bibfnamefont {C.~S.}\
  \bibnamefont {{Frenk}}}, \ and\ \bibinfo {author} {\bibfnamefont {S.~D.~M.}\
  \bibnamefont {{White}}},\ }\href {\doibase 10.1111/j.1365-2966.2008.14066.x}
  {\bibfield  {journal} {\bibinfo  {journal} {\mnras}\ }\textbf {\bibinfo
  {volume} {391}},\ \bibinfo {pages} {1685} (\bibinfo {year} {2008})},\ \Eprint
  {http://arxiv.org/abs/0809.0898} {arXiv:0809.0898} \BibitemShut {NoStop}%
\bibitem [{\citenamefont {{Einasto}}(1965)}]{Einasto1965}%
  \BibitemOpen
  \bibfield  {author} {\bibinfo {author} {\bibfnamefont {J.}~\bibnamefont
  {{Einasto}}},\ }\href {https://ui.adsabs.harvard.edu/abs/1965TrAlm...5...87E}
  {\bibfield  {journal} {\bibinfo  {journal} {Trudy Astrofizicheskogo Instituta
  Alma-Ata}\ }\textbf {\bibinfo {volume} {5}},\ \bibinfo {pages} {87} (\bibinfo
  {year} {1965})}\BibitemShut {NoStop}%
\bibitem [{\citenamefont {{Navarro}}\ \emph {et~al.}(2004)\citenamefont
  {{Navarro}}, \citenamefont {{Hayashi}}, \citenamefont {{Power}},
  \citenamefont {{Jenkins}}, \citenamefont {{Frenk}}, \citenamefont {{White}},
  \citenamefont {{Springel}}, \citenamefont {{Stadel}},\ and\ \citenamefont
  {{Quinn}}}]{NavarroEtAl2004}%
  \BibitemOpen
  \bibfield  {author} {\bibinfo {author} {\bibfnamefont {J.~F.}\ \bibnamefont
  {{Navarro}}}, \bibinfo {author} {\bibfnamefont {E.}~\bibnamefont
  {{Hayashi}}}, \bibinfo {author} {\bibfnamefont {C.}~\bibnamefont {{Power}}},
  \bibinfo {author} {\bibfnamefont {A.~R.}\ \bibnamefont {{Jenkins}}}, \bibinfo
  {author} {\bibfnamefont {C.~S.}\ \bibnamefont {{Frenk}}}, \bibinfo {author}
  {\bibfnamefont {S.~D.~M.}\ \bibnamefont {{White}}}, \bibinfo {author}
  {\bibfnamefont {V.}~\bibnamefont {{Springel}}}, \bibinfo {author}
  {\bibfnamefont {J.}~\bibnamefont {{Stadel}}}, \ and\ \bibinfo {author}
  {\bibfnamefont {T.~R.}\ \bibnamefont {{Quinn}}},\ }\href {\doibase
  10.1111/j.1365-2966.2004.07586.x} {\bibfield  {journal} {\bibinfo  {journal}
  {\mnras}\ }\textbf {\bibinfo {volume} {349}},\ \bibinfo {pages} {1039}
  (\bibinfo {year} {2004})},\ \Eprint {http://arxiv.org/abs/astro-ph/0311231}
  {astro-ph/0311231} \BibitemShut {NoStop}%
\bibitem [{\citenamefont {{Navarro}}\ \emph {et~al.}(2010)\citenamefont
  {{Navarro}}, \citenamefont {{Ludlow}}, \citenamefont {{Springel}},
  \citenamefont {{Wang}}, \citenamefont {{Vogelsberger}}, \citenamefont
  {{White}}, \citenamefont {{Jenkins}}, \citenamefont {{Frenk}},\ and\
  \citenamefont {{Helmi}}}]{NavarroEtAl2010}%
  \BibitemOpen
  \bibfield  {author} {\bibinfo {author} {\bibfnamefont {J.~F.}\ \bibnamefont
  {{Navarro}}}, \bibinfo {author} {\bibfnamefont {A.}~\bibnamefont {{Ludlow}}},
  \bibinfo {author} {\bibfnamefont {V.}~\bibnamefont {{Springel}}}, \bibinfo
  {author} {\bibfnamefont {J.}~\bibnamefont {{Wang}}}, \bibinfo {author}
  {\bibfnamefont {M.}~\bibnamefont {{Vogelsberger}}}, \bibinfo {author}
  {\bibfnamefont {S.~D.~M.}\ \bibnamefont {{White}}}, \bibinfo {author}
  {\bibfnamefont {A.}~\bibnamefont {{Jenkins}}}, \bibinfo {author}
  {\bibfnamefont {C.~S.}\ \bibnamefont {{Frenk}}}, \ and\ \bibinfo {author}
  {\bibfnamefont {A.}~\bibnamefont {{Helmi}}},\ }\href {\doibase
  10.1111/j.1365-2966.2009.15878.x} {\bibfield  {journal} {\bibinfo  {journal}
  {\mnras}\ }\textbf {\bibinfo {volume} {402}},\ \bibinfo {pages} {21}
  (\bibinfo {year} {2010})},\ \Eprint {http://arxiv.org/abs/0810.1522}
  {arXiv:0810.1522} \BibitemShut {NoStop}%
\bibitem [{\citenamefont {{Pontzen}}\ and\ \citenamefont
  {{Governato}}(2012)}]{PontzenEtAl2012}%
  \BibitemOpen
  \bibfield  {author} {\bibinfo {author} {\bibfnamefont {A.}~\bibnamefont
  {{Pontzen}}}\ and\ \bibinfo {author} {\bibfnamefont {F.}~\bibnamefont
  {{Governato}}},\ }\href {\doibase 10.1111/j.1365-2966.2012.20571.x}
  {\bibfield  {journal} {\bibinfo  {journal} {\mnras}\ }\textbf {\bibinfo
  {volume} {421}},\ \bibinfo {pages} {3464} (\bibinfo {year} {2012})},\ \Eprint
  {http://arxiv.org/abs/1106.0499} {arXiv:1106.0499 [astro-ph.CO]} \BibitemShut
  {NoStop}%
\bibitem [{\citenamefont {{Governato}}\ \emph {et~al.}(2012)\citenamefont
  {{Governato}}, \citenamefont {{Zolotov}}, \citenamefont {{Pontzen}},
  \citenamefont {{Christensen}}, \citenamefont {{Oh}}, \citenamefont
  {{Brooks}}, \citenamefont {{Quinn}}, \citenamefont {{Shen}},\ and\
  \citenamefont {{Wadsley}}}]{GovernatoEtAl2012}%
  \BibitemOpen
  \bibfield  {author} {\bibinfo {author} {\bibfnamefont {F.}~\bibnamefont
  {{Governato}}}, \bibinfo {author} {\bibfnamefont {A.}~\bibnamefont
  {{Zolotov}}}, \bibinfo {author} {\bibfnamefont {A.}~\bibnamefont
  {{Pontzen}}}, \bibinfo {author} {\bibfnamefont {C.}~\bibnamefont
  {{Christensen}}}, \bibinfo {author} {\bibfnamefont {S.~H.}\ \bibnamefont
  {{Oh}}}, \bibinfo {author} {\bibfnamefont {A.~M.}\ \bibnamefont {{Brooks}}},
  \bibinfo {author} {\bibfnamefont {T.}~\bibnamefont {{Quinn}}}, \bibinfo
  {author} {\bibfnamefont {S.}~\bibnamefont {{Shen}}}, \ and\ \bibinfo {author}
  {\bibfnamefont {J.}~\bibnamefont {{Wadsley}}},\ }\href {\doibase
  10.1111/j.1365-2966.2012.20696.x} {\bibfield  {journal} {\bibinfo  {journal}
  {\mnras}\ }\textbf {\bibinfo {volume} {422}},\ \bibinfo {pages} {1231}
  (\bibinfo {year} {2012})},\ \Eprint {http://arxiv.org/abs/1202.0554}
  {arXiv:1202.0554} \BibitemShut {NoStop}%
\bibitem [{\citenamefont {{Di Cintio}}\ \emph {et~al.}(2014)\citenamefont {{Di
  Cintio}}, \citenamefont {{Brook}}, \citenamefont {{Macci{\`o}}},
  \citenamefont {{Stinson}}, \citenamefont {{Knebe}}, \citenamefont
  {{Dutton}},\ and\ \citenamefont {{Wadsley}}}]{DiCintioEtAl2014a}%
  \BibitemOpen
  \bibfield  {author} {\bibinfo {author} {\bibfnamefont {A.}~\bibnamefont {{Di
  Cintio}}}, \bibinfo {author} {\bibfnamefont {C.~B.}\ \bibnamefont {{Brook}}},
  \bibinfo {author} {\bibfnamefont {A.~V.}\ \bibnamefont {{Macci{\`o}}}},
  \bibinfo {author} {\bibfnamefont {G.~S.}\ \bibnamefont {{Stinson}}}, \bibinfo
  {author} {\bibfnamefont {A.}~\bibnamefont {{Knebe}}}, \bibinfo {author}
  {\bibfnamefont {A.~A.}\ \bibnamefont {{Dutton}}}, \ and\ \bibinfo {author}
  {\bibfnamefont {J.}~\bibnamefont {{Wadsley}}},\ }\href {\doibase
  10.1093/mnras/stt1891} {\bibfield  {journal} {\bibinfo  {journal} {\mnras}\
  }\textbf {\bibinfo {volume} {437}},\ \bibinfo {pages} {415} (\bibinfo {year}
  {2014})},\ \Eprint {http://arxiv.org/abs/1306.0898} {arXiv:1306.0898
  [astro-ph.CO]} \BibitemShut {NoStop}%
\bibitem [{\citenamefont {Wegg}\ \emph {et~al.}(2019)\citenamefont {Wegg},
  \citenamefont {Gerhard},\ and\ \citenamefont {Bieth}}]{WeggEtAl2019}%
  \BibitemOpen
  \bibfield  {author} {\bibinfo {author} {\bibfnamefont {C.}~\bibnamefont
  {Wegg}}, \bibinfo {author} {\bibfnamefont {O.}~\bibnamefont {Gerhard}}, \
  and\ \bibinfo {author} {\bibfnamefont {M.}~\bibnamefont {Bieth}},\ }\href
  {\doibase 10.1093/mnras/stz572} {\bibfield  {journal} {\bibinfo  {journal}
  {\mnras}\ }\textbf {\bibinfo {volume} {485}},\ \bibinfo {pages} {3296}
  (\bibinfo {year} {2019})},\ \Eprint {http://arxiv.org/abs/1806.09635}
  {arXiv:1806.09635} \BibitemShut {NoStop}%
\bibitem [{\citenamefont {Cautun}\ \emph {et~al.}(2020)\citenamefont {Cautun},
  \citenamefont {Ben{\'\i}tez-Llambay}, \citenamefont {Deason}, \citenamefont
  {Frenk}, \citenamefont {Fattahi}, \citenamefont {G{\'o}mez}, \citenamefont
  {Grand}, \citenamefont {Oman}, \citenamefont {Navarro},\ and\ \citenamefont
  {Simpson}}]{CautunEtAl2020}%
  \BibitemOpen
  \bibfield  {author} {\bibinfo {author} {\bibfnamefont {M.}~\bibnamefont
  {Cautun}}, \bibinfo {author} {\bibfnamefont {A.}~\bibnamefont
  {Ben{\'\i}tez-Llambay}}, \bibinfo {author} {\bibfnamefont {A.~J.}\
  \bibnamefont {Deason}}, \bibinfo {author} {\bibfnamefont {C.~S.}\
  \bibnamefont {Frenk}}, \bibinfo {author} {\bibfnamefont {A.}~\bibnamefont
  {Fattahi}}, \bibinfo {author} {\bibfnamefont {F.~A.}\ \bibnamefont
  {G{\'o}mez}}, \bibinfo {author} {\bibfnamefont {R.~J.~J.}\ \bibnamefont
  {Grand}}, \bibinfo {author} {\bibfnamefont {K.~A.}\ \bibnamefont {Oman}},
  \bibinfo {author} {\bibfnamefont {J.~F.}\ \bibnamefont {Navarro}}, \ and\
  \bibinfo {author} {\bibfnamefont {C.~M.}\ \bibnamefont {Simpson}},\ }\href
  {\doibase 10.1093/mnras/staa1017} {\bibfield  {journal} {\bibinfo  {journal}
  {\mnras}\ }\textbf {\bibinfo {volume} {494}},\ \bibinfo {pages} {4291}
  (\bibinfo {year} {2020})},\ \Eprint {http://arxiv.org/abs/1911.04557}
  {arXiv:1911.04557 [astro-ph.GA]} \BibitemShut {NoStop}%
\bibitem [{\citenamefont {{Blanchet}}\ and\ \citenamefont {{Lavalle}}(2012)}]{GaiaCollab2018}%
  \BibitemOpen
  \bibfield  {author} {\bibinfo {author} {\bibnamefont {{The Gaia
  Collaboration}}},\ }\href {\doibase 10.1051/0004-6361/201833051} {\bibfield
  {journal} {\bibinfo  {journal} {\aap}\ }\textbf {\bibinfo {volume} {616}},\
  \bibinfo {eid} {A1} (\bibinfo {year} {2018})},\ \Eprint
  {http://arxiv.org/abs/1804.09365} {arXiv:1804.09365} \BibitemShut {NoStop}%
\bibitem [{\citenamefont {Eilers}\ \emph {et~al.}(2019)\citenamefont {Eilers},
  \citenamefont {Hogg}, \citenamefont {Rix},\ and\ \citenamefont
  {Ness}}]{EilersEtAl2019}%
  \BibitemOpen
  \bibfield  {author} {\bibinfo {author} {\bibfnamefont {A.-C.}\ \bibnamefont
  {Eilers}}, \bibinfo {author} {\bibfnamefont {D.~W.}\ \bibnamefont {Hogg}},
  \bibinfo {author} {\bibfnamefont {H.-W.}\ \bibnamefont {Rix}}, \ and\
  \bibinfo {author} {\bibfnamefont {M.~K.}\ \bibnamefont {Ness}},\ }\href
  {\doibase 10.3847/1538-4357/aaf648} {\bibfield  {journal} {\bibinfo
  {journal} {\apj}\ }\textbf {\bibinfo {volume} {871}},\ \bibinfo {eid} {120}
  (\bibinfo {year} {2019})},\ \Eprint {http://arxiv.org/abs/1810.09466}
  {arXiv:1810.09466 [astro-ph.GA]} \BibitemShut {NoStop}%
\bibitem [{\citenamefont {Hogg}\ \emph {et~al.}(2019)\citenamefont {Hogg},
  \citenamefont {Eilers},\ and\ \citenamefont {Rix}}]{HoggEtAl2019}%
  \BibitemOpen
  \bibfield  {author} {\bibinfo {author} {\bibfnamefont {D.~W.}\ \bibnamefont
  {Hogg}}, \bibinfo {author} {\bibfnamefont {A.-C.}\ \bibnamefont {Eilers}}, \
  and\ \bibinfo {author} {\bibfnamefont {H.-W.}\ \bibnamefont {Rix}},\ }\href
  {\doibase 10.3847/1538-3881/ab398c} {\bibfield  {journal} {\bibinfo
  {journal} {\aj}\ }\textbf {\bibinfo {volume} {158}},\ \bibinfo {eid} {147}
  (\bibinfo {year} {2019})},\ \Eprint {http://arxiv.org/abs/1810.09468}
  {arXiv:1810.09468 [astro-ph.GA]} \BibitemShut {NoStop}%
\bibitem [{\citenamefont {{D'Onghia}}\ \emph {et~al.}(2010)\citenamefont
  {{D'Onghia}}, \citenamefont {{Springel}}, \citenamefont {{Hernquist}},\ and\
  \citenamefont {{Keres}}}]{DOnghiaEtAl2010}%
  \BibitemOpen
  \bibfield  {author} {\bibinfo {author} {\bibfnamefont {E.}~\bibnamefont
  {{D'Onghia}}}, \bibinfo {author} {\bibfnamefont {V.}~\bibnamefont
  {{Springel}}}, \bibinfo {author} {\bibfnamefont {L.}~\bibnamefont
  {{Hernquist}}}, \ and\ \bibinfo {author} {\bibfnamefont {D.}~\bibnamefont
  {{Keres}}},\ }\href {\doibase 10.1088/0004-637X/709/2/1138} {\bibfield
  {journal} {\bibinfo  {journal} {\apj}\ }\textbf {\bibinfo {volume} {709}},\
  \bibinfo {pages} {1138} (\bibinfo {year} {2010})},\ \Eprint
  {http://arxiv.org/abs/0907.3482} {arXiv:0907.3482 [astro-ph.CO]} \BibitemShut
  {NoStop}%
\bibitem [{\citenamefont {{Zhu}}\ \emph {et~al.}(2016)\citenamefont {{Zhu}},
  \citenamefont {{Marinacci}}, \citenamefont {{Maji}}, \citenamefont {{Li}},
  \citenamefont {{Springel}},\ and\ \citenamefont {{Hernquist}}}]{ZhuEtAl2016}%
  \BibitemOpen
  \bibfield  {author} {\bibinfo {author} {\bibfnamefont {Q.}~\bibnamefont
  {{Zhu}}}, \bibinfo {author} {\bibfnamefont {F.}~\bibnamefont {{Marinacci}}},
  \bibinfo {author} {\bibfnamefont {M.}~\bibnamefont {{Maji}}}, \bibinfo
  {author} {\bibfnamefont {Y.}~\bibnamefont {{Li}}}, \bibinfo {author}
  {\bibfnamefont {V.}~\bibnamefont {{Springel}}}, \ and\ \bibinfo {author}
  {\bibfnamefont {L.}~\bibnamefont {{Hernquist}}},\ }\href {\doibase
  10.1093/mnras/stw374} {\bibfield  {journal} {\bibinfo  {journal} {\mnras}\
  }\textbf {\bibinfo {volume} {458}},\ \bibinfo {pages} {1559} (\bibinfo {year}
  {2016})},\ \Eprint {http://arxiv.org/abs/1506.05537} {arXiv:1506.05537}
  \BibitemShut {NoStop}%
\bibitem [{\citenamefont {{Molin{\'e}}}\ \emph {et~al.}(2017)\citenamefont
  {{Molin{\'e}}}, \citenamefont {{S{\'a}nchez-Conde}}, \citenamefont
  {{Palomares-Ruiz}},\ and\ \citenamefont {{Prada}}}]{MolineEtAl2017}%
  \BibitemOpen
  \bibfield  {author} {\bibinfo {author} {\bibfnamefont {{\'A}.}~\bibnamefont
  {{Molin{\'e}}}}, \bibinfo {author} {\bibfnamefont {M.~A.}\ \bibnamefont
  {{S{\'a}nchez-Conde}}}, \bibinfo {author} {\bibfnamefont {S.}~\bibnamefont
  {{Palomares-Ruiz}}}, \ and\ \bibinfo {author} {\bibfnamefont
  {F.}~\bibnamefont {{Prada}}},\ }\href {\doibase 10.1093/mnras/stx026}
  {\bibfield  {journal} {\bibinfo  {journal} {\mnras}\ }\textbf {\bibinfo
  {volume} {466}},\ \bibinfo {pages} {4974} (\bibinfo {year} {2017})},\ \Eprint
  {http://arxiv.org/abs/1603.04057} {arXiv:1603.04057} \BibitemShut {NoStop}%
\bibitem [{\citenamefont {{S{\'a}nchez-Conde}}\ and\ \citenamefont
  {{Prada}}(2014)}]{Sanchez-CondeEtAl2014}%
  \BibitemOpen
  \bibfield  {author} {\bibinfo {author} {\bibfnamefont {M.~A.}\ \bibnamefont
  {{S{\'a}nchez-Conde}}}\ and\ \bibinfo {author} {\bibfnamefont
  {F.}~\bibnamefont {{Prada}}},\ }\href {\doibase 10.1093/mnras/stu1014}
  {\bibfield  {journal} {\bibinfo  {journal} {\mnras}\ }\textbf {\bibinfo
  {volume} {442}},\ \bibinfo {pages} {2271} (\bibinfo {year} {2014})},\ \Eprint
  {http://arxiv.org/abs/1312.1729} {arXiv:1312.1729} \BibitemShut {NoStop}%
\bibitem [{\citenamefont {{Hayashi}}\ \emph {et~al.}(2003)\citenamefont
  {{Hayashi}}, \citenamefont {{Navarro}}, \citenamefont {{Taylor}},
  \citenamefont {{Stadel}},\ and\ \citenamefont {{Quinn}}}]{HayashiEtAl2003}%
  \BibitemOpen
  \bibfield  {author} {\bibinfo {author} {\bibfnamefont {E.}~\bibnamefont
  {{Hayashi}}}, \bibinfo {author} {\bibfnamefont {J.~F.}\ \bibnamefont
  {{Navarro}}}, \bibinfo {author} {\bibfnamefont {J.~E.}\ \bibnamefont
  {{Taylor}}}, \bibinfo {author} {\bibfnamefont {J.}~\bibnamefont {{Stadel}}},
  \ and\ \bibinfo {author} {\bibfnamefont {T.}~\bibnamefont {{Quinn}}},\ }\href
  {\doibase 10.1086/345788} {\bibfield  {journal} {\bibinfo  {journal} {\apj}\
  }\textbf {\bibinfo {volume} {584}},\ \bibinfo {pages} {541} (\bibinfo {year}
  {2003})},\ \Eprint {http://arxiv.org/abs/astro-ph/0203004} {astro-ph/0203004}
  \BibitemShut {NoStop}%
\bibitem [{\citenamefont {van~den Bosch}\ and\ \citenamefont
  {Ogiya}(2018)}]{BoschEtAl2018}%
  \BibitemOpen
  \bibfield  {author} {\bibinfo {author} {\bibfnamefont {F.~C.}\ \bibnamefont
  {van~den Bosch}}\ and\ \bibinfo {author} {\bibfnamefont {G.}~\bibnamefont
  {Ogiya}},\ }\href {\doibase 10.1093/mnras/sty084} {\bibfield  {journal}
  {\bibinfo  {journal} {\mnras}\ }\textbf {\bibinfo {volume} {475}},\ \bibinfo
  {pages} {4066} (\bibinfo {year} {2018})},\ \Eprint
  {http://arxiv.org/abs/1801.05427} {arXiv:1801.05427} \BibitemShut {NoStop}%
\bibitem [{\citenamefont {{Weinberg}}(1994)}]{Weinberg1994}%
  \BibitemOpen
  \bibfield  {author} {\bibinfo {author} {\bibfnamefont {M.~D.}\ \bibnamefont
  {{Weinberg}}},\ }\href {\doibase 10.1086/117161} {\bibfield  {journal}
  {\bibinfo  {journal} {\aj}\ }\textbf {\bibinfo {volume} {108}},\ \bibinfo
  {pages} {1398} (\bibinfo {year} {1994})},\ \Eprint
  {http://arxiv.org/abs/astro-ph/9404015} {astro-ph/9404015} \BibitemShut
  {NoStop}%
\bibitem [{\citenamefont {{Gnedin}}\ and\ \citenamefont
  {{Ostriker}}(1999)}]{GnedinEtAl1999b}%
  \BibitemOpen
  \bibfield  {author} {\bibinfo {author} {\bibfnamefont {O.~Y.}\ \bibnamefont
  {{Gnedin}}}\ and\ \bibinfo {author} {\bibfnamefont {J.~P.}\ \bibnamefont
  {{Ostriker}}},\ }\href {\doibase 10.1086/306864} {\bibfield  {journal}
  {\bibinfo  {journal} {\apj}\ }\textbf {\bibinfo {volume} {513}},\ \bibinfo
  {pages} {626} (\bibinfo {year} {1999})},\ \Eprint
  {http://arxiv.org/abs/astro-ph/9902326} {astro-ph/9902326} \BibitemShut
  {NoStop}%
\bibitem [{\citenamefont {Pe{\~n}arrubia}\ \emph {et~al.}(2010)\citenamefont
  {Pe{\~n}arrubia}, \citenamefont {Benson}, \citenamefont {Walker},
  \citenamefont {Gilmore}, \citenamefont {McConnachie},\ and\ \citenamefont
  {Mayer}}]{PenarrubiaEtAl2010}%
  \BibitemOpen
  \bibfield  {author} {\bibinfo {author} {\bibfnamefont {J.}~\bibnamefont
  {Pe{\~n}arrubia}}, \bibinfo {author} {\bibfnamefont {A.~J.}\ \bibnamefont
  {Benson}}, \bibinfo {author} {\bibfnamefont {M.~G.}\ \bibnamefont {Walker}},
  \bibinfo {author} {\bibfnamefont {G.}~\bibnamefont {Gilmore}}, \bibinfo
  {author} {\bibfnamefont {A.~W.}\ \bibnamefont {McConnachie}}, \ and\ \bibinfo
  {author} {\bibfnamefont {L.}~\bibnamefont {Mayer}},\ }\href {\doibase
  10.1111/j.1365-2966.2010.16762.x} {\bibfield  {journal} {\bibinfo  {journal}
  {\mnras}\ }\textbf {\bibinfo {volume} {406}},\ \bibinfo {pages} {1290}
  (\bibinfo {year} {2010})},\ \Eprint {http://arxiv.org/abs/1002.3376}
  {arXiv:1002.3376} \BibitemShut {NoStop}%
\bibitem [{\citenamefont {Drakos}\ \emph {et~al.}(2017)\citenamefont {Drakos},
  \citenamefont {Taylor},\ and\ \citenamefont {Benson}}]{DrakosEtAl2017}%
  \BibitemOpen
  \bibfield  {author} {\bibinfo {author} {\bibfnamefont {N.~E.}\ \bibnamefont
  {Drakos}}, \bibinfo {author} {\bibfnamefont {J.~E.}\ \bibnamefont {Taylor}},
  \ and\ \bibinfo {author} {\bibfnamefont {A.~J.}\ \bibnamefont {Benson}},\
  }\href {\doibase 10.1093/mnras/stx652} {\bibfield  {journal} {\bibinfo
  {journal} {\mnras}\ }\textbf {\bibinfo {volume} {468}},\ \bibinfo {pages}
  {2345} (\bibinfo {year} {2017})},\ \Eprint {http://arxiv.org/abs/1703.07836}
  {arXiv:1703.07836} \BibitemShut {NoStop}%
\bibitem [{\citenamefont {{Sheth}}\ \emph {et~al.}(2001)\citenamefont
  {{Sheth}}, \citenamefont {{Mo}},\ and\ \citenamefont
  {{Tormen}}}]{ShethEtAl2001}%
  \BibitemOpen
  \bibfield  {author} {\bibinfo {author} {\bibfnamefont {R.~K.}\ \bibnamefont
  {{Sheth}}}, \bibinfo {author} {\bibfnamefont {H.~J.}\ \bibnamefont {{Mo}}}, \
  and\ \bibinfo {author} {\bibfnamefont {G.}~\bibnamefont {{Tormen}}},\ }\href
  {\doibase 10.1046/j.1365-8711.2001.04006.x} {\bibfield  {journal} {\bibinfo
  {journal} {\mnras}\ }\textbf {\bibinfo {volume} {323}},\ \bibinfo {pages} {1}
  (\bibinfo {year} {2001})},\ \Eprint {http://arxiv.org/abs/astro-ph/9907024}
  {astro-ph/9907024} \BibitemShut {NoStop}%
\bibitem [{\citenamefont {{Zentner}}(2007)}]{Zentner2007}%
  \BibitemOpen
  \bibfield  {author} {\bibinfo {author} {\bibfnamefont {A.~R.}\ \bibnamefont
  {{Zentner}}},\ }\href {\doibase 10.1142/S0218271807010511} {\bibfield
  {journal} {\bibinfo  {journal} {International Journal of Modern Physics D}\
  }\textbf {\bibinfo {volume} {16}},\ \bibinfo {pages} {763} (\bibinfo {year}
  {2007})},\ \Eprint {http://arxiv.org/abs/astro-ph/0611454} {astro-ph/0611454}
  \BibitemShut {NoStop}%
\bibitem [{\citenamefont {{Blanchet}}\ and\ \citenamefont {{Lavalle}}(2012)}]{PlanckCollab2020}%
  \BibitemOpen
  \bibfield  {author} {\bibinfo {author} {\bibnamefont {{The Planck
  Collaboration}}},\ }\href
  {\doibase 10.1051/0004-6361/201833910} {\bibfield  {journal} {\bibinfo
  {journal} {\aap}\ }\textbf {\bibinfo {volume} {641}},\ \bibinfo {eid} {A6}
  (\bibinfo {year} {2020})},\ \Eprint {http://arxiv.org/abs/1807.06209}
  {arXiv:1807.06209 [astro-ph.CO]} \BibitemShut {NoStop}%
\bibitem [{\citenamefont {Parkinson}\ \emph {et~al.}(2008)\citenamefont
  {Parkinson}, \citenamefont {Cole},\ and\ \citenamefont
  {Helly}}]{ParkinsonEtAl2008}%
  \BibitemOpen
  \bibfield  {author} {\bibinfo {author} {\bibfnamefont {H.}~\bibnamefont
  {Parkinson}}, \bibinfo {author} {\bibfnamefont {S.}~\bibnamefont {Cole}}, \
  and\ \bibinfo {author} {\bibfnamefont {J.}~\bibnamefont {Helly}},\ }\href
  {\doibase 10.1111/j.1365-2966.2007.12517.x} {\bibfield  {journal} {\bibinfo
  {journal} {\mnras}\ }\textbf {\bibinfo {volume} {383}},\ \bibinfo {pages}
  {557} (\bibinfo {year} {2008})},\ \Eprint {http://arxiv.org/abs/0708.1382}
  {arXiv:0708.1382} \BibitemShut {NoStop}%
\bibitem [{\citenamefont {{Bergstr{\"o}m}}\ \emph {et~al.}(1998)\citenamefont
  {{Bergstr{\"o}m}}, \citenamefont {{Ullio}},\ and\ \citenamefont
  {{Buckley}}}]{BergstroemEtAl1998a}%
  \BibitemOpen
  \bibfield  {author} {\bibinfo {author} {\bibfnamefont {L.}~\bibnamefont
  {{Bergstr{\"o}m}}}, \bibinfo {author} {\bibfnamefont {P.}~\bibnamefont
  {{Ullio}}}, \ and\ \bibinfo {author} {\bibfnamefont {J.~H.}\ \bibnamefont
  {{Buckley}}},\ }\href {\doibase 10.1016/S0927-6505(98)00015-2} {\bibfield
  {journal} {\bibinfo  {journal} {Astroparticle Physics}\ }\textbf {\bibinfo
  {volume} {9}},\ \bibinfo {pages} {137} (\bibinfo {year} {1998})},\ \Eprint
  {http://arxiv.org/abs/astro-ph/9712318} {astro-ph/9712318} \BibitemShut
  {NoStop}%
\bibitem [{\citenamefont {{Buckley}}\ and\ \citenamefont
  {{Hooper}}(2010)}]{BuckleyEtAl2010}%
  \BibitemOpen
  \bibfield  {author} {\bibinfo {author} {\bibfnamefont {M.~R.}\ \bibnamefont
  {{Buckley}}}\ and\ \bibinfo {author} {\bibfnamefont {D.}~\bibnamefont
  {{Hooper}}},\ }\href {\doibase 10.1103/PhysRevD.82.063501} {\bibfield
  {journal} {\bibinfo  {journal} {\prd}\ }\textbf {\bibinfo {volume} {82}},\
  \bibinfo {eid} {063501} (\bibinfo {year} {2010})},\ \Eprint
  {http://arxiv.org/abs/1004.1644} {arXiv:1004.1644 [hep-ph]} \BibitemShut
  {NoStop}%
\bibitem [{\citenamefont {{Blanchet}}\ and\ \citenamefont {{Lavalle}}(2012)}]{FermiLATEtAl2009}%
  \BibitemOpen
  \bibfield  {author} {\bibinfo {author} {\bibnamefont {{The Fermi-LAT
  Collaboration}}},\ }\href {\doibase
  10.1088/0004-637X/697/2/1071} {\bibfield  {journal} {\bibinfo  {journal}
  {\apj}\ }\textbf {\bibinfo {volume} {697}},\ \bibinfo {pages} {1071}
  (\bibinfo {year} {2009})},\ \Eprint {http://arxiv.org/abs/0902.1089}
  {arXiv:0902.1089 [astro-ph.IM]} \BibitemShut {NoStop}%
\bibitem [{\citenamefont {{Ackermann}}\ \emph {et~al.}(2012)}]{AckermannEtAl2012}%
  \BibitemOpen
  \bibfield  {author} {\bibinfo {author} {\bibfnamefont {{The Fermi-LAT Collaboration}}},\ }\href
  {\doibase 10.1088/0004-637X/750/1/3} {\bibfield  {journal} {\bibinfo
  {journal} {\apj}\ }\textbf {\bibinfo {volume} {750}},\ \bibinfo {eid} {3}
  (\bibinfo {year} {2012})},\ \Eprint {http://arxiv.org/abs/1202.4039}
  {arXiv:1202.4039 [astro-ph.HE]} \BibitemShut {NoStop}%
\bibitem [{\citenamefont {{Cirelli}}\ \emph {et~al.}(2011)\citenamefont
  {{Cirelli}}, \citenamefont {{Corcella}}, \citenamefont {{Hektor}},
  \citenamefont {{H{\"u}tsi}}, \citenamefont {{Kadastik}}, \citenamefont
  {{Panci}}, \citenamefont {{Raidal}}, \citenamefont {{Sala}},\ and\
  \citenamefont {{Strumia}}}]{CirelliEtAl2011}%
  \BibitemOpen
  \bibfield  {author} {\bibinfo {author} {\bibfnamefont {M.}~\bibnamefont
  {{Cirelli}}}, \bibinfo {author} {\bibfnamefont {G.}~\bibnamefont
  {{Corcella}}}, \bibinfo {author} {\bibfnamefont {A.}~\bibnamefont
  {{Hektor}}}, \bibinfo {author} {\bibfnamefont {G.}~\bibnamefont
  {{H{\"u}tsi}}}, \bibinfo {author} {\bibfnamefont {M.}~\bibnamefont
  {{Kadastik}}}, \bibinfo {author} {\bibfnamefont {P.}~\bibnamefont {{Panci}}},
  \bibinfo {author} {\bibfnamefont {M.}~\bibnamefont {{Raidal}}}, \bibinfo
  {author} {\bibfnamefont {F.}~\bibnamefont {{Sala}}}, \ and\ \bibinfo {author}
  {\bibfnamefont {A.}~\bibnamefont {{Strumia}}},\ }\href {\doibase
  10.1088/1475-7516/2011/03/051} {\bibfield  {journal} {\bibinfo  {journal}
  {\jcap}\ }\textbf {\bibinfo {volume} {3}},\ \bibinfo {eid} {051} (\bibinfo
  {year} {2011})},\ \Eprint {http://arxiv.org/abs/1012.4515} {arXiv:1012.4515
  [hep-ph]} \BibitemShut {NoStop}%
\bibitem [{\citenamefont {Atwood}\ \emph {et~al.}(2013)\citenamefont {Atwood},
  \citenamefont {Albert}, \citenamefont {Baldini}, \citenamefont {Tinivella},
  \citenamefont {Bregeon}, \citenamefont {Pesce-Rollins}, \citenamefont
  {Sgr{\`o}}, \citenamefont {Bruel}, \citenamefont {Charles}, \citenamefont
  {Drlica-Wagner}, \citenamefont {Franckowiak}, \citenamefont {Jogler},
  \citenamefont {Rochester}, \citenamefont {Usher}, \citenamefont {Wood},
  \citenamefont {Cohen-Tanugi},\ and\ \citenamefont {Zimmer}}]{AtwoodEtAl2013}%
  \BibitemOpen
  \bibfield  {author} {\bibinfo {author} {\bibfnamefont {W.}~\bibnamefont
  {Atwood} \ {\em et al.}},\ }\href {https://ui.adsabs.harvard.edu/abs/2013arXiv1303.3514A}
  {\bibfield  {journal} {\bibinfo  {journal} {arXiv e-prints}\ ,\ \bibinfo
  {eid} {arXiv:1303.3514}} (\bibinfo {year} {2013})},\ \Eprint
  {http://arxiv.org/abs/1303.3514} {arXiv:1303.3514 [astro-ph.IM]} \BibitemShut
  {NoStop}%
\bibitem [{\citenamefont {Bruel}\ \emph {et~al.}(2018)\citenamefont {Bruel},
  \citenamefont {Burnett}, \citenamefont {Digel}, \citenamefont {Johannesson},
  \citenamefont {Omodei},\ and\ \citenamefont {Wood}}]{BruelEtAl2018}%
  \BibitemOpen
  \bibfield  {author} {\bibinfo {author} {\bibfnamefont {P.}~\bibnamefont
  {Bruel}}, \bibinfo {author} {\bibfnamefont {T.~H.}\ \bibnamefont {Burnett}},
  \bibinfo {author} {\bibfnamefont {S.~W.}\ \bibnamefont {Digel}}, \bibinfo
  {author} {\bibfnamefont {G.}~\bibnamefont {Johannesson}}, \bibinfo {author}
  {\bibfnamefont {N.}~\bibnamefont {Omodei}}, \ and\ \bibinfo {author}
  {\bibfnamefont {M.}~\bibnamefont {Wood}},\ }\href
  {https://ui.adsabs.harvard.edu/abs/2018arXiv181011394B} {\bibfield  {journal}
  {\bibinfo  {journal} {arXiv e-prints}\ ,\ \bibinfo {eid} {arXiv:1810.11394}}
  (\bibinfo {year} {2018})},\ \Eprint {http://arxiv.org/abs/1810.11394}
  {arXiv:1810.11394 [astro-ph.IM]} \BibitemShut {NoStop}%
\bibitem [{\citenamefont {{Cerde{\~n}o}}\ \emph {et~al.}(2012)\citenamefont
  {{Cerde{\~n}o}}, \citenamefont {{Delahaye}},\ and\ \citenamefont
  {{Lavalle}}}]{CerdenoEtAl2012}%
  \BibitemOpen
  \bibfield  {author} {\bibinfo {author} {\bibfnamefont {D.~G.}\ \bibnamefont
  {{Cerde{\~n}o}}}, \bibinfo {author} {\bibfnamefont {T.}~\bibnamefont
  {{Delahaye}}}, \ and\ \bibinfo {author} {\bibfnamefont {J.}~\bibnamefont
  {{Lavalle}}},\ }\href {\doibase 10.1016/j.nuclphysb.2011.09.020} {\bibfield
  {journal} {\bibinfo  {journal} {Nuclear Physics B}\ }\textbf {\bibinfo
  {volume} {854}},\ \bibinfo {pages} {738} (\bibinfo {year} {2012})},\ \Eprint
  {http://arxiv.org/abs/1108.1128} {arXiv:1108.1128 [hep-ph]} \BibitemShut
  {NoStop}%
\bibitem [{\citenamefont {{Steigman}}\ \emph {et~al.}(2012)\citenamefont
  {{Steigman}}, \citenamefont {{Dasgupta}},\ and\ \citenamefont
  {{Beacom}}}]{SteigmanEtAl2012}%
  \BibitemOpen
  \bibfield  {author} {\bibinfo {author} {\bibfnamefont {G.}~\bibnamefont
  {{Steigman}}}, \bibinfo {author} {\bibfnamefont {B.}~\bibnamefont
  {{Dasgupta}}}, \ and\ \bibinfo {author} {\bibfnamefont {J.~F.}\ \bibnamefont
  {{Beacom}}},\ }\href {\doibase 10.1103/PhysRevD.86.023506} {\bibfield
  {journal} {\bibinfo  {journal} {\prd}\ }\textbf {\bibinfo {volume} {86}},\
  \bibinfo {eid} {023506} (\bibinfo {year} {2012})},\ \Eprint
  {http://arxiv.org/abs/1204.3622} {arXiv:1204.3622 [hep-ph]} \BibitemShut
  {NoStop}%
\bibitem [{\citenamefont {Li}\ and\ \citenamefont {Ma}(1983)}]{LiEtAl1983}%
  \BibitemOpen
  \bibfield  {author} {\bibinfo {author} {\bibfnamefont {T.~P.}\ \bibnamefont
  {Li}}\ and\ \bibinfo {author} {\bibfnamefont {Y.~Q.}\ \bibnamefont {Ma}},\
  }\href {\doibase 10.1086/161295} {\bibfield  {journal} {\bibinfo  {journal}
  {\apj}\ }\textbf {\bibinfo {volume} {272}},\ \bibinfo {pages} {317} (\bibinfo
  {year} {1983})}\BibitemShut {NoStop}%
\bibitem [{\citenamefont {Zhang}\ and\ \citenamefont
  {Ramsden}(1990)}]{ZhangEtAl1990}%
  \BibitemOpen
  \bibfield  {author} {\bibinfo {author} {\bibfnamefont {S.~N.}\ \bibnamefont
  {Zhang}}\ and\ \bibinfo {author} {\bibfnamefont {D.}~\bibnamefont
  {Ramsden}},\ }\href {\doibase 10.1007/BF00462037} {\bibfield  {journal}
  {\bibinfo  {journal} {Experimental Astronomy}\ }\textbf {\bibinfo {volume}
  {1}},\ \bibinfo {pages} {145} (\bibinfo {year} {1990})}\BibitemShut {NoStop}%
\bibitem [{\citenamefont {Mattox}\ \emph {et~al.}(1996)\citenamefont {Mattox},
  \citenamefont {Bertsch}, \citenamefont {Chiang}, \citenamefont {Dingus},
  \citenamefont {Digel}, \citenamefont {Esposito}, \citenamefont {Fierro},
  \citenamefont {Hartman}, \citenamefont {Hunter}, \citenamefont {Kanbach},
  \citenamefont {Kniffen}, \citenamefont {Lin}, \citenamefont {Macomb},
  \citenamefont {Mayer-Hasselwander}, \citenamefont {Michelson}, \citenamefont
  {von Montigny}, \citenamefont {Mukherjee}, \citenamefont {Nolan},
  \citenamefont {Ramanamurthy}, \citenamefont {Schneid}, \citenamefont
  {Sreekumar}, \citenamefont {Thompson},\ and\ \citenamefont
  {Willis}}]{MattoxEtAl1996}%
  \BibitemOpen
  \bibfield  {author} {\bibinfo {author} {\bibfnamefont {J.~R.}\ \bibnamefont
  {Mattox}\ {\em et al.}},\ }\href {\doibase
  10.1086/177068} {\bibfield  {journal} {\bibinfo  {journal} {\apj}\ }\textbf
  {\bibinfo {volume} {461}},\ \bibinfo {pages} {396} (\bibinfo {year}
  {1996})}\BibitemShut {NoStop}%
\bibitem [{\citenamefont {{Cousins}}\ \emph {et~al.}(2008)\citenamefont
  {{Cousins}}, \citenamefont {{Linnemann}},\ and\ \citenamefont
  {{Tucker}}}]{CousinsEtAl2008}%
  \BibitemOpen
  \bibfield  {author} {\bibinfo {author} {\bibfnamefont {R.~D.}\ \bibnamefont
  {{Cousins}}}, \bibinfo {author} {\bibfnamefont {J.~T.}\ \bibnamefont
  {{Linnemann}}}, \ and\ \bibinfo {author} {\bibfnamefont {J.}~\bibnamefont
  {{Tucker}}},\ }\href {\doibase 10.1016/j.nima.2008.07.086} {\bibfield
  {journal} {\bibinfo  {journal} {Nuclear Instruments and Methods in Physics
  Research A}\ }\textbf {\bibinfo {volume} {595}},\ \bibinfo {pages} {480}
  (\bibinfo {year} {2008})},\ \Eprint {http://arxiv.org/abs/physics/0702156}
  {physics/0702156} \BibitemShut {NoStop}%
\bibitem [{\citenamefont {{Blanchet}}\ and\ \citenamefont {{Lavalle}}(2012)}]{AckermannEtAl2015c}%
  \BibitemOpen
  \bibfield  {author} {\bibinfo {author} {\bibnamefont {{The Fermi-LAT
  Collaboration}}},\ }\href
  {\doibase 10.1103/PhysRevLett.115.231301} {\bibfield  {journal} {\bibinfo
  {journal} {\prl}\ }\textbf {\bibinfo {volume} {115}},\ \bibinfo {eid}
  {231301} (\bibinfo {year} {2015}{\natexlab{b}})},\ \Eprint
  {http://arxiv.org/abs/1503.02641} {arXiv:1503.02641 [astro-ph.HE]}
  \BibitemShut {NoStop}%
\bibitem [{\citenamefont {{Blanchet}}\ and\ \citenamefont {{Lavalle}}(2012)}]{AlbertEtAl2017}%
  \BibitemOpen
  \bibfield  {author} {\bibinfo {author}
    {\bibfnamefont {The Fermi-LAT Collaboration}},\ }\href {\doibase
  10.3847/1538-4357/834/2/110} {\bibfield  {journal} {\bibinfo  {journal}
  {\apj}\ }\textbf {\bibinfo {volume} {834}},\ \bibinfo {eid} {110} (\bibinfo
  {year} {2017})},\ \Eprint {http://arxiv.org/abs/1611.03184}
  {arXiv:1611.03184} \BibitemShut {NoStop}%
\bibitem [{\citenamefont {Wilks}(1938)}]{Wilks1938}%
  \BibitemOpen
  \bibfield  {author} {\bibinfo {author} {\bibfnamefont {S.~S.}\ \bibnamefont
  {Wilks}},\ }\href {\doibase 10.1214/aoms/1177732360} {\bibfield  {journal}
  {\bibinfo  {journal} {Ann. Math. Statist.}\ }\textbf {\bibinfo {volume}
  {9}},\ \bibinfo {pages} {60} (\bibinfo {year} {1938})}\BibitemShut {NoStop}%
\bibitem [{\citenamefont {Wilks}(1962)}]{Wilks1962}%
  \BibitemOpen
  \bibfield  {author} {\bibinfo {author} {\bibfnamefont {S.~S.}\ \bibnamefont
  {Wilks}},\ }\href@noop {} {\emph {\bibinfo {title} {Mathematical
  Statistics}}}\ (\bibinfo  {publisher} {Wiley},\ \bibinfo {address} {New York,
  NY},\ \bibinfo {year} {1962})\BibitemShut {NoStop}%
\bibitem [{\citenamefont {Cowan}\ \emph {et~al.}(2011)\citenamefont {Cowan},
  \citenamefont {Cranmer}, \citenamefont {Gross},\ and\ \citenamefont
  {Vitells}}]{CowanEtAl2011}%
  \BibitemOpen
  \bibfield  {author} {\bibinfo {author} {\bibfnamefont {G.}~\bibnamefont
  {Cowan}}, \bibinfo {author} {\bibfnamefont {K.}~\bibnamefont {Cranmer}},
  \bibinfo {author} {\bibfnamefont {E.}~\bibnamefont {Gross}}, \ and\ \bibinfo
  {author} {\bibfnamefont {O.}~\bibnamefont {Vitells}},\ }\href {\doibase
  10.1140/epjc/s10052-011-1554-0} {\bibfield  {journal} {\bibinfo  {journal}
  {European Physical Journal C}\ }\textbf {\bibinfo {volume} {71}},\ \bibinfo
  {eid} {1554} (\bibinfo {year} {2011})},\ \Eprint
  {http://arxiv.org/abs/1007.1727} {arXiv:1007.1727 [physics.data-an]}
  \BibitemShut {NoStop}%
\bibitem [{\citenamefont {{Macci{\`o}}}\ \emph {et~al.}(2008)\citenamefont
  {{Macci{\`o}}}, \citenamefont {{Dutton}},\ and\ \citenamefont {{van den
  Bosch}}}]{MaccioEtAl2008}%
  \BibitemOpen
  \bibfield  {author} {\bibinfo {author} {\bibfnamefont {A.~V.}\ \bibnamefont
  {{Macci{\`o}}}}, \bibinfo {author} {\bibfnamefont {A.~A.}\ \bibnamefont
  {{Dutton}}}, \ and\ \bibinfo {author} {\bibfnamefont {F.~C.}\ \bibnamefont
  {{van den Bosch}}},\ }\href {\doibase 10.1111/j.1365-2966.2008.14029.x}
  {\bibfield  {journal} {\bibinfo  {journal} {\mnras}\ }\textbf {\bibinfo
  {volume} {391}},\ \bibinfo {pages} {1940} (\bibinfo {year} {2008})},\ \Eprint
  {http://arxiv.org/abs/0805.1926} {arXiv:0805.1926} \BibitemShut {NoStop}%
\bibitem [{\citenamefont {{Prada}}\ \emph {et~al.}(2012)\citenamefont
  {{Prada}}, \citenamefont {{Klypin}}, \citenamefont {{Cuesta}}, \citenamefont
  {{Betancort-Rijo}},\ and\ \citenamefont {{Primack}}}]{PradaEtAl2012}%
  \BibitemOpen
  \bibfield  {author} {\bibinfo {author} {\bibfnamefont {F.}~\bibnamefont
  {{Prada}}}, \bibinfo {author} {\bibfnamefont {A.~A.}\ \bibnamefont
  {{Klypin}}}, \bibinfo {author} {\bibfnamefont {A.~J.}\ \bibnamefont
  {{Cuesta}}}, \bibinfo {author} {\bibfnamefont {J.~E.}\ \bibnamefont
  {{Betancort-Rijo}}}, \ and\ \bibinfo {author} {\bibfnamefont
  {J.}~\bibnamefont {{Primack}}},\ }\href {\doibase
  10.1111/j.1365-2966.2012.21007.x} {\bibfield  {journal} {\bibinfo  {journal}
  {\mnras}\ }\textbf {\bibinfo {volume} {423}},\ \bibinfo {pages} {3018}
  (\bibinfo {year} {2012})},\ \Eprint {http://arxiv.org/abs/1104.5130}
  {arXiv:1104.5130} \BibitemShut {NoStop}%
\bibitem [{\citenamefont {{Dutton}}\ and\ \citenamefont
  {{Macci{\`o}}}(2014)}]{DuttonEtAl2014}%
  \BibitemOpen
  \bibfield  {author} {\bibinfo {author} {\bibfnamefont {A.~A.}\ \bibnamefont
  {{Dutton}}}\ and\ \bibinfo {author} {\bibfnamefont {A.~V.}\ \bibnamefont
  {{Macci{\`o}}}},\ }\href {\doibase 10.1093/mnras/stu742} {\bibfield
  {journal} {\bibinfo  {journal} {\mnras}\ }\textbf {\bibinfo {volume} {441}},\
  \bibinfo {pages} {3359} (\bibinfo {year} {2014})},\ \Eprint
  {http://arxiv.org/abs/1402.7073} {arXiv:1402.7073} \BibitemShut {NoStop}%
\bibitem [{\citenamefont {S{\'a}nchez-Conde}\ \emph {et~al.}(2011)\citenamefont
  {S{\'a}nchez-Conde}, \citenamefont {Cannoni}, \citenamefont {Zandanel},
  \citenamefont {G{\'o}mez},\ and\ \citenamefont
  {Prada}}]{Sanchez-CondeEtAl2011}%
  \BibitemOpen
  \bibfield  {author} {\bibinfo {author} {\bibfnamefont {M.~A.}\ \bibnamefont
  {S{\'a}nchez-Conde}}, \bibinfo {author} {\bibfnamefont {M.}~\bibnamefont
  {Cannoni}}, \bibinfo {author} {\bibfnamefont {F.}~\bibnamefont {Zandanel}},
  \bibinfo {author} {\bibfnamefont {M.~E.}\ \bibnamefont {G{\'o}mez}}, \ and\
  \bibinfo {author} {\bibfnamefont {F.}~\bibnamefont {Prada}},\ }\href
  {\doibase 10.1088/1475-7516/2011/12/011} {\bibfield  {journal} {\bibinfo
  {journal} {\jcap}\ }\textbf {\bibinfo {volume} {12}},\ \bibinfo {eid} {011}
  (\bibinfo {year} {2011})},\ \Eprint {http://arxiv.org/abs/1104.3530}
  {arXiv:1104.3530 [astro-ph.HE]} \BibitemShut {NoStop}%
\bibitem [{\citenamefont {{Berezinsky}}\ \emph {et~al.}(2013)\citenamefont
  {{Berezinsky}}, \citenamefont {{Dokuchaev}},\ and\ \citenamefont
  {{Eroshenko}}}]{BerezinskyEtAl2013}%
  \BibitemOpen
  \bibfield  {author} {\bibinfo {author} {\bibfnamefont {V.~S.}\ \bibnamefont
  {{Berezinsky}}}, \bibinfo {author} {\bibfnamefont {V.~I.}\ \bibnamefont
  {{Dokuchaev}}}, \ and\ \bibinfo {author} {\bibfnamefont {Y.~N.}\ \bibnamefont
  {{Eroshenko}}},\ }\href {\doibase 10.1088/1475-7516/2013/11/059} {\bibfield
  {journal} {\bibinfo  {journal} {\jcap}\ }\textbf {\bibinfo {volume} {11}},\
  \bibinfo {eid} {059} (\bibinfo {year} {2013})},\ \Eprint
  {http://arxiv.org/abs/1308.6742} {arXiv:1308.6742 [astro-ph.CO]} \BibitemShut
  {NoStop}%
\bibitem [{\citenamefont {Carr}\ \emph {et~al.}(2016)\citenamefont {Carr},
  \citenamefont {K{\"u}hnel},\ and\ \citenamefont {Sandstad}}]{CarrEtAl2016}%
  \BibitemOpen
  \bibfield  {author} {\bibinfo {author} {\bibfnamefont {B.}~\bibnamefont
  {Carr}}, \bibinfo {author} {\bibfnamefont {F.}~\bibnamefont {K{\"u}hnel}}, \
  and\ \bibinfo {author} {\bibfnamefont {M.}~\bibnamefont {Sandstad}},\ }\href
  {\doibase 10.1103/PhysRevD.94.083504} {\bibfield  {journal} {\bibinfo
  {journal} {\prd}\ }\textbf {\bibinfo {volume} {94}},\ \bibinfo {eid} {083504}
  (\bibinfo {year} {2016})},\ \Eprint {http://arxiv.org/abs/1607.06077}
  {arXiv:1607.06077 [astro-ph.CO]} \BibitemShut {NoStop}%
\end{thebibliography}%

\end{document}